\documentclass[12pt]{article}
\usepackage{amsmath}
\usepackage{amsfonts}
\usepackage{graphics}
\usepackage{epsfig}
\usepackage{amssymb}
\usepackage{amscd}
\usepackage{latexsym}
\usepackage{graphicx}
\usepackage{multirow}
\usepackage{geometry}
\usepackage{hyperref}
\usepackage{color}
\usepackage[french,english]{babel}
\usepackage[utf8]{inputenc}

 \newtheorem{theorem}{Theorem}[section]

 \newtheorem{proposition}[theorem]{Proposition}

  \newtheorem{definition}[theorem]{Definition}
 \newtheorem{assumption}[theorem]{Assumption}

 \newtheorem{remark}[theorem]{Remark}
 \numberwithin{equation}{section}
 \numberwithin{table}{section}

\def\to{\rightarrow}
\def\td{\tilde}

\def\Tr{{\rm Tr}}

\def\DD{\mathcal D}
\def\RR{\mathcal R}
\newcommand{\beq}{\begin{equation}}
\newcommand{\eeq}{\end{equation}}
\newcommand{\bea}{\begin{eqnarray}}
\newcommand{\eea}{\end{eqnarray}}
\newcommand{\beqq}{\begin{equation*}}
\newcommand{\eeqq}{\end{equation*}}
\newcommand{\beaa}{\begin{eqnarray*}}
\newcommand{\eeaa}{\end{eqnarray*}}
\newcommand{\proof}[1]{{\noindent \bf proof:}\par {#1} $\square$}

\def\PVI{(P_{\rm VI})}

\newcommand{\Res}{\mathop{\rm Res}}

\begin{document}
% \maketitle
\sloppy

\pagestyle{empty}
\vspace{10pt}
\begin{center}
{\Large \bf {Painlev\'{e} equations, topological type property and \\[+.3em] reconstruction by the topological recursion}}
\end{center}
\vspace{-2pt}

\begin{center}
\textbf{K. Iwaki$^\dagger$, O. Marchal$^\ddagger$, 
A. Saenz$^\star$}\footnote{iwaki@math.nagoya-u.ac.jp, 
olivier.marchal@univ-st-etienne.fr, ais6a@virginia.edu}
\end{center}
\vspace{7pt}

$^\dagger$ 
\textit{Graduate School of Mathematics, Nagoya University, 
Furocho, Nagoya, Japan}

$^\ddagger$ \textit{Universit\'{e} de Lyon, 
CNRS UMR 5208, Universit\'{e} Jean Monnet, 
Institut Camille Jordan, France}

$^\star$\textit{Department of Mathematics, 
University of Virginia, U.S.A.}

\vspace{10pt}

\begin{abstract} In this article we prove that Lax pairs associated with $\hbar$-dependent six Painlev\'e equations satisfy the topological type property proposed by Berg\`ere, Borot and Eynard for any generic choice of the monodromy parameters. Consequently we show that one can reconstruct the formal $\hbar$-expansion of the isomonodromic $\tau$-function and of the determinantal formulas by applying the so-called topological recursion to the spectral curve attached to the Lax pair in all six Painlev\'e cases. Finally we illustrate the former results with the explicit computations of the first orders of the six $\tau$-functions.
\end{abstract}

%%% ----------------------------------------------------------------------
{\small \smallskip
Mathematics Subject Classification (MSC) codes: 34M55, 34M56, 34E20, 60B20

}
\vspace{-1.em}

%%% ----------------------------------------------------------------------
\tableofcontents
\pagestyle{plain}
%%%%%%%%%%%%%%%%%%%%%%%%%%%%%%%%%
%%%%%%%%%%%%%%%%%%%%%%%%%%%%%%%%%
\section{Introduction}

It is now well known that the study of Hermitian random matrices is intrinsically related to integrable systems and Painlev\'e equations. In particular, it is proved that the partition functions of Hermitian matrix models coincide with the $\tau$-functions of certain integrable systems \cite{Iso1,Iso2}. Moreover, it is also well understood that the local correlations between eigenvalues in Hermitian random matrices exhibit universal behaviors when the size of the matrix goes to infinity. For example, the gap probability in the bulk of the distribution of eigenvalues can be directly connected to the Painlev\'e V equation, the so-called Hastings McLeod solution of the Painlev\'e II equation is related to the gap probability at the edge, and many similar results in different locations of the eigenvalues distribution are now available in relation with the other Painlev\'e equations. See \cite{WitteForrester, MehtaBook, TW}  for example.

In a more algebraic perspective, it was realized that the connection between integrable systems and Hermitian matrix models can be understood because of the existence of loop equations (also known as Schwinger-Dyson equations) that can be solved perturbatively (under additional assumptions like convex potentials or genus $0$ spectral curve) via the topological recursion introduced by Eynard and Orantin in \cite{EO}. Since the scope of the topological recursion has been proved to go much beyond matrix models, it is natural to wonder if one could define the equivalence of correlation functions arising in the formalism of the topological recursion directly into the integrable systems formalism. In \cite{BBEnew} and \cite{Deter}, Berg\`ere, Borot and Eynard suggested determinantal formulas that associate to any Lax pair (in fact any finite dimensional linear differential system) a set of correlation functions that satisfy the same loop equations as the one arising in Hermitian matrix models. Consequently at the perturbative level, one may expect these correlation functions to be reconstructed by the topological recursion. However, since loop equations may have many solutions it is not obvious why the functions generated by the topological recursion (that are one set of solutions to the loop equations) should necessarily identify with the determinantal formulas (that are also one set of solutions to the loop equations). 

\begin{table} 
\begin{itemize} 
\item (Painlev\'e I)
\beq \label{P1eq} \hbar^2\ddot{q}=6q^2+t. \eeq
\item (Painlev\'e II)
\beq \label{P2eq} \hbar^2\ddot{q}=2q^3+t q-\theta +\frac{\hbar}{2}. \eeq
\item (Painlev\'e III)
\beq \label{P3eq}\hbar^2\ddot{q}=\frac{\hbar^2}{q}\dot{q}^2-\frac{\hbar^2}{t}\dot{q}+\frac{4}{t}\left(\theta_0 q^2-\theta_\infty+\hbar\right)+4 q^3-\frac{4}{q}. \eeq
\item (Painlev\'e IV)
\beq \label{P4eq}\hbar^2\ddot{q}=\frac{\hbar^2}{2q}\dot{q}^2+2\left(3q^3+4t q^2+\left(t^2-2\theta_\infty+\hbar\right)q-\frac{\theta_0^2}{q}\right). \eeq
\item (Painlev\'e V)
\bea \label{P5eq}\hbar^2\ddot{q}&=&\left(\frac{1}{2q}+\frac{1}{q-1}\right)(\hbar\dot{q})^2-\hbar^2\frac{\dot{q}}{t}+\frac{(q-1)^2}{t^2}\left(\alpha q+\frac{\beta}{q}\right)+\frac{\gamma q}{t}+\frac{\delta q(q+1)}{q-1},
\eea
where 
\[
\alpha=\frac{(\theta_0-\theta_1-\theta_\infty)^2}{8}\,,
\beta=-\frac{(\theta_0-\theta_1+\theta_\infty)^2}{8}\,,
\,\,\gamma=\theta_0+\theta_1-\hbar \text{ and } \delta=-\frac{1}{2}.
\]
\item (Painlev\'e VI)
\bea\label{P6eq}\hbar^2\ddot{q}&=&\frac{\hbar^2}{2}\left(\frac{1}{q}+\frac{1}{q-1}+\frac{1}{q-t}\right)\dot{q}^2-\hbar^2\left(\frac{1}{t}+\frac{1}{t-1}+\frac{1}{q-t}\right)\dot{q}\cr
&&+\frac{q(q-1)(q-t)}{t^2(t-1)^2}\left[\alpha+\beta \frac{t}{q^2}+\gamma \frac{t-1}{(q-1)^2}+\delta \frac{t(t-1)}{(q-t)^2}\right],
\eea
where the parameters are
\beqq \alpha=\frac{1}{2}(\theta_\infty-\hbar)^2\,,\, \beta=-\frac{\theta_0^2}{2}\,,\, \gamma=\frac{\theta_1^2}{2} \text{ and } \delta=\frac{\hbar^2-\theta_t^2}{2}.\eeqq
\end{itemize}
\caption{List of ($\hbar$-dependent) Painlev\'e equations. ($\dot{q} = \frac{dq}{dt}$.)}
\label{table:P-eq}

\end{table}

In \cite{BBEnew} and \cite{Deter} the authors discussed about sufficient conditions on the Lax pair with a small parameter $\hbar$, known as the topological type property, on the Lax pair to prove that both sets are identical. The purpose of this article is to show that the $2\times 2$  Lax pairs associated with the six Painlev\'e equations with generic monodromy parameters satisfies the topological type property (see Definition \ref{def:TT-property-Lax}).  

We introduce the small parameter $\hbar$ in the Lax pairs presented by Jimbo, Miwa and Ueno in \cite{JMI} through a rescaling of the parameters, providing a $\hbar$-dependent version of the Painlev\'e equations in Table \ref{table:P-eq} (whose standard forms can be recovered by setting $\hbar=1$). Note that the Painlev\'e equations with a formal parameter has been discussed in terms of the WKB method in \cite{KT-painleve}. 
We also provide $\hbar$-dependent versions of the Hamiltonian structures underlying the Painlev\'e equations (cf. \cite{Okamoto}). 
Then, we introduce and compute the spectral curve through the semi-classical limit of the Lax matrix. It turns out that the spectral curves arising in all six Lax matrices are of genus $0$, and hence they admit a rational parametrization. Under some genericity assumptions, we verify that the spectral curves are generic so that the topological recursion algorithm can be applied.  Then, our main results are formulated as follows (Theorem \ref{MainTheo}): 
\begin{itemize}
\item
Under some genericity assumption (Assumptions \ref{assumption:non-singular-times} and \ref{NonSingMono}), we prove that the all Lax pairs associated with the six Painlev\'e equations are of topological type. 
\item 
Consequently, for all six Painlev\'e equations, Berg\`ere-Borot-Eynard correlation functions defined from these Lax pairs coincide with the Eynard-Orantin differentials obtained from the application of the topological recursion to the corresponding spectral curve. This implies that the $\tau$-function of the isomonodromy Lax system coincides with the generating function of the symplectic invariants of the spectral curve.  
\end{itemize}
Our results extend similar results developed for the Painlev\'e II equation in \cite{P2} as well as partial results for the Painlev\'e V equation in \cite{P5}.  Our analysis is sufficiently general to cover all six Painlev\'e equations. 

We note that our approach is purely formal. The $\tau$-functions reconstructed by the topological recursion correspond to formal power series solutions of the Painlev\'e equation. Such solutions are known to be divergent in general, and hence, our formal solutions are asymptotic expansion of exact solutions. Analytic properties of exact solution of Painlev\'e equations are widely studied; see \cite{FIKN} for example. 

Our paper is organized as follows: 
In Section \ref{section:Lax} we present the six Lax pairs, Hamiltonians and the $\tau$-functions, describing the six Painlev\'e equations. Since our gauge choices and notations are slightly different from the historical ones given by Jimbo and Miwa in \cite{JMII}, we also provide the explicit correspondences in Appendix \ref{AppendixConnectionJM}.  We will consider the $\hbar$-formal series solution of the Painlev\'e equations, and we summarize several of their properties in Section \ref{section:formal-solution}. In Section \ref{SpectralCurves}, we present the computation and analysis of the spectral curves. Section \ref{DeterminantalFormSection} is dedicated to the presentation of the determinantal formulas and the topological type property. Finally, in Section \ref{SectionMainTheorem}, we get to the statement of our main theorem (i.e. that all six Lax pairs satisfy the topological type property). The technical elements of proof of the topological type property are postponed to the Appendices \ref{AppendixParityProof}, \ref{AppendixProofPole} and \ref{AppendixLeadingOrder}. The proof follows the same strategy as the one developed in \cite{P2}. Eventually, conclusions and outlooks are presented in Section \ref{SectionConclusion} and computations of the special cases $F^{(0)}$ and $F^{(1)}$ are presented in Appendix \ref{AppendixSymplecticInvariants}.

%%%%%%%%%%%%%%%%%%%%%%%%%%%%%%%%%
%%%%%%%%%%%%%%%%%%%%%%%%%%%%%%%%%
\section*{Acknowledgement}
The authors would like to thank R. Conte for discussions about the choice of Lax pairs. K. Iwaki would like to thank M. Manabe and M. Mulase who gave explanations on the theory of the topological recursion. K. Iwaki is supported by JSPS Grant 16K17613, 16H06337 and the JSPS project: Advancing Strategic International Networks to Accelerate the Circulation of Talented Researchers ``Mathematical Science of Symmetry, Topology and Moduli, Evolution of International Research Network based on OCAMI".
A. Saenz would also like to thank M. Mulase for many fruitful discussions. A. Saenz is supported by the UC Davis Dissertation Year Fellowship. O. Marchal would like to thank Universit\'e Lyon $1$, Universit\'e Jean Monnet and Institut Camille Jordan for the opportunity to make this research possible. O. Marchal would also like to thank B. Eynard for fruitful discussions about the project.

%%%%%%%%%%%%%%%%%%%%%%%%%%%%%%%%%
%%%%%%%%%%%%%%%%%%%%%%%%%%%%%%%%%
\section{Painlev\'e equations, Lax pairs, Hamiltonians and $\tau$-functions} \label{section:Lax}
\subsection{Lax pairs \label{LaxPairs}}

Painlev\'e equations play an important role in the theory of integrable systems. They were studied originally by Gambier and Painlev\'e, and their connections with integrable systems was detailed later by Jimbo, Miwa and Ueno in a series of papers \cite{JMII, JMIII, JMI}. In \cite{JMII}, Jimbo and Miwa gave a list of the Lax pairs whose compatibility equations are equivalent to the six Painlev\'e equations (with $\hbar = 1$). In this paper we will use an $\hbar$-dependent version of Painlev\'e equations given in table \ref{table:P-eq} and associated Lax pairs which are obtained from the previous Lax pairs in \cite{JMII} by certain rescaling of parameters together with a certain gauge transformations (see Appendix \ref{AppendixConnectionJM}).

\begin{definition}[Lax pair] 
A ($\hbar$-dependent) Lax pair is a pair $(\mathcal{D}(x,t), \mathcal{R}(x,t))$ of $n \times n$ matrices, whose entries are rational functions of x and holomorphic in $t$ on some domain, so that the system of partial differential equations
\begin{equation} \label{eq:Lax-pair}
\hbar \partial_x \Psi(x,t)=\mathcal{D}(x,t)\Psi(x,t) \,\,\,,\,\,\, 
\hbar \partial_t \Psi(x,t)=\mathcal{R}(x,t)\Psi(x,t)
\end{equation}
is compatible (i.e. $\partial_x \partial_t = \partial_t \partial_x $). We call the system \eqref{eq:Lax-pair} a Lax system, or an isomonodromy system. In the literature, $x$ is usually called the spatial parameter (or the spectral parameter) while $t$ provides the time parameter. 
\end{definition}

Lax pairs that we consider in this paper are only $2 \times 2$. Also, we note that the compatibility of the previous two differential equations is not obvious, and we recall a standard result:

\begin{proposition}[Compatibility conditions] The compatibility of the isomonodromy system is equivalent to the following equation (known as the zero-curvature equation):
\begin{equation}  \label{eq:compatibility}
\hbar  \partial_t \mathcal{D}(x,t) - \hbar \partial_x \mathcal{R}(x,t) +\left[\mathcal{D}(x,t),\mathcal{R}(x,t)\right]=0.
\end{equation}
\end{proposition}  

The following is the list of the $\hbar$-dependent Lax pair $\left(\mathcal{D}_{J}(x,t),\mathcal{R}_{J}(x,t)\right)$ whose compatibility condition \eqref{eq:compatibility} is equivalent to the $J^{\rm th}$ Painlev\'e equation ($J={\rm I}, \dots, {\rm VI}$): 

\begin{center}
\begin{itemize}
\item (Painlev\'e I)
\beq \label{P1}\DD_{\rm I}(x,t) = \begin{pmatrix} -p&x^2+qx+q^2+\frac{t}{2}\\4(x-q)&p\end{pmatrix} \,\,,\,\, \RR_{\rm I}(x,t) = \begin{pmatrix} 0&\frac{x}{2}+q\\2 &0\end{pmatrix}. \eeq
\item (Painlev\'e II)
\beq\label{P2}\DD_{\rm II}(x,t) = \begin{pmatrix} x^2+p+\frac{t}{2}&x-q\\-2\left(xp+qp+\theta\right)& -\left(x^2+p+\frac{t}{2}\right)\end{pmatrix} \,\,,\,\, 
\RR_{\rm II}(x,t) = \begin{pmatrix} \frac{x+q}{2}& \frac{1}{2}\\-p& -\frac{x+q}{2}\end{pmatrix}.
\eeq
\item (Painlev\'e III)
\begin{eqnarray}
\label{P3} \DD_{\rm III}(x,t) & = & \frac{t}{2} \sigma_3 
+ \frac{1}{x} \begin{pmatrix} 
- \frac{\theta_\infty}{2} & - pq \\ -q(p-t) - \theta_\infty + \frac{t(\theta_0+\theta_\infty)}{2p} & \frac{\theta_\infty}{2}
 \end{pmatrix} \nonumber \\
 & & + \frac{1}{x^2} \begin{pmatrix} p - \frac{t}{2} & - p \\ p - t & -p + \frac{t}{2} \end{pmatrix} \nonumber \\
 \RR_{\rm III}(x,t) & = & \frac{x}{2} \sigma_3 - \frac{1}{tx}  \begin{pmatrix} p - \frac{t}{2} & - p \\ p - t & -p + \frac{t}{2} \end{pmatrix} + \frac{1}{t} \begin{pmatrix} r_{11} & r_{12} \\ r_{21} & r_{22} \end{pmatrix},
\end{eqnarray}
where $\sigma_3 = \begin{pmatrix} 1 & 0 \\ 0 & -1 \end{pmatrix}$ 
is the Pauli matrix, and 
\begin{eqnarray*} 
r_{11} & = & - \frac{\theta_\infty}{2} + t q + \frac{t(\theta_0+\theta_\infty)}{2p} \\
r_{12} & = & -pq, \quad r_{21} =  -q(p-t) - \theta_\infty + \frac{t(\theta_0+\theta_\infty)}{2p}.
\end{eqnarray*}

\item (Painlev\'e IV)
\begin{eqnarray}
\label{P4} \DD_{\rm IV}(x,t)&=& x \sigma_3 
+ \begin{pmatrix} t & 1 \\ -2(pq+\theta_0+\theta_\infty) & -t \end{pmatrix} 
+ \frac{1}{x} \begin{pmatrix} pq+\theta_0 & -q \\ p(pq+2\theta_0) & -pq-\theta_0 \end{pmatrix} 
\nonumber \\
\RR_{\rm IV}(x,t) &  =  & x \sigma_3 + \begin{pmatrix} q+t & 1 \\ -2(pq+\theta_0+\theta_\infty) & -q-t \end{pmatrix} .
\end{eqnarray}

\item (Painlev\'e V)
\begin{eqnarray}
\label{P5}\DD_{\rm V}(x,t) & = & \frac{t}{2} \sigma_3
+ \frac{1}{x} \begin{pmatrix} pq + \frac{\theta_0}{2} & -pq-\theta_0 \\ pq & -pq - \frac{\theta_0}{2} \end{pmatrix} 
  \nonumber \\ 
& & + \frac{1}{x-1} \begin{pmatrix} -pq - \frac{\theta_0+\theta_1}{2} & p + \frac{\theta_0 - \theta_1 + \theta_\infty}{2q} \\ 
-q \left( pq + \frac{\theta_0+\theta_1+\theta_\infty}{2} \right) & pq + \frac{\theta_0+\theta_\infty}{2} \end{pmatrix}
 \nonumber \\
\RR_{\rm V}(x,t) & = & \frac{x}{2} \sigma_3  + \frac{1}{2t} \begin{pmatrix} r_{11} & r_{12} \\ r_{21} & -r_{11}  \end{pmatrix},
\end{eqnarray}
where 
\begin{eqnarray*}
r_{11} & = & - p(q-1)^2 + \theta_0 - \frac{q(\theta_0+\theta_1+\theta_\infty)}{2} - \frac{\theta_0-\theta_1+\theta_\infty}{2q} \\
r_{12} & = & 2p(q-1) + 2\theta_0 - \frac{\theta_0 - \theta_1 + \theta_\infty}{q} \\
r_{21} & = & -2pq(q-1) + q(\theta_0+\theta_1+\theta_\infty).
\end{eqnarray*}

\item (Painlev\'e VI)
\beq \label{P6}\DD_{\rm VI}(x,t) = \frac{A_0(t)}{x}+\frac{A_1(t)}{x-1}+\frac{A_t(t)}{x-t}\,\,,\,\, \RR_{\rm VI}(x,t) =-\frac{A_t(t)}{x-t}-\frac{(q-t)(\theta_\infty-\hbar)}{2t(t-1)}\sigma_3, \eeq
where
\beaa  A_0&=&\begin{pmatrix} z_0+\frac{\theta_0}{2}&-\frac{q}{t}\\[+.3em] \frac{tz_0(z_0+\theta_0)}{q}&-\left(z_0+\frac{\theta_0}{2}\right)\end{pmatrix} \,\,,\,\, A_1=\begin{pmatrix} z_1+\frac{\theta_1}{2}&\frac{q-1}{t-1}\\[+.2em] -\frac{(t-1)z_1(z_1+\theta_1)}{q-1}&-\left(z_1+\frac{\theta_1}{2}\right)\end{pmatrix}\cr
A_t&=&\begin{pmatrix} z_t+\frac{\theta_t}{2}&-\frac{q-t}{t(t-1)}\\ \frac{t(t-1)z_t(z_t+\theta_t)}{q-t}&-\left(z_t+\frac{\theta_t}{2}\right)\end{pmatrix},
\eeaa
and the parameter $\theta_\infty$ is chosen so that 
\[
A_\infty=-(A_0+A_1+A_t) = \begin{pmatrix} \frac{\theta_\infty}{2}&0\\0&-\frac{\theta_\infty}{2}\end{pmatrix}.
\]
Here, $z_0(t), z_1(t)$ and $z_t(t)$ are auxiliary functions of $t$ that can be expressed in terms $q(t)$ and a function $p(t)$ defined by
\beq p=\frac{z_0+\theta_0}{q}+\frac{z_1+\theta_1}{q-1}+\frac{z_t+\theta_t}{q-t}.\eeq
The explicit expression of $(z_0,z_1,z_t)$ in terms of $(p,q)$ is given by
\footnotesize{\bea \label{z0z1zt} z_0&=&\frac{1}{\theta_\infty t}\Big[ q^2(q-1)(q-t)p^2-pq((\theta_0+\theta_1+\theta_t-\theta_\infty)q^2-((\theta_0+\theta_1-\theta_\infty)t+\theta_0+\theta_t-\theta_\infty)q+(\theta_0-\theta_\infty)t)\cr
&&+\frac{1}{4}(\theta_0+\theta_1+\theta_t-\theta_\infty)^2q^2-\frac{1}{4}(\theta_0+\theta_1+\theta_t-\theta_\infty)( (\theta_0+\theta_1-\theta_t-\theta_\infty)t+\theta_0-\theta_1-\theta_\infty+\theta_t)q-t\theta_0\theta_\infty\Big]\cr
z_1&=&\frac{1}{(t-1)\theta_\infty}\Big[ -q(q-1)^2(q-t)p^2+p(q-1)( (\theta_0+\theta_1+\theta_t-\theta_\infty)^2q^2-q((\theta_0+\theta_1-\theta_\infty)t+\theta_0+\theta_t)+\theta_0 t)\cr
&&-\frac{1}{4}(\theta_0+\theta_1+\theta_t-\theta_\infty)^2(q-1)^2+\frac{1}{4}(\theta_0+\theta_1+\theta_t-\theta_\infty)(t(\theta_0+\theta_1-\theta_t-\theta_\infty)+2\theta_\infty-2\theta_1)(q-1)\cr
&&-\theta_1\theta_\infty(t-1)\cr
z_t&=&-z_0-z_1-\frac{1}{2}(\theta_0+\theta_1+\theta_t+\theta_\infty).
\eea}\normalsize{}
\end{itemize}
\end{center}

For each Lax pair $(\mathcal{D}_J(x,t) , \mathcal{R}_J(x,t))$, $p=p(t)$ and $q=q(t)$ are functions that depend on the time parameter $t$ and $\hbar$ such that the function $q(t)$ satisfies the $J^{\rm th}$ $\hbar$-dependent Painlev\'e equation in Table \ref{table:P-eq} and the function $p(t)$ is directly related to the Hamiltonian formulation of the Painlev\'e equations that will be given in the next section. In these six Painlev\'e equations and Lax pairs the monodromy parameters $\theta, \theta_0, \theta_1, \theta_\infty, \theta_t \in \mathbb{C}$ are assumed to be generic in the sense of Assumption \ref{NonSingMono} detailed below. 

As one can notice, our Lax pairs slightly differ from the original ones given by Jimbo and Miwa \cite{JMII}, but the Lax pairs end up being equivalent via a guage transformation. For completeness, we provide the correspondence in Appendix \ref{AppendixConnectionJM}. 

\subsection{Hamiltonians, $\tau$-functions and Okamoto's $\sigma$-form of the Painlev\'e equations}\label{OkamotoForm}

It is well known, since the works of Okamoto \cite{Okamoto}, that the Painlev\'e equations (with $\hbar = 1$) can be represented as Hamiltonian systems. In this section, we provide the corresponding Hamiltonians associated to our Painlev\'e equations as well as the $\tau$-functions and Okamoto's $\sigma$-functions, in our $\hbar$-dependent setting. We derive the $\hbar$-dependent Painlev\'e equations in Table 1.1 from the $\hbar$-dependent Lax pairs, and the details are presented in Appendix \ref{AppendixDerivationPainleve}.

\begin{theorem}[Hamiltonian formulation]\label{HamiltonianFormulation} For all $J = {\rm I}, \dots, {\rm VI}$, the $\hbar$-dependent versions of the $J^{\text{th}}$ Painlev\'e equation in Table \ref{table:P-eq} can be described as a (time-dependent) Hamiltonian system $H_J(p,q,t)$ as
\beq \label{eq:HJ}
\hbar \dot{q} =\frac{\partial H_J}{\partial p} (p,q,t) ~~\text{and}~~
\hbar \dot{p} =-\frac{\partial H_J}{\partial q} (p,q,t). \eeq
\end{theorem}

We list here the various Hamiltonians $H_{J}$ as well as their relations to the $\tau$-function $\tau_{J}$ and Okamoto's $\sigma$-function $\sigma_J$.

\begin{itemize}
\item (Painlev\'e I) 
% The Hamiltonian is given by
\beq \label{H1} H_{\rm I}(p,q,t)=\frac{1}{2}p^2-2q^3-t q.\eeq
% Moreover, we have
\beq \label{Tau1} \hbar^2 \frac{d}{d t}\ln \tau_{\rm I}(t)=\sigma_{\rm I}(t) 
~~\text{ and }~~ \sigma_{\rm I}(t)=H_{\rm I}(p(t),q(t),t).\eeq
Okamoto $\sigma$-function satisfies the following differential equation:
\beq \label{sigma1} \hbar^2\ddot{\sigma}_{\rm I}^2+4\dot{\sigma}_{\rm I}^3+2t\dot{\sigma}_{\rm I}-2\sigma_{\rm I}=0.\eeq
\item (Painlev\'e II) 
% The Hamiltonian is given by
\beq \label{H2} H_{\rm II}(p,q,t)=\frac{1}{2}p^2+(q^2+\frac{t}{2})p+\theta q.\eeq
% Moreover, we have
\beq \label{Tau2} \hbar^2 \frac{d}{d t}\ln \tau_{\rm II}(t)= 
\sigma_{\rm II}(t) ~~\text{ and } ~~ \sigma_{\rm II}(t)=H_{\rm II}(p(t),q(t),t).\eeq
Okamoto $\sigma$-function satisfies the following differential equation:
\beq \label{sigma2} \hbar^2\ddot{\sigma}_{\rm II}^2+4\dot{\sigma}_{\rm II}^3+2t\dot{\sigma}_{\rm II}^2-2\sigma_{\rm II}\dot{\sigma}_{\rm II}-\frac{\theta^2}{4}=0.\eeq
\item (Painlev\'e III) 
% The Hamiltonian is given by
\beq \label{H3} H_{\rm III}(p,q,t,\hbar)=\frac{1}{t}\left[2q^2p^2+2(-tq^2+\theta_\infty q+t)p-(\theta_0+\theta_\infty)tq-t^2-\frac{1}{4}(\theta_0^2-\theta_\infty^2)-\hbar pq\right].\eeq
% Moreover, we have
\bea \label{Tau3} \hbar^2 \frac{d}{d t}\ln \tau_{\rm III}(t)&=&H_{\rm III}(p(t),q(t),t,\hbar)+\hbar\frac{p q}{t}\cr
&=&\frac{1}{t}\left[2q^2p^2+2(-t q^2+\theta_\infty q+t)p-(\theta_0+\theta_\infty)t q-t^2-\frac{1}{4}(\theta_0^2-\theta_\infty^2)\right]. \nonumber \\
\eea
Okamoto $\sigma$-function is directly related to the $\tau$-function by $\sigma_{\rm III}(t)= \hbar^2 t \frac{d}{d t}\ln \tau_{\rm III}(t)$. It satisfies the following differential equation:
\bea \label{sigma3} \hbar^2\left(t\ddot{\sigma}_{\rm III}-\dot{\sigma}_{\rm III}\right)^2-4(2\sigma_{\rm III}-t\dot{\sigma}_{\rm III})(\dot{\sigma}_{\rm III}^2-4t^2)-2(\theta_0^2+\theta_\infty^2)(\dot{\sigma}_{\rm III}^2+4t^2)+16\theta_0\theta_\infty t \dot{\sigma}_{\rm III}=0. \notag \\ \eea

\item (Painlev\'e IV) 
% The Hamiltonian is given by
\beq \label{H4} H_{\rm IV}(p,q,t)=q p^2+2(q^2+t q+\theta_0)p+2(\theta_0+\theta_\infty)q.\eeq
% Moreover, we have
\bea \label{Tau4} \hbar^2 \frac{d}{d t}\ln \tau_{\rm IV}(t)=H_{\rm IV}(p(t),q(t),t) ~~\text{ and }~~\sigma_{\rm IV}(t)=H_{\rm IV}(p(t),q(t),t)+2t\left(\theta_0+\frac{1}{3}\theta_\infty\right). \nonumber \\ \eea 
Okamoto $\sigma$-function satisfies the following differential equation:
\beq \label{sigma4} \hbar^2 \ddot{\sigma}_{\rm IV}^2-4(t\dot{\sigma}_{\rm IV}-\sigma_{\rm IV})^2+4(\dot{\sigma}_{\rm IV}+\alpha)(\dot{\sigma}_{\rm IV}+\beta)(\dot{\sigma}_{\rm IV}+\gamma)=0\eeq
with $\alpha=-2\theta_0-\frac{2}{3}\theta_\infty$, $\beta=2\theta_0-\frac{2}{3}\theta_\infty$ and $\gamma=\frac{4}{3}\theta_\infty$.
\item (Painlev\'e V) 
% The Hamiltonian is given by
\begin{eqnarray} \label{H5} 
H_{\rm V}(p,q,t)&=&\frac{1}{t}\biggl[ q(q-1)^2p^2+\left(\frac{\theta_0-\theta_1+\theta_\infty}{2}(q-1)^2+(\theta_0+\theta_1)q(q-1)-tq\right)p \nonumber \\
& & +\frac{1}{2}\theta_0(\theta_0+\theta_1+\theta_\infty)q \biggr].
\end{eqnarray}
% The $\tau$-function is directly connected to the Hamiltonian by
\beq \label{Tau5} \hbar^2 \frac{d}{d t}\ln \tau_{\rm V}=H_{\rm V}(p(t),q(t),t)-\frac{\theta_0+\theta_\infty}{2}-\frac{1}{4t}(\theta_0-\theta_1+\theta_\infty)(\theta_0+\theta_1+\theta_\infty).\eeq 
Okamoto $\sigma$-function is defined by $\sigma_{\rm V}(t)=t H_{\rm V}(p(t),q(t),t)$ and satisfies
\bea \label{sigma5} \hbar^2 t^2 \ddot{\sigma}_{\rm V}^2-\left(\sigma_{\rm V}-t\dot{\sigma}_{\rm V}+2\dot{\sigma}_{\rm V}^2+(\nu_1+\nu_2+\nu_3)\dot{\sigma}_{\rm V}\right)^2+4\dot{\sigma}_{\rm V}(\dot\sigma_{\rm V}+\nu_1)(\dot\sigma_{\rm V}+\nu_2)(\dot\sigma_{\rm V}+\nu_3)=0, \nonumber \\ \eea
where $\left(\nu_1,\nu_2,\nu_3\right)=\left(-\frac{\theta_0-\theta_1+\theta_\infty}{2},-\theta_0,-\frac{\theta_0+\theta_1+\theta_\infty}{2}\right)$.

\item (Painlev\'e VI) 
\[
H_{\rm VI}(p,q,t,\hbar) = \hspace{+35.em} \vspace{-1.em}
\] 
\begin{multline}
\frac{1}{t(t-1)}\bigg[ q(q-1)(q-t)p^2 - \Bigl( \theta_0(q-1)(q-t)+\theta_1q(q-t)+(\theta_t-\hbar)q(q-1)\Bigr)p \\
+\frac{1}{4}(\theta_0+\theta_1+\theta_t-\theta_\infty)(\theta_0+\theta_1+\theta_t+\theta_\infty-\hbar)(q-t) 
+\frac{1}{2}((t-1)\theta_0+t\theta_1)(\theta_t-\hbar)\biggr].
\end{multline}

The $\tau$-function is defined by
\bea \label{Tau6} \hbar^2 \frac{d}{d t}\ln \tau_{\rm VI}&=& 
H_{\rm VI}(p(t),q(t),t,\hbar=0)\cr &=& 
\frac{1}{t(t-1)}\Big[ q(q-1)(q-t)p^2-p\left(\theta_0(q-1)(q-t)+\theta_1q(q-t)+\theta_tq(q-1)\right)\cr
&&+\frac{1}{4}(\theta_0+\theta_1+\theta_t-\theta_\infty)(\theta_0+\theta_1+\theta_t+\theta_\infty)(q-t)+\frac{1}{2}((t-1)\theta_0+t\theta_1)\theta_t\Big].\nonumber \\
\eea

% \newpage
Okamoto $\sigma$-function 
\beq \sigma_{\rm VI}(t)=\hbar^2 t(t-1)\frac{d}{d t}\ln \tau_{\rm VI}(t)+\frac{1}{4}(\theta_t^2-\theta_\infty^2)t-\frac{1}{8}(\theta_0^2+\theta_t^2-\theta_1^2-\theta_\infty^2)\eeq
satisfies the following differential equation:
\bea \label{sigma6}0&=&\hbar^2\dot{\sigma}_{\rm VI}t^2(t-1)^2\ddot{\sigma}_{\rm VI}^2+\left(2\dot{\sigma}_{\rm VI}(t\dot{\sigma}_{\rm VI}-\sigma_{\rm VI})-\dot{\sigma}_{\rm VI}^2+\frac{1}{16}(\theta_t^2-\theta_\infty^2)(\theta_1^2-\theta_0^2)\right)^2\cr
&&-\left(\dot{\sigma}_{\rm VI}+\frac{(\theta_t+\theta_\infty)^2}{4}\right)\left(\dot{\sigma}_{\rm VI}+\frac{(\theta_t-\theta_\infty)^2}{4}\right)\left(\dot{\sigma}_{\rm VI}+\frac{(\theta_0+\theta_1)^2}{4}\right)\left(\dot{\sigma}_{\rm VI}+\frac{(\theta_0-\theta_1)^2}{4}\right).\nonumber \\
\eea
% Note that we can also define $y(t)=t(t-1)H_{\rm VI}(p(t),q(t),\hbar)+\frac{1}{4}((\theta_t-\hbar)^2-(\theta_\infty-\hbar)^2)t-\frac{1}{8}(\theta_0^2-\theta_1^2+(\theta_t-\hbar)^2-(\theta_\infty-\hbar)^2)$ and observe that it satisfies:
% \bea 0&=&\hbar^2t^2(t-1)^2\dot{y}\ddot{y}^2+\left(2\dot{y}(t\dot{y}-y)-\dot{y}^2+\frac{1}{16}(\theta_t-\theta_\infty)(\theta_t+\theta_\infty-2\hbar)(\theta_1^2-\theta_0^2)\right)^2\cr
% &&-\left(\dot{y}+\frac{(\theta_t+\theta_\infty-2\hbar)^2}{4}\right)\left(\dot{y}+\frac{(\theta_t-\theta_\infty)^2}{4}\right)\left(\dot{y}+\frac{(\theta_0+\theta_1)^2}{4}\right)\left(\dot{y}+\frac{(\theta_0-\theta_1)^2}{4}\right).\nonumber \\
% \eea
\end{itemize}

One can verify that the equations of motion \eqref{eq:HJ} for these Hamiltonians respectively recover \eqref{CompatibilityP1}, \eqref{CompatibilityP2}, \eqref{CompatibilityP3}, \eqref{CompatibilityP4}, \eqref{CompatibilityP5}, \eqref{CompatibilityP6}. Note that the $\tau$-functions $\ln \tau_J$ are defined up to constants (with respect to $t$) since they are defined only through their time derivatives. We also remark that the $\sigma$-functions always satisfy differential equations that only involve $\hbar^2$ but not directly $\hbar$.

%%%%%%%%%%%%%%%%%%%%%%%%%%%%%%%%%%%%%%%%%%%%%%
\section{Formal series expansion in $\hbar$} \label{section:formal-solution}
\subsection{Formal series solution of Painlev\'e equations and singular times}
In this paper, we are interested in the formal power series solution of the $J^{\rm th}$ Painlev\'e equation of the form
\beq \label{Serieshbar} q(t)=\sum_{k=0}^\infty q^{(k)}(t)\hbar^k.
\eeq
If we ignore the convergence issue, such an expansion can be computed in a recursive manner. (Actually, \eqref{Serieshbar} is an asymptotic expansion of a solution of the Painlev\'e equation. See Remark \ref{remark:Borel-sum} below.) Let us write the $J^{\text{th}}$ Painlev\'e equation in Table \ref{table:P-eq} as 
\beq \hbar^2\ddot{q}(t)=\hbar B_J(q,\dot{q},t, \hbar) +A_J(q,t), \eeq
where $A_J$, which is independent of $\dot{q}$ and $\hbar$, 
and $B_J$ are rational in $q,\dot{q},t$ and linear in $\hbar$.
Then, the leading term $q^{(0)}(t)$ of the expansion \eqref{Serieshbar} must satisfies 
\begin{equation} \label{eq:leading-relation-in-PJ}
A_J(q^{(0)}(t), t) = 0.
\end{equation}
In order to have more compact notation, we will denote $q^{(0)}(t)$ by $q_0(t)$ in the rest of the paper. Note that $q_0$ is a multivalued function of $t$ branching along the following set of times:

\begin{definition}[Singular times] \label{SingularTimes}
We call a time $t$ singular for the $J^{\rm th}$ Painlev\'e equation  
if it belongs to the set
\beq
\Delta_{J} = \left\{ t\in \mathbb{C} \mid \exists q \text{ so that } A_J(q,t)=0 ~\text{ and } ~\frac{\partial A_J}{\partial q}(q,t)=0\right\}.
\eeq
\end{definition}

In what follows we assume 
\begin{assumption} \label{assumption:non-singular-times}
The time variable $t$ lies in a domain $U$ satisfying $U \cap \Delta_J = \emptyset$.
\end{assumption}

The singular times are also called (non-linear analogue of) turning points in the WKB analysis for the Painlev\'e equations (see Definition 2.1 in \cite{KT-painleve}). 
It is known that 
\begin{proposition}[Existence of formal series solution: 
Proposition 1.1 in  \cite{KT-painleve}]
Under Assumption \ref{assumption:non-singular-times}, there exists a formal series solution \eqref{Serieshbar} of the Painlev\'e equation whose coefficients $\left( q^{(k)}(t) \right)_{k \ge 0}$ are holomorphic on $U$.
\end{proposition}
In fact we can verify that the higher coefficients $\left( q^{(k)}(t) \right)_{k \ge 1}$ are recursively determined from $q_0(t)$, but are (multivalued) functions with singularities on $\Delta_{J}$. 
Hence, we must avoid these points when we deal with a formal series expansion in $\hbar$. 

We list here the corresponding conditions for singular times:
\begin{itemize} 
\item (Painlev\'e I) 
The leading term $q_0$ satisfies the algebraic relation $A_{\rm I}(q_0,t) = 6q_0^2+t=0$. The condition of singular times imposes the relation $12q_0=0$. Thus, we have $\Delta_{\rm I} = \{ 0 \}$. The singular time $t=0$ corresponds to $q_0=0$.
\item (Painlev\'e II) 
The leading term $q_0$ satisfies the algebraic relation $A_{\rm II}(q_0,t) = 2q_0^3+tq_0-\theta=0$. The condition of singular times imposes the relation $6q_0^2+t=0$, giving us that singular times correspond to times for which we have the algebraic relation:
\beq \label{SingP2} 4q_0^3+\theta=0.\eeq
\item (Painlev\'e III) The leading term $q_0$ satisfies the algebraic relation $t q_0^4+\theta_0q_0^3-\theta_\infty q_0-t=0$. The condition of singular times imposes the relation $4t q_0^3+3\theta_0q_0^2-\theta_\infty=0$, giving us that singular times correspond to times for which we have the algebraic relation:
\beq \label{SingP3}\theta_0 q_0^6-3\theta_\infty q_0^4+3\theta_0 q_0^2-\theta_\infty=0.\eeq
\item (Painlev\'e IV) The leading term $q_0$ satisfies the algebraic relation $3q_0^4+4tq_0^3+(t^2-2\theta_\infty)q_0^2-\theta_0^2=0$. The condition of singular times imposes the relation $6q_0^2+6t q_0+t^2-2\theta_\infty=0$, giving us that singular times correspond to times for which we have the algebraic relation:
 \beq \label{SingP4}3q_0^8+8\theta_\infty q_0^6+6\theta_0^2q_0^4+\theta_0^4=0.\eeq
\item (Painlev\'e V) The leading term $q_0$ satisfies the algebraic relation:
\bea  \label{Q0P5}
0&=&(\theta_0-\theta_1-\theta_\infty)^2q_0^5-3(\theta_0-\theta_1-\theta_\infty)^2q_0^4\cr
&&-2(2t^2-4(\theta_0+\theta_1)t-\theta_0^2-\theta_1^2-\theta_\infty^2+4\theta_\infty(\theta_0-\theta_1)+2\theta_0\theta_1)q_0^3\cr
&  &-2(2t^2+4(\theta_0+\theta_1)t-\theta_0^2-\theta_1^2-\theta_\infty^2-4\theta_\infty(\theta_0-\theta_1)+2\theta_0\theta_1)q_0^2\cr
&&-3(\theta_0-\theta_1+\theta_\infty)^2q_0+(\theta_0-\theta_1+\theta_\infty)^2.
\eea
The condition of singular times imposes the relation:
\small{\beqq t=-\frac{(q_0-1)^2\left((\theta_0-\theta_1-\theta_\infty)^2q_0^4+2(\theta_0-\theta_1-\theta_\infty)^2q_0^3-2(\theta_0-\theta_1+\theta_\infty)^2q_0-(\theta_0-\theta_1+\theta_\infty)^2\right)}{8q_0^3(\theta_0+\theta_1)}.\eeqq}\normalsize{}
Thus, singular times correspond to the solutions of the algebraic relation:
\small{\bea \label{SingP5} 0&=&(\theta_0-\theta_1-\theta_\infty)^4q_0^9+3(\theta_0-\theta_1-\theta_\infty)^4q_0^8+8(\theta_0-\theta_1-\theta_\infty)^2(\theta_0^2+6\theta_0\theta_1+\theta_1^2-\theta_\infty^2)q_0^6\cr
&&-6(\theta_0-\theta_1-\theta_\infty)^2(\theta_0-\theta_1+\theta_\infty)^2q_0^5+6(\theta_0-\theta_1-\theta_\infty)^2(\theta_0-\theta_1+\theta_\infty)^2q_0^4\cr
&&-8(\theta_0-\theta_1+\theta_\infty)^2(\theta_0^2+6\theta_0\theta_1+\theta_1^2-\theta_\infty^2)q_0^3 -3(\theta_0-\theta_1+\theta_\infty)^4q_0-(\theta_0-\theta_1+\theta_\infty)^4.\nonumber \\
\eea}\normalsize{}
\item (Painlev\'e VI) The leading term $q_0(t)$ satisfies the equation:
\beq \label{Q0P6} 
\theta_\infty^2-\frac{\theta_0^2t}{q_0^2}+\frac{(t-1)\theta_1^2}{(q_0-1)^2}-\frac{\theta_t^2t(t-1)}{(q_0-t)^2}=0.\eeq
This is equivalent to a polynomial equation of degree $6$:
\bea \label{q0} 0&=&\theta_\infty^2q_0^6-2(t+1)\theta_\infty^2q_0^5 +\left(-t\theta_0^2+(t-1)\theta_1^2+(t^2+4t+1)\theta_\infty^2-t(t-1)\theta_t^2\right)q_0^4\cr
&&+2t\left((t+1)\theta_0^2-(t-1)\theta_1^2-(t+1)\theta_\infty^2+(t-1)\theta_t^2\right)q_0^3\cr
&&-t\left( (t^2+4t+1)\theta_0^2-t(t-1)\theta_1^2-t\theta_\infty^2+(t-1)\theta_t^2\right)q_0^2\cr
&&+2t^2(t+1)\theta_0^2q_0-t^3\theta_0^2.
\eea
The condition of singular times imposes the relation:
\beq \label{SingP6} \frac{\theta_0^2t}{q_0^3}-\frac{(t-1)\theta_1^2}{(q_0-1)^3}+\frac{t(t-1)\theta_t^2}{(q_0-t)^3}=0.\eeq
\end{itemize}

\begin{remark} \label{remark:Borel-sum}
It is known that the formal power series solution \eqref{Serieshbar} is not convergent for generic monodoromy parameters, but Borel summable under a certain condition (see \cite{Kam-Ko}). In other words, the expansion \eqref{Serieshbar} is the asymptotic expansion of an exact solution of the Painlev\'e equation for $\hbar \to 0$. As is discussed in \cite{Kam-Ko}, to give a criterion of the Borel summability, the singular times (together with the non-linear analogue of Stokes curves introduced in Definition 2.1 \cite{KT-painleve}) play an important role. The Riemann-Hilbert method was also traditionary used to describe analytic properties of such solutions; see \cite{FIKN}.  These results are important, but we omit its description because we are only interested in properties of the coefficients of the formal series solution \eqref{Serieshbar}.

It is also known that the space of solutions of Painlev\'e  equations are 2-dimensional space (the Okamoto space of initial conditions \cite{Okamoto-initial}), and the solution obtained as the Borel sum of \eqref{Serieshbar} corresponds to a specific point in the space; that is, the generic solutions of the Painlev\'e equations do not have an $\hbar$-expansion. More precisely, general (formal) solutions contain not only power series terms but also some non-perturbative exponential terms (see \cite{Kawai}). But the analytic properties (Borel summability etc.) are not established for such solutions in full generality. 
\end{remark}

\subsection{Formal series expansion of Lax matrices and $\tau$-functions}

Since we are considering the solution $q(t)$ of the Painlev\'e equation with a series expansion \eqref{Serieshbar} in $\hbar$, the functions $p(t)$ in the Hamiltonian system \eqref{eq:HJ} also have a series expansion. As we mentioned above, we are interested in properties of the coefficients of the series expansion in $\hbar$. We will compare the coefficients with the correlation functions or symplectic invariants of topological recursion introduced in the next section. To clarify our point of view, in what follows we consider the situation: 

\begin{itemize}
\item 
The pair of functions $(q,p)$ in the Lax pairs are regarded as the formal series solution
\begin{equation} \label{eq:Serieshbar-hamiltonian-system}
(q(t),p(t)) = \left(  \sum_{k=0}^{\infty} q^{(k)}(t) \hbar^k, 
\sum_{k=0}^{\infty} p^{(k)}(t) \hbar^k \right)
\end{equation}
of the Hamiltonian system \eqref{eq:HJ}, where $q(t)$ coincides with \eqref{Serieshbar}. 

\item 
Accordingly, the Lax pairs $({\mathcal D}_J, {\mathcal R}_J)$ given in \eqref{P1}-\eqref{P6} are regarded as formal power series 
\beq \label{Dhbar} \mathcal{D}_J(x,t)=\sum_{k=0}^\infty \mathcal{D}_J^{(k)}(x,t)\hbar^k ~~\text{ and } ~~ \mathcal{R}_J(x,t)=\sum_{k=0}^\infty \mathcal{R}_J^{(k)}(x,t)\hbar^k
\eeq
in $\hbar$. 

\item 
(The logarithm of) $\tau$-functions and $\sigma$-functions are also regarded as the formal power series in $\hbar$, which can be computed from \eqref{eq:Serieshbar-hamiltonian-system} through the relations given in Section \ref{OkamotoForm}. 
\end{itemize}

\begin{remark} 
For our purpose, the existence of a power series expansion \eqref{Dhbar} of the Lax matrices will be important when we investigate the WKB formal solution (i.e., formal power series solution in $\hbar$) of the Lax system \eqref{eq:Lax-pair} in Section \ref{DeterminantalFormSection}. We note that the property \eqref{Dhbar} does not hold in the gauge choice by Jimbo-Miwa \cite{JMII}. Indeed, in \cite{JMII} (and as we summarized in Appendix \ref{AppendixConnectionJM}), most of the Lax pairs involved off-diagonal factors $u(t)$ satisfying differential equations of the form $\hbar \frac{d}{dt} \ln u=f (q,p)$ after introducing the parameter $\hbar$. Here $f(q,p)$ is some explicit function of $(q(t),p(t))$ which has a series expansion starting at $\hbar^0$. Thus $u(t)$ does not have a power series expansion; instead it has an exponential behavior of the form 
\[
u(t) = \exp\left( \frac{1}{\hbar} \sum_{n = 0}^{\infty} u_n(t) \hbar^n \right). 
\]
This remains in the off-diagonal components in the original Jimbo-Miwa's Lax matrix after introduction of $\hbar$, and hence we cannot have the series expansion of the form \eqref{Dhbar}. Fortunately, in all six cases, we found a gauge transformation $\Psi_J \mapsto U_J \Psi_J$ so that the exponential factors disappear. Details can be found in Appendix \ref{AppendixConnectionJM} for all six cases.  
\end{remark}

\subsection{Symmetry $\hbar\leftrightarrow -\hbar$ }

Here we investigate the effect when we replace the parameter $\hbar$ to $-\hbar$ and discussing the consequences on the functions $p(t), q(t), \ln \tau(t), \sigma(t)$. We denote $\,\dagger$ the involution operator that changes $\hbar$ to $-\hbar$. We would like to connect $(p^\dagger,q^\dagger)$ and $(p,q)$. For example, in Painlev\'e II, $q^\dagger(t)$ corresponds to the solution of the differential equation $\hbar^2\ddot{y}=2y^3+t y(-\frac{\hbar}{2}-\theta)$ while $q(t)$ corresponds to the solution of the differential equation $\hbar^2\ddot{y}=2y^3+ty(\frac{\hbar}{2}-\theta)$. There are many ways to compute the connection: one can use the fact that $\sigma(t)$ satisfies a differential equation only involving $\hbar^2$ but not directly $\hbar$ and show recursively that odd coefficients of the series vanish. We choose here a simpler way starting directly from the Hamiltonian formalism. More specifically, we use the fact that by definition of the Hamiltonian system:
\beq \label{ParityDetermination} H_J(p^\dagger,q^\dagger,t,-\hbar)=H_J(p,q,t,\hbar) ~~\text{ and }~~ \frac{\partial H_J}{\partial t}(p^\dagger,q^\dagger,t,-\hbar)=\frac{\partial H_J}{\partial t}(p,q,t,\hbar). \eeq
The last condition is sufficient to determine $p^\dagger$ and $q^\dagger$. Then, other quantities $(\ln \tau)^\dagger$ or $\sigma^\dagger$ can easily be obtained. We find the following results:

\begin{itemize}
\item (Painlev\'e I)
\beq \label{ParityP1} q^\dagger=q\,,\,p^\dagger=-p\,,\, \sigma_{\rm I}^\dagger=\sigma_{\rm I}\,\,,\,\,(\ln \tau_{\rm I})^\dagger=\ln \tau_{\rm I}.\eeq
\item (Painlev\'e II)
\beq \label{ParityP2} q^\dagger=-q-\frac{\theta}{p}\,,\,p^\dagger=p\,,\, \sigma_{\rm II}^\dagger=\sigma_{\rm II}\,\,,\,\,(\ln \tau_{\rm II})^\dagger=\ln \tau_{\rm II}. \eeq
\item (Painlev\'e III)
\beq \label{ParityP3} q^\dagger=\frac{-2q p^2+2(t q-\theta_\infty)p+t(\theta_0+\theta_\infty)}{2(p-t)p}\,,\,p^\dagger=p\,,\, \sigma_{\rm III}^\dagger=\sigma_{\rm III}\,\,,\,\,(\ln \tau_{\rm III})^\dagger=\ln \tau_{\rm III}. \eeq
\item (Painlev\'e IV)
\beq \label{ParityP4} q^\dagger=\frac{p(p q+2\theta_0)}{2(p q+\theta_0+\theta_\infty)} \,,\,\, p^\dagger=\frac{2q(p q+\theta_0+\theta_\infty)}{p q+2\theta_0}\,,\, \sigma_{\rm IV}^\dagger=\sigma_{\rm IV}\,\,,\,\,(\ln \tau_{\rm IV})^\dagger=\ln \tau_{\rm IV}. \eeq
\item (Painlev\'e V)
\bea \label{ParityP5} q^\dagger&=&\frac{p(2p q+\theta_0-\theta_1+\theta_\infty)}{(p q+\theta_0)(2p q+\theta_0+\theta_1+\theta_\infty)} \,,\,\, p^\dagger=\frac{q(p q+\theta_0)(2p q+\theta_0+\theta_1+\theta_\infty)}{2p q+\theta_0-\theta_1+\theta_\infty}\cr
\sigma_{\rm V}^\dagger&=&\sigma_{\rm V}\,\,,\,\,(\ln \tau_{\rm V})^\dagger=\ln \tau_{\rm V}.\eea
\item (Painlev\'e VI)
\bea \label{ParityP6} 
q^\dagger&=&\frac{t^2z_0(z_0+\theta_0)(q-1)}{t^2z_0(z_0+\theta_0)(q-1)-(t-1)^2z_1(z_1+\theta_1)q} \,\,, \,\, p^\dagger=\frac{z_0+\theta_0}{q^\dagger}+\frac{z_1+\theta_1}{q^\dagger-1}+\frac{z_t+\theta_t}{q^\dagger-t}\cr
z_0^\dagger&=&z_0 \,\,,\,\, z_1^\dagger=z_1\,\,,\,\, z_t^\dagger=z_t\,\,,\, \sigma_{\rm VI}^\dagger=\sigma_{\rm VI}\,\,,\,\,(\ln \tau_{\rm VI})^\dagger=\ln \tau_{\rm VI},
\eea
where $z_0$, $z_1$ and $z_t$ are given in \eqref{z0z1zt}
\end{itemize}

We then observe in all six cases that 
\begin{proposition}
The series expansions of $\tau$-functions and the $\sigma$-functions may only involve even powers of $\hbar$, and we may write:
\beq \label{Orderlogtau}  \ln \tau_J(t)=\sum_{g=0}^\infty \tau_J^{(g)} \hbar^{2g-2} ~~\text{ and }~~ 
\sigma_J(t)=\sum_{g=0}^\infty \sigma_J^{(g)} \hbar^{2g}. \eeq
\end{proposition}

This is of course consistent with the fact that the differential equations for $\sigma_J(t)$ (equations \eqref{sigma1}-\eqref{sigma6}) only involve $\hbar^2$ but not directly $\hbar$. This is also coherent with the fact that we want to match the $\tau$-function with the symplectic invariants (that only involve even powers of $\hbar$) arising from the topological recursion.

%%%%%%%%%%%%%%%%%%%%%%%%%%%%%%%%%%%%%%%%%%%%%%%%%%%%%%%%%%%%%%%%%%%%%%%%%%%%%%%%
\section{Spectral curves and topological recursion\label{SpectralCurves}}
\subsection{Computation of the spectral curves} \label{subsection:spec-compute}

Following the ideas developed by Berg\`ere and Eynard in \cite{Deter} for $2\times 2$ traceless Lax pairs, we introduce 
\begin{definition} \label{def:spec-curve}
The spectral curve associated to the Lax pair \eqref{eq:Lax-pair} is defined by an algebraic equation:
\beq \label{eq:spec-curve} 
y^2= E_J(x,t) ~~\text{where}~~ E_J(x,t) = - \det \mathcal{D}^{(0)}_J(x,t). \eeq
Here $\mathcal{D}^{(0)}_J(x,t)$ is the leading term of the series expansion \eqref{Dhbar} of the Lax matrix ${\mathcal D}_J(x,t)$. 
\end{definition}

In other words, the spectral curve is given by the leading order in $\hbar$ of the characteristic polynomial of $\mathcal{D}_J(x,t)$. 
% Computing the spectral curves is relatively straightforward. 
Using the equation \eqref{eq:leading-relation-in-PJ} satisfied by $q_0$, we can find the explicit form of $E_J$, which is a rational function of $x$, as shown in Table \ref{table:spectral-curve-equation}.

\begin{table}[h] 
\begin{itemize}
\item (Painlev\'e I)
\begin{equation}
E_{\rm I}(x,t)=4(x+2q_0)(x-q_0)^2. 
\end{equation}

\item (Painlev\'e II)
\beq
E_{\rm II}(x,t)=(x-q_0)^2\left(x^2+2q_0x+q_0^2+\frac{\theta}{q_0}\right).
\eeq

\item 
(Painlev\'e III)
\beq
E_{\rm III}(x,t)=\frac{(\theta_\infty-\theta_0q_0^2)^2(q_0x+1)^2
(x^2+\frac{2q_0(\theta_\infty q_0^2-\theta_0)}{\theta_0q_0^2-\theta_\infty}x+q_0^2)}{4x^4(q_0^4-1)^2}. 
\eeq
% where $t=\frac{q_0(\theta_\infty-\theta_0q_0^2)}{q_0^4-1}$.

\item 
(Painlev\'e IV)
\beq 
E_{\rm IV}(x,t)=\frac{\left(x-q_0\right)^2\left(x^2+2(q_0+t)x+\frac{\theta_0^2}{q_0^2}\right)}{x^2}, 
\eeq
where $t=-2q_0+\sqrt{q_0^2+2\theta_\infty+\frac{\theta_0^2}{q_0^2}}$.

\item 
(Painlev\'e V)
\beq E_{\rm V}(x,t)=\frac{t^2(x-Q_0)^2(x-Q_1)(x-Q_2)}{4x^2(x-1)^2}.\eeq 
See Appendix \ref{AppendixSpecP56} for formulas connecting $(Q_0,Q_1,Q_2)$ with $q_0$, $p_0$ and $t$.

\item 
(Painlev\'e VI)
\beq E_{\rm VI}(x,t) = \frac{\theta_{\infty}^2(x-q_0)^2 P_2(x)}{4x^2(x-1)^2(x-t)^2}, \eeq
where $P_2(x)=x^2+\left(-1-\frac{\theta_0^2t^2}{\theta_\infty^2 q_0^2}+\frac{\theta_1^2(t-1)^2}{\theta_\infty^2(q_0-1)^2}\right)x
+\frac{\theta_0^2t^2}{\theta_\infty^2q_0^2}$.

\end{itemize}

\caption{List of spectral curves.}
\label{table:spectral-curve-equation}

\end{table}

Here we observe that all spectral curves contain a double point singularity where the singularity can be found at $x = q_0$ for $J = {\rm I}, {\rm II}, {\rm IV}, {\rm VI}$, at $x=-\frac{1}{q_0}$ for ${J} = {\rm III}$, and at $x=Q_0$ for $J = {\rm V}$. Remarkably, all these spectral curves are of genus $0$; this is non-trivial because a naive degree counting suggests that \eqref{eq:spec-curve} is an elliptic curve.  We note that the algebraic equation \eqref{eq:leading-relation-in-PJ} satisfied by $q_0$ plays a crucial role in the factorization property of $E_J(x)$. This fact was already observed in a work of Kawai and Takei (cf. Proposition 1.3 in \cite{KT-painleve}).

Precisely speaking, the curve \eqref{eq:spec-curve} is a family of algebraic curves parametrized by the time $t$ and the monodromy parameters $\theta_{\ast}$. (We will omit the dependence of the parameters for simplicity.) As presented in \cite{Deter} and \cite{EO}, the spectral curve is the key element to implement the topological recursion. It is also connected with the WKB expansion of the matrix $\Psi(x,t)$ since its semi-classical limit is described by some integral over the curve \eqref{eq:spec-curve} (see Section \ref{subsection:correlation-function} below).

\subsection{General features of the spectral curves} 
\label{subsection:properties-of-spectral-curves}

For our purpose, we also introduce the notion of genericity on the monodromy parameters in the Painlev\'e equations to avoid situations where the geometry of the spectral curve  degenerates. In addition to Assumption \ref{assumption:non-singular-times} given in previous section for the time parameter, we further assume that

\begin{assumption}[Non-singular monodromy parameters]\label{NonSingMono} 
The monodromy parameters of the Painlev\'e equations are assumed to be generic in the following sense:
\begin{itemize} 
\item (Painlev\'e I) No assumption is needed since there is no monodoromy parameter.
\item (Painlev\'e II) $\theta\neq 0$.
\item (Painlev\'e III) $\theta_\infty\neq 0$, $\theta_0\neq 0$, and $\theta_\infty^2\neq\theta_0^2$.
\item (Painlev\'e IV) $\theta_\infty\neq 0$, $\theta_0\neq 0$, and $\theta_\infty^2\neq\theta_0^2$.
\item (Painlev\'e V) $\theta_0,\theta_1,\theta_\infty\neq 0$, and $\theta_\infty+\epsilon_0\theta_0+\epsilon_1\theta_1\neq 0$ for all possible choice of $(\epsilon_0,\epsilon_1) \in \{-1,1\}^2$.
\item (Painlev\'e VI) $\theta_0,\theta_1,\theta_t,\theta_\infty\neq 0$, $\theta_0^2\neq \theta_1^2$, and $\theta_\infty+\epsilon_0\theta_0+\epsilon_1\theta_1\neq 0$ for all possible choice of $(\epsilon_0,\epsilon_1) \in \{-1,1\}^2$ and $\theta_\infty+\epsilon_0\theta_0+\epsilon_1\theta_1+\epsilon_t\theta_t\neq 0$ for all possible choice of $(\epsilon_0,\epsilon_1,\epsilon_t) \in \{-1,1\}^3$.
\end{itemize}
\end{assumption}

In the topological recursion, the number and type of branch points of the spectral curve are crucial and thus we need to exclude all possible degenerate cases that may arise in the previous list of spectral curves. 
In this subsection, we detail why we can exclude all these cases for our spectral curves.

\begin{proposition}[Non degeneracy of the spectral curve]\label{NonSingSpectralCurve}
For any $J = {\rm I}, \dots, {\rm VI}$, the functions $E_J(x)$ generically have a double zero and at most two simple zeros. 
Moreover: 
\begin{itemize}
\item[(i)]  As long as the monodromy parameters are non-singular in the sense of Assumption \ref{NonSingMono}, the zeros of $E_J(x)$ are different from its poles and the simple zeros are distinct.
\item[(ii)]  The function $E_J(x)$ admits a triple zero if and only if the time $t$ is singular in the sense of Definition \ref{SingularTimes}. In other words the curve acquires a singularity more severe than a node only at singular times  given in Definition \ref{SingularTimes}. 
(Cf. Proposition 2.1 in \cite{KT-painleve})
\end{itemize}
\end{proposition}

The proof of Proposition \ref{NonSingSpectralCurve} is done by case-by-case checking, as follows (see also Appendix \ref{AppendixSpecP56} for computational details in Painlev\'e V and VI).
\begin{itemize} 
\item (Painlev\'e I) $E_{\rm I}(x)$ presents a double zero at $x=q_0$ and a simple zero at $x=-2q_0$. Consequently as soon as $q_0\neq 0$, i.e. $t\notin \Delta_{\rm I} = \{ 0 \}$ (since $6q_0^2+t=0$) these zeros are distinct. This exceptional situation precisely corresponds to the singular time.
\item (Painlev\'e II) $E_{\rm II}(x)$ presents a double zero at $x=q_0$ and two simple zeros at $x=-q_0\pm \sqrt{\frac{\theta}{q_0}}$. Thus we observe that these simple zeros are always distinct as soon as the monodromy parameter $\theta$ is not vanishing. Moreover, these simple zeros can never equal $q_0$ as soon as $t\notin \Delta_{\rm II}$. Indeed, saying that one of the simple zero equals the double zero $q_0$ at time $t$ is equivalent to say that $4q_0^3+\theta=0$ which precisely corresponds to the singular times.
\item (Painlev\'e III) $E_{\rm III}(x)$ has a double zero at $x=-\frac{1}{q_0}$ and generically two simple zeros solutions of 
$ x^2+\frac{2q_0(\theta_\infty q_0^2-\theta_0)}{\theta_0q_0^2-\theta_\infty}x+q_0^2 = 0$. The spectral curve is also singular at $x=0$. Requiring that we have at least a triple zero is equivalent to having a singular time \eqref{SingP3}. Moreover, requiring that coincidence of two simple zeros is equivalent to $q_0^2(q_0^4-1) (\theta_\infty^2-\theta_0^2)=0$. This precisely corresponds to a singular monodromy parameters and therefore can be discarded.
\item (Painlev\'e IV) $E_{\rm IV}(x)$ presents a double zero at $x=q_0$, a double pole at $x=0$ and generically two simple zeros satisfying $x^2+2(q_0+t)x+\frac{\theta_0^2}{q_0^2} = 0$. Requiring that a zero occurs at $x=0$ is equivalent to having the singular monodromy $\theta_0=0$. It may happen that for some time $t$, the two simple zeros coincide. This is equivalent to requiring that $q_0^2(q_0+t)^2=\theta_0^2$. Combining the equation together with the defining equation $t^2q_0^2+4tq_0^3+3q_0^4-2\theta_\infty q_0^2-\theta_0=0$ of $q_0$ leads to:
\beqq (\theta_0+ \theta_\infty)q_0^2=0 \text{ or } (\theta_0- \theta_\infty)q_0^2=0. \eeqq
Since $q_0$ can never vanish, it is equivalent to $\theta_0^2 =  \theta_\infty^2$ i.e. that we have singular monodromy parameters. Eventually, requiring that for some time $t$, one of the simple zero coincide with the double zero $q_0$ is equivalent to require that $3q_0^3+2tq_0+\frac{\theta_0^2}{q_0}=0$ which is precisely equivalent to \eqref{SingP4}, i.e. that we have a singular time.
\item (Painlev\'e V) The generic case corresponds to $E_{\rm V}(x)$ having two double poles at $x\in\{0,1\}$ with a double zero at $x=Q_0$ and two simple zeros at $Q_1$ and $Q_2$. In Appendix \ref{AppendixSpecP56}, the discussion leads to the fact that $Q_0$, $Q_1$ and $Q_2$ cannot be equal to $0$ or $1$ as soon as the monodromies $\theta_0, \theta_1$ and $\theta_\infty$ are non-vanishing. We also prove that the simple zeros can never coincide as long as the monodromy parameters are non-singular. Moreover, we show that requiring a triple zero is equivalent to having a singular time.
\item (Painlev\'e VI) The generic case corresponds to a $E_{\rm VI}(x)$ having a double zero at $x=q_0$ and two simple zeros of $P_2(x)=x^2+\left(-1-\frac{\theta_0^2t^2}{\theta_\infty^2 q_0^2}+\frac{\theta_1^2(t-1)^2}{\theta_\infty^2(q_0-1)^2}\right)x+\frac{\theta_0^2t^2}{\theta_\infty^2q_0^2}$. Note that the polynomial $P_2(x)$ is equivalently defined through the following conditions:
\beq  P_2(0)=\frac{\theta_0^2t^2}{\theta_\infty^2 q_0^2} \,\,,\,\, P_2(1)=\frac{(t-1)^2\theta_1^2}{\theta_\infty^2(q_0-1)^2}\text{ and } P_2(t)=\frac{t^2(t-1)^2\theta_t^2}{\theta_\infty^2(q_0-t)^2}.\eeq
So that we have:
\beq P_2(x)=\frac{\theta_0^2t}{\theta_\infty^2q_0^2}(x-1)(x-t)-\frac{(t-1)\theta_1^2}{\theta_\infty^2(q_0-1)^2}x(x-t)+\frac{t(t-1)\theta_t^2}{\theta_\infty^2(q_0-t)^2}x(x-1). \eeq
$E_{\rm VI}(x)$ also has double poles at $x\in\{0,1,t\}$. The discussion in Appendix \ref{AppendixSpecP56} leads to the fact that as long as the monodromy parameters are non-singular then the zeros can never equal the poles $\{0,1,t\}$ and that the simple zeros may never coincide. We show that requiring a triple zero is equivalent to having a singular time \eqref{SingP6}.
\end{itemize}

\subsection{Parametrization of spectral curves}
Since the spectral curves \eqref{eq:spec-curve} arising from Painlev\'e equations are of genus $0$ (as we observed in Section \ref{subsection:spec-compute}), they can be parametrized by a pair $(x(z), y(z))$ of rational functions, where $z$ represents a coordinate of ${\mathbb P}^1$.  Under Assumptions \ref{assumption:non-singular-times} and \ref{NonSingMono}, we observe that the function $E_J(x)$ always has at most two simple zeros, and other zeros or poles are not ramification points. Thus we may employ the Zhukovsky parametrization, which depends on the number of branch points of the covering of surfaces:
\begin{itemize}
\item (Painlev\'e I) This is the single branch point case. We have the parametrization
\begin{eqnarray} \label{ParametrizationP1}
\begin{cases}x(z) = z^{2} -2q_0 \\
y(z) = 2z (z^2 - 3q_0).
\end{cases} 
\end{eqnarray} 
\item (Painlev\'e II - VI) These are the two branch points cases. Denote generically $E_J(x)=(x-a)(x-b)C_{J}^2(x)$ with a rational function $C_J(x)$ of $x$. Then, we have the general parametrization
\begin{eqnarray} \label{eq:par-rep2}
\begin{cases}
\displaystyle x(z) = \frac{a+b}{2} + \frac{b-a}{4}\left( z+\frac{1}{z} \right) \\[+.7em]
\displaystyle y(z) = \frac{b-a}{4}\left( z-\frac{1}{z} \right) \, C_J(x(z)).
\end{cases} 
\end{eqnarray}
\end{itemize}
In the case of the Painlev\'e equations II through VI, the functions $C_J(x)$ are given by:
\bea \label{eq:expression-CJ}
C_{\rm II}(x)&=&(x-q_0) \,\,,\,\, C_{\rm III}(x)=\frac{(\theta_\infty-\theta_0q_0^2)(q_0 x+1)}{2x^2(q_0^4-1)} \,\,,\,\, 
C_{\rm IV}(x)=\frac{x-q_0}{x}, \nonumber \\[+.3em]
C_{\rm V}(x)&=&\frac{t(x-q_0)}{2x(x-1)} \,\,,\,\, C_{\rm VI}(x)=\frac{\theta_\infty (x-q_0)}{2x(x-1)(x-t)}.
\eea

Let $R = \{ z \in {\mathbb P^1}~|~ dx(z) = 0 \}$ be the set of branch point 
of the spectral curve. Generically, these points correspond to simple zeros of the function $E_J(x)$. 
\begin{proposition} \label{prop:generic-spectral-curve}
Under Assumptions \ref{assumption:non-singular-times} and \ref{NonSingMono}, we have the following:
\begin{itemize}
\item[(i)]
$R = \{0 \}$ (resp. $R=\{ +1, -1 \}$) in the case of 
\eqref{ParametrizationP1} (resp. \eqref{eq:par-rep2}).
For both cases, the points in $R$ are simple zeros of $dx(z)$. 
\item[(ii)]
$dy(z)$ never vanishes on $R$.
\end{itemize} 
\end{proposition}
\proof{
The first claim (i) is obvious in the case of \eqref{ParametrizationP1}. For the case \eqref{eq:par-rep2}, the claim is also true since the two simple zeros $a, b$ of $E_J(x)$ are distinct under the assumption, due to Proposition \ref{NonSingSpectralCurve}.

The second claim (ii) is also obvious for \eqref{ParametrizationP1} because Assumption \ref{assumption:non-singular-times} implies $q_0 \ne 0$. For \eqref{eq:par-rep2}, our claim is equivalent to the fact that $C_J(x)$ never vanish at $x = a$ and $b$, which can also be checked by  due to Proposition \ref{NonSingSpectralCurve}.  
}

\bigskip
Proposition \ref{prop:generic-spectral-curve} shows that the pair $(x(z), y(z))$ 
satisfies the regularity condition (e.g., Definition 2.3 in \cite{EO09}) so that 
we can apply the topological recursion of \cite{EO}. 

\subsection{Topological recursion \label{TopRec}}
Here we apply the topological recursion to the spectral curves \eqref{ParametrizationP1} or  \eqref{eq:par-rep2}. In both cases, there exists a global involution $z\mapsto \bar{z}$ satisfying:
\beqq x(\bar{z})=x(z) ~\text{  and  }~ y(\bar{z})= - y(z). \eeqq
In the parameterizations presented above, the involution is respectively $\bar{z}=-z$ for Painlev\'e ${\rm I}$ and $\bar{z}=z^{-1}$ for the other cases. Note that in the case of a genus $0$ curve, the involution is not only local 
%(i.e. valid around branch points) 
but global (i.e. valid on the whole ${\mathbb P}^1$). We now recall the definition of correlation functions and symplectic invariants as introduced by Eynard and Orantin in \cite{EO}.

\begin{definition}[Definition 4.2 of \cite{EO}]\label{EynardOrantin}For $g \ge 0$ and $n \ge 1$, the Eynard-Orantin differentials (known also as correlation functions) $\omega^{(g)}_{n}(z_{1},\dots,z_{n})$ of type $(g,n)$ associated to the spectral curve $(x(z),y(z))$ are defined by the following recursive relations:
\begin{eqnarray} \label{eq:top-rec}
\omega^{(0)}_{1}(z_{1}) & = & y(z_1) dx(z_1), \\
\omega^{(0)}_{2}(z_{1},z_{2}) & = & \frac{dz_{1}dz_{2}}{(z_{1}-z_{2})^{2}}, \\ 
\omega^{(g)}_{n+1}(z_{0},z_{1},\dots,z_{n}) & = & \sum_{r \in R} \underset{z\to r}{\Res} \,K(z_{0},z) \Bigl[ \omega^{(g-1)}_{n+1}(z,\bar{z},z_{1},\dots,z_{n})  \label{eq:top-rec-higher}\\
& & + \sum'_{\substack{g_{1} + g_{2} = g \cr 
I \cup J = \{1,\dots,n\} }} \omega^{(g_{1})}_{1+|I|}(z,z_{I})\omega^{(g_{2})}_{1+|J|}(\bar{z},z_{J}) \Bigr].  \nonumber
\end{eqnarray}
Here 
\begin{equation}
K(z_{0},z) = \frac{\int^{\bar{z}}_{z} \omega^{(0)}_{2}(\cdot, z_{0})}{(y(z)-y(\bar{z}))dx(z)}
\end{equation}
is called the recursion kernel, and the $\,'$ in the last line of \eqref{eq:top-rec-higher} means that the cases $(g_{1}, I) = (0, \emptyset)$ and $(g_{2}, J) = (0, \emptyset)$ are excluded from the sum. 
\end{definition}

The Eynard-Orantin differentials $\omega^{(g)}_{n}$'s are meromorphic multi-differentials and are known to be holomorphic in each variable $z_i$ except at the branch points if $(g,n) \ne (0,1), (0,2)$. In \cite{EO}, the authors also introduced symplectic invariants $F^{(g)}$ defined by 

\begin{definition}[Definition 4.3 of \cite{EO}] \label{def:symp-inv}The $g^{\text{th}}$ symplectic invariant of the spectral curve is defined by 
\begin{equation} \label{eq:def-symplectic-invariant}
F^{(g)} = \frac{1}{2-2g} \, \underset{r \in R}{\sum} \Res_{z \to r} \Phi(z) \omega^{(g)}_{1}(z) \quad \text{for $g \ge 2$}, 
\end{equation}
where 
\begin{equation}
\Phi(z) = \int^{z}_{z_{o}} y(\tilde{z})dx(\tilde{z}) \quad\text{($z_{o}$ is a generic point)}.
\end{equation}
% Note here that the $g^{\rm th}$ free energy is usually denoted by $F^{(g)}$ or $F_{g}$ in the literature (See \cite{EO}). We choose to use a different labeling since in this paper superscripts correspond to the order in $\hbar$ in the series expansion. 
$F^{(0)}$ and $F^{(1)}$ are defined with distinct and specific formulas (see \S 4.2.2 and \S 4.2.3 of \cite{EO} with a different sign convention). 
\end{definition}

We denote by $F^{(g)}_J$ the symplectic invariants defined from the spectral curve \eqref{eq:spec-curve} associated to the $J^{\rm th}$ Painlev\'e equation. 

% Note that this definition extends to the case $n\geq0$ (with the identification $\omega_0^{(g)}=F^{(g)}$). The following property is satisfied by the Eynard-Orantin differentials:
% \[
% \omega_n^{(g)}(z_1,\dots,z_n) = \frac{1}{2-2g-n}\underset{r \in R}{\sum} \Res_{z \to r} \Phi(z) \omega^{(g)}_{n+1}(z,z_1,\dots,z_n).
%  \quad \text{for $g \ge 0$}
% \]

\section{Determinantal formulas and topological type property}\label{DeterminantalFormSection}

\subsection{Determinantal formula} \label{subsection:correlation-function}

In this section we review the formalism of determinantal formulas (developed in \cite{BBEnew},\cite{Deter}) that connect the WKB solution of isomonodromy systems to the topological recursion correlation functions. 

Let us consider the differential equation 
\begin{equation} \label{eq:x-diff-eq}
\hbar \partial_x \Psi(x) = {\mathcal D}(x) \Psi(x),
\end{equation}
which is the first equation in our Lax system \eqref{eq:Lax-pair}. Here we are omitting the $t$-dependence and $\hbar$-dependence for simplicity. We consider the case that the matrix ${\mathcal D}$ in \eqref{eq:x-diff-eq} is one of ${\mathcal D}_J$ given in \eqref{P1}--\eqref{P6} which admits a series expansion \eqref{Dhbar}.  Take a matrix version of the formal WKB formal solution  
\begin{equation} \label{eq:WKB-solution}
\Psi(x) =
\begin{pmatrix} \psi(x) & \phi(x) \\ \td{\psi}(x) & \td{\phi}(x) 
\end{pmatrix} = \left( \sum_{k=0}^{\infty} \Psi^{(k)}(x) \hbar^k \right) \exp\left(\frac{T(x)}{\hbar} \right)
\end{equation}
of the system \eqref{eq:Lax-pair}, where the matrix $T$ is written in terms of $E_J$ given in Table \ref{table:spectral-curve-equation} as 
\begin{equation}
T(x) = {\rm diag}(s(x), -s(x)) ~~\text{with}~~
\frac{\partial s}{\partial x}(x) = \sqrt{E_J(x)}.
\end{equation}
We also impose a normalization condition: 
\begin{equation} \label{eq:normalization-condition}
\det \Psi(x) = 1. 
\end{equation}
Since the Lax matrices ${\mathcal D}_J$ are traceless, we can always find such a formal solution (see \S 2 in \cite{IS} for a construction of the WKB solution). Note that these conditions do not specify the WKB solution uniquely; it still has ambiguity of right-multiplication by a constant diagonal matrix with determinant one. However, such ambiguity disappears in the definition of correlation functions below.

Determinantal formulas are obtained from the Christoffel-Darboux kernel
\beq \label{PKK} 
K(x_1,x_2)=\frac{\psi(x_1)\td{\phi}(x_2)-
\td{\psi}(x_1)\phi(x_2)}{x_1-x_2}
\eeq
with the following definition:
\begin{definition}[Definition 2.3 of {\cite{Deter}}]\label{DeterFormm} The (connected) correlation functions are defined by:
\bea \label{eq:W1}
W_1(x)&=&\frac{\partial \psi}{\partial x}(x)\td{\phi}(x)-\frac{\partial \td{\psi}}{\partial x}(x)\phi(x),\\ 
\label{eq:Wn}
W_n(x_1,\dots,x_n)&=&-\frac{\delta_{n,2}}{(x_1-x_2)^2}+(-1)^{n+1}\sum_{\sigma: \text{$n$-cycles}}\prod_{i=1}^n K(x_i,x_{\sigma(i)}) ~\quad\text{for $n \ge 2$}. \eea 
\end{definition}

By definition, the correlation functions $W_n$ is a formal power series in $\hbar$ of the form 
\begin{equation} \label{eq:correlation-series-pre}
W_1(x) = \sum_{g=0}^{\infty} w_1^{(g)}(x) \hbar^{g-1}~~,~~
W_n(x_1,\dots,x_n) = \sum_{g=0}^{\infty} w_n^{(g)}(x_1,\dots,x_n) \hbar^{g}.
\end{equation}
Here $w_n^{(k)}(x_1,\dots,x_n)$ is a symmetric function of $x_{1}, \dots, x_{n}$. From the view point of the WKB method, $w_n^{(k)}$ possibly has singularities at zeros and poles of $E_J(x)$. 

Note that there exists an alternative expression for the correlation functions in terms of a rank $1$ projector:
\beq \label{defM} 
M(x)=\Psi(x) \begin{pmatrix}1&0\\0&0\end{pmatrix} \Psi^{-1}(x)=\begin{pmatrix}\psi\td{\phi}&-\psi\phi\\ \td{\psi}\td{\phi}&-\phi\td{\psi}\end{pmatrix}.
\eeq
In fact this is the canonical projector on the first coordinate taken into the basis defined by $\Psi(x)$. The rank $1$ projector satisfies:
\beq M^2=M \,\,,\,\, \Tr M=1 \,\,,\,\, \det M=0. \eeq
% We remark that, under a general gauge transformation $\td{\Psi}(x,t)=U(x,t)\Psi(x,t)$, $M$ is only affected by a simple change of basis: $\td{M}(x,t)=U^{-1}(x,t)M(x,t)U(x,t)$. 
Theorem $2.1$ of \cite{Deter} gives an alternative expression for $W_n(x_1,\dots,x_n)$ in terms of the matrix $M(x)$:
\bea \label{alternative} 
W_1(x)&=&-\frac{1}{\hbar} \Tr (\mathcal{D}(x)M(x)),\\
W_2(x_1,x_2)&=&\frac{\Tr (M(x_1)M(x_2)) -1}{(x_1-x_2)^2},\\
\label{eq:alternative-Wn}
W_n(x_1,\dots,x_n)&=& (-1)^{n+1}\Tr \sum_{\sigma: \text{$n$-cycles}} \prod_{i=1}^n \frac{M(x_{\sigma(i)})}{x_{\sigma(i)}-x_{\sigma(i+1)}} \nonumber \\
&=&\frac{(-1)^{n+1}}{n}\sum_{\sigma \in S_n} \frac{ \Tr M(x_{\sigma(1)})\dots M(x_{\sigma(n)})}{(x_{\sigma(1)}-x_{\sigma(2)})\dots (x_{\sigma(n-1)}-x_{\sigma(n)})(x_{\sigma(n)}-x_{\sigma(1)})} \quad  \text{for $n \ge 3$}. \nonumber \\
\eea
From this alternate expression, it turns out that the correlation functions are invariant under a certain class of gauge transformations (see \S 2.4 in \cite{BBEnew}).

Note that the definition of the correlation functions only involves the matrix $\mathcal{D}(x)$. In fact these definitions apply for any $2\times 2$ linear system of the form \eqref{eq:x-diff-eq} even if it does not come from a Lax pair. We also mention that as presented in \cite{BBEnew} and \cite{Deter}, the correlation functions satisfy the so-called loop equations (i.e. an infinite set of relations connecting the various functions). The loop equations will be used in Appendix \ref{AppendixLeadingOrder} to prove our main results.

\subsection{Topological type property}

Following the work of Berg\`ere, Borot and Eynard, we now give the definition of the topological type property (TT property for short). We employ the definition which is valid when the spectral curve of \eqref{eq:x-diff-eq} given in Definition \ref{def:spec-curve} is of genus $0$ (see \cite{BBEnew, Deter} for general case). This is enough for our purpose since the spectral curves arising from the Lax pair for Painlev\'e equations are of genus $0$ as we have seen in Section \ref{subsection:spec-compute}.  

\begin{definition}[Definition 3.3 of \cite{BBEnew}, \S 2.5 of \cite{Deter}] \label{def:TT-property-Lax}
The differential system \eqref{eq:x-diff-eq} is said to be of topological type, if the correlation functions $W_{n}$ given in definition \eqref{eq:W1} and \eqref{eq:Wn}
(or \eqref{alternative}--\eqref{eq:alternative-Wn}) satisfy the following conditions: 
\begin{enumerate} 
% \item[(0)] \underline{Existence of a series expansion in $\hbar$}: The correlation functions admit a series expansion in $\hbar$ of the form:
% \beq W_n(x_1,\dots,x_n)=\sum_{g=0}^\infty W_n^{(g)}(x_1,\dots,x_n)\hbar^g\eeq
\item[(1)] \underline{Parity property}: $W_{n}|_{\hbar \hspace{+.1em} \mapsto - \hbar} = (-1)^{n} W_{n}$ holds for $n \ge 1$. This is equivalent to say that the series expansion \eqref{eq:correlation-series-pre} contains only even (resp. odd) degree terms in $\hbar$ when $n$ is even (resp. odd).
\item[(2)] \underline{Pole structure}: The functions $w_n^{(g)}(x_1,\dots,x_n)$ in \eqref{eq:correlation-series-pre} only have poles at the branch points of the spectral curve \eqref{eq:spec-curve} when $(g,n) \neq (0,1), (0,2)$, and $w^{(0)}_2(x_1,x_2)$ has a double pole at $x_1 = x_2$ with no other poles. In fact for genus $0$ curves we must have $w_2^{(0)}(x(z_1),x(z_2))dx(z_1)dx(z_2)=\frac{dz_1 dz_2}{(z_1-z_2)^2}$.
\item[(3)] \underline{Leading order}: The leading order of the series expansion of the correlation function $W_n$ is at least of order $\hbar^{n-2}$.
\end{enumerate}
% We also say that the Lax system \eqref{eq:Lax-pair} is of topological type if the differential equation \eqref{eq:x-diff-eq} in the Lax pair is of topological type in the above sense.
\end{definition}

In summary, these conditions are necessary and sufficient to prove that the functions $W_n$ admit a series expansion of the form:
\beq \label{Exponents} W_n(x_{1},\dots,x_{n})=\sum_{g=0}^\infty \hbar^{2g-2+n}W_n^{(g)}(x_1,\dots,x_n) \quad \text{for $n \ge 1$} \eeq
with $W_n^{(g)}(x_1,\dots,x_n)$ regular at the even zeros of the function $E_J(x)$. Note that, in the formulation of TT property, we do not need the time evolution of the Lax pair. However, in many examples (including the Lax pairs in this paper), the time evolution equation is effectively used to prove the TT property.  (Cf. Remark \ref{remark:advantage-of-Lax-pair} for example).

%\begin{remark}
%If the spectral curve $Y^{2} = E_{\rm J}(x)$ is of genus greater than one, we must add a condition on filling fractions (period integrals of $W_{n}^{(g)}$ along %closed cycles on the spectral curve). See Section $2.5$ of \cite{Deter} for details.
%\end{remark}

The main interest of the TT property is that it is a sufficient condition to prove the connection with the topological recursion. Indeed, it is proved in \cite{BBEnew} and \cite{Deter} that:

\begin{theorem}[Theorem $2.1$ of \cite{Deter}, Theorem 3.1 and Corollary $4.2$ of \cite{BBEnew}]
\label{thm:TT}
Suppose that the differential equation \eqref{eq:x-diff-eq} satisfies the TT property. Then, we have the following: 
\begin{itemize}
\item[(i)]  
The functions $W_n^{(g)}(x_{1},\dots,x_{n})$ appearing in the formal expansion \eqref{Exponents} of the correlation functions $W_{n}$ are identical to the Eynard-Orantin differentials $\omega^{(g)}_{n}(z_1,\dots,z_n)$ obtained from the topological recursion applied on the spectral curve in the following way:
\begin{equation} \label{eq:EO-and-Wgn}
W^{(g)}_{n}(x(z_{1}),\dots,x(z_{n}))dx(z_{1})\cdots dx(z_{n}) = \omega^{(g)}_{n}(z_{1},\dots,z_{n}) \quad \text{for $g \ge 0$ and $n \ge 1$},
\end{equation}
where $x(z)$ appears in the parametrization \eqref{ParametrizationP1}--\eqref{eq:par-rep2} of the spectral curve. 

\item[(ii)] 
Moreover, if the equation \eqref{eq:x-diff-eq} is a part of the Lax system \eqref{eq:Lax-pair}, then the generating function 
\begin{equation} \label{eq:fg-sigma}
\ln \tau = - \sum_{g=0}^{\infty} \hbar^{2g-2} F^{(g)}
%\frac{dF^{(2g)}}{dt}(t) = \quad \text{for $g \ge 0$} \text{ where } \frac{d}{dt}\tau^{(2g)} \text{ is defined in }\eqref{Orderlogtau}
% \hbar^2 \frac{d}{dt} \Bigl( \sum_{g=0}^{\infty} \hbar^{2g-2} 
% F_{\rm J}^{(g)} \Bigr) = \sigma_{\rm J}(t,\hbar)
\end{equation}
of the symplectic invariants obtained from the topological recursion applied to the spectral curve \eqref{eq:spec-curve} gives the isomonodromic $\tau$-function for the Lax system \eqref{eq:Lax-pair} in the sense of Jimbo-Miwa-Ueno \cite{JMI}. 
%The $\tau$-function for the $J^{\rm th}$ Painlev\'e equation matches with the generating function of the symplectic invariants obtained from the topological recursion applied to the spectral curve \eqref{eq:spec-curve}. That is, if we denote by $F_J^{(g)}$ the $g^{\rm th}$ symplectic invariants of the spectral curve \eqref{eq:spec-curve}, then the formal series
% satisfies the defining equation of the $\tau$-function given in Section \ref{OkamotoForm}.
\end{itemize}
\end{theorem}

Note that the minus sign arising in \eqref{eq:fg-sigma} is just a matter of convention regarding the definition of the $F^{(g)}$ (we followed definition of \cite{EO}). The previous theorem is particularly interesting since it shows that the topological recursion reconstructs the formal series expansions of the determinantal formulas and the $\tau$-function of the Lax system. 
% Note that the TT property is a sufficient condition for a Lax system to be reconstructed from the topological recursion. However at the moment it is not known whether the TT property is really necessary but so far all known cases in which the topological recursion is known to reconstruct the determinantal formulas are of topological type. 

%%%%%%%%%%%%%%%%%%%%%%%%%%%%%%%%%
\subsection{Main theorem}\label{SectionMainTheorem}

Our main theorem is formulated as follows:

\begin{theorem}\label{MainTheo} For each $J = {\rm I}, \dots, {\rm VI}$, 
under Assumptions \ref{assumption:non-singular-times} and \ref{NonSingMono}, the first equation $\hbar \partial_x \Psi = {\mathcal D}_J \Psi$ in the Lax system \eqref{eq:Lax-pair} associated to the $J^{\text{th}}$ Painlev\'e equation (with the Lax matrices in \eqref{P1}--\eqref{P6}) satisfies the TT property. Therefore, the $\hbar$-expansion of the $\tau$-function and correlation functions $W_n$ are respectively identified with the generating functions of symplectic invariants $F_{J}^{(g)}$ and the correlation functions of the differentials $\omega_{n}^{(g)}$ computed from the topological recursion applied to the corresponding spectral curve \eqref{ParametrizationP1}--\eqref{eq:par-rep2} as follows:
\begin{eqnarray} \label{eq:main-thm1} \ln \tau_J(t,\hbar) & = & - \sum_{g=0}^{\infty} \hbar^{2g-2} F^{(g)}_J(t), \\
\label{eq:main-thm2}W_n\bigl( x(z_1),\dots,x(z_n) \bigr) dx(z_1) \cdots dx(z_n)& = & \sum_{g=0}^{\infty} \hbar^{2g-2+n} \omega^{(g)}_n(z_1,\dots,z_n) \quad \text{for $n \ge 1$}.
\end{eqnarray}
\end{theorem}

To support and complete our theorem we provide in Appendix \ref{AppendixSymplecticInvariants} the specific computations of $F^{(0)}$ and $F^{(1)}$ and prove that they match up with $-\tau^{(0)}$ and $-\tau^{(1)}$ up to some additive $t$-independent constants. (Recall that the $\tau$-function is only defined up to constants, and only the matching of the derivatives makes sense). The general proof of the TT property for all Lax pairs listed in Section \ref{LaxPairs} is provided in Appendices \ref{AppendixParityProof}, \ref{AppendixProofPole}, \ref{AppendixLeadingOrder} where we will prove that the three conditions (1)--(3) of Definition \ref{def:TT-property-Lax} hold.

%%%%%%%%%%%%%%%%%%%%%%%%%%%%%%%%%
%%%%%%%%%%%%%%%%%%%%%%%%%%%%%%%%%
\section{Conclusion}\label{SectionConclusion}
In this article, we showed how to introduce a small formal $\hbar$ parameter by rescaling of variables $x$, $t$ and the monodromy parameters $\theta_\ast$ in the formalism of the six Painlev\'e equations and associated Lax pairs. In this formalism we then presented a proof of the TT property for all six Painlev\'e cases as well as the various connections with the $\tau$-function, Okamoto's $\sigma$-functions and Hamiltonians of the underlying problems. The proof of the TT property implies that for the six Painlev\'e equations, we can match the $\hbar$-formal series expansions of the correlation functions (i.e. determinantal formulas) $W_n$ and the $\tau$-functions $\ln \tau$ with the Eynard-Orantin differentials $\omega_n^{(g)}$ and symplectic invariants $F^{(g)}$ computed from the topological recursion applied on the spectral curve associated to the Lax pair, respectively. 

Several questions arise from this work that would deserve further study:
\begin{itemize}
\item 
We have introduced the $\hbar$ parameter with a rescaling of the monodromy parameters but it would be interesting to see if other interesting rescalings lead to regimes for which a spectral curve can be defined.
\item 
The approach developed in this article is purely formal since we assumed that solutions of Painlev\'e equations had a series expansion in $\hbar$. However it is well known in matrix models and in perturbation theory that these series expansions may not be convergent. As is shown in \cite{Kam-Ko}, the formal series solution of the Painlev\'e equations are Borel summable under some conditions. Study of non-perturbative effect (i.e. non-linear Stokes phenomenon; see \cite{FIKN, Kapaev} for example) with the aid of topological recursion is an important open problem.
% \item 
% Connections between the topological recursion, matrix models, and integrable systems are now well documented in the literature, and this article corroborates this point. However, since the strategy presented in this article (selection of a good gauge, introduction of $\hbar$, computation of the spectral curve and proof of the topological type property) is quite general, it seems that it could be successfully attempted on other integrable systems.
% \item 
% In this article, we introduced the $\hbar$ parameter from the Lax pair (by rescaling the parameters) and deduced the corresponding modifications on the Hamiltonians underlying the Painlev\'e equations. Surprisingly we found that for Painlev\'e equations $1,2,4$ and $5$, the introduction of the $\hbar$ parameter does not change the explicit expression of the Hamiltonians (see theorem \ref{HamiltonianFormulation}). For Painlev\'e equations $3$ and $6$, only a linear term in $\hbar$ appears, and we can always obtain the tau-function (see \eqref{Relation}) by taking $\ln \tau_J(t)= H_J(p(t),q(t),t,\hbar=0)$. This suggests that our parameter $\hbar$ may have some nice interpretation in the Hamiltonian formulation. 
\item 
As we mentioned in Remark \ref{remark:Borel-sum}, the formal solution discussed in this paper is very specific one. General solutions contain two free parameters since the Painlev\'e equation is second order (these free parameters are regarded as a coordinate of the Okamoto space of initial conditions). In 2012, Gamayun-Iorgov-Lisovyy gave an explicit expression of general $\tau$-function of Painlev\'e VI which contains two free parameters \cite{GIL}. They propose a new method to compute the Painlev\'e $\tau$-function via the Virasoro conformal blocks (with central charge = 1), and the Alday-Gaiotto-Tachikawa correspondence \cite{AGT} gives an explicit combinatorial expression of the conformal blocks. Their results are generalized to some class of Painlev\'e equations and $q$-Painlev\'e equations (see \cite{GL,ILT, JNS}). It is interesting to find a new method to compute the general $\tau$-function based on the topological recursion, and a relation to the results based on CFT and AGT. Natural candidates in this direction are described in \cite{BorEyn, Eyn17}. The paper \cite{BLMST} also gives some ideas of how we can build general expansions of general solutions of Painlev\'e equations as Fourier transform of some functions that do have a topological expansion.

\end{itemize} 

\section*{Appendix}

\appendix

%%%%%%%%%%%%%%%%%%%%%%%%%%%%%%%%%
%%%%%%%%%%%%%%%%%%%%%%%%%%%%%%%%%
%%%%%%%%%%%%%%%%%%%%%%%%%%%%%%%%%
\section{\label{AppendixConnectionJM}Connection with Jimbo-Miwa Lax pairs}

In \cite{JMII}, the authors produced a list of Lax pairs corresponding to all six $\hbar$-independent Painlev\'e equations (i.e., the Painlev\'e equations in Table \ref{table:P-eq} with $\hbar = 1$). In this paper we employed slightly different Lax pairs, and for completeness, we describe here the various transformations connecting both sets. 
To avoid confusion, we denote with a label ``$JM$" all quantities appearing in Jimbo-Miwa paper \cite{JMII}. The Jimbo-Miwa systems are of the form
\begin{equation} \label{LaxJM} 
\frac{\partial }{\partial x}Y_{JM}(x,t)=A_{JM}(x,t)Y_{JM}(x,t), 
\quad \frac{\partial }{\partial t}Y_{JM}(x,t)=B_{JM}(x,t)Y_{JM}(x,t),
\end{equation}
where $A_{JM}(x,t)$ and $B_{JM}(x,t)$ are $2\times 2$ matrices.

%%%%%%%%%%%%%%%%%%%%%%%%%%%%%%%%%
\subsection{Gauge transformation}
First, we describe a relation between the Jimbo-Miwa system \eqref{LaxJM} and 
\begin{equation} \label{Lax-hbar-is-1}
\frac{\partial }{\partial x} \tilde{\Psi}(x,t)= 
\tilde{\mathcal D}(x,t) \tilde{\Psi}(x,t), 
\quad 
\frac{\partial }{\partial t}\tilde{\Psi}(x,t)=
\tilde{\mathcal R}(x,t) \tilde{\Psi}(x,t)
\end{equation}
which is obtained from our system \eqref{eq:Lax-pair} by just setting $\hbar = 1$. Namely, $(\tilde{\mathcal D}, \tilde{\mathcal R}) = ({\mathcal D}, {\mathcal R})\bigl|_{\hbar=1}$, where ${\mathcal D}$ and ${\mathcal R}$ are the Lax matrices given in \eqref{P1}--\eqref{P6}. (Here we omit the label $J$ of the Painlev\'e equations for simplicity.) 

The Lax systems \eqref{LaxJM} and \eqref{Lax-hbar-is-1} are related by an appropriate gauge transformation. A gauge transformation $\td{\Psi}(x,t)=U(x,t) Y_{JM}(x,t)$ yields the following relation between Lax matrices:
\bea \label{Gauge}\td{\mathcal{D}}(x,t)&=&U(x,t)A_{JM}(x,t)U^{-1}(x,t)+ \frac{\partial U}{\partial x}(x,t)U^{-1}(x,t) \notag \\[+.3em]
\td{\mathcal{R}}(x,t)&=&U(x,t)B_{JM}(x,t)U^{-1}(x,t)+\frac{\partial U}{\partial t}(x,t)U^{-1}(x,t).
\eea

The gauge transform between \eqref{LaxJM} and \eqref{Lax-hbar-is-1} are given as follows.

\begin{itemize}
\item 
In the case of Painlev\'e I, the Jimbo-Miwa Lax pair ($A_{JM}, B_{JM}$) coincides with ($\tilde{\mathcal D}, \tilde{\mathcal R}$) under the identification $(p,q)=(z_{JM},y_{JM})$.

\item 
The Lax pair for Painlev\'e II proposed by Jimbo and Miwa is
\beaa A_{JM}(x,t)&=&\begin{pmatrix}x^2+z_{JM}+\frac{t}{2}&u_{JM}(x-y_{JM})\\ -2u_{JM}^{-1}\left(z_{JM}x+z_{JM}y_{JM}+\theta\right) &-\left(x^2+z_{JM}+\frac{t}{2}\right)\end{pmatrix}\cr
B_{JM}(x,t)&=&\frac{1}{2}\begin{pmatrix}x&u_{JM}\\ -2u_{JM}^{-1}z_{JM}&-x\end{pmatrix},
\eeaa
where $y_{JM}(t)$, $z_{JM}(t)$ and $u_{JM}(t)$ are functions of $t$. Our Lax pair \eqref{Lax-hbar-is-1} is connected to the former one with
\beqq \tilde{\Psi}(x,t)=\begin{pmatrix} u_{JM}^{-\frac{1}{2}}(t)&0\\0&u_{JM}^{\frac{1}{2}}(t)\end{pmatrix} Y_{JM}(x,t) ~~\text{ and }~~ 
\left(p,q\right)=\left(z_{JM},y_{JM}\right).\eeqq 
Here we used $\frac{d}{dt} \ln u_{JM} = -y_{JM}$.
\item The Lax pair for Painlev\'e III proposed by Jimbo and Miwa is given by
\beaa A_{JM}(x,t)&=&\frac{t}{2}\sigma_3+\frac{1}{x}\begin{pmatrix}-\frac{\theta_\infty}{2}&u_{JM}\\ v_{JM} &\frac{\theta_\infty}{2}\end{pmatrix}+\frac{1}{x^2}\begin{pmatrix} z_{JM}-\frac{t}{2}&-w_{JM}z_{JM}\\ w_{JM}^{-1}(z_{JM}-t)& -z_{JM}+\frac{t}{2}\end{pmatrix}\cr
B_{JM}(x,t)&=&\frac{x}{2}\sigma_3+ \frac{1}{t}\begin{pmatrix}0&u_{JM}\\ v_{JM}&0\end{pmatrix}-\frac{1}{tx}\begin{pmatrix} z_{JM}-\frac{t}{2}&-w_{JM}z_{JM}\\ w_{JM}^{-1}(z_{JM}-t)& -z_{JM}+\frac{t}{2}\end{pmatrix},
\eeaa
where $u_{JM}=-z_{JM}y_{JM}w_{JM}$ and $v_{JM}=\frac{-(z_{JM}-t)y_{JM}-\theta_\infty+\frac{t}{2z_{JM}}(\theta_0+\theta_\infty)}{w_{JM}}$. The connection with our Lax pair \eqref{Lax-hbar-is-1} is given by the gauge transformation
\beqq \tilde{\Psi}(x,t)=\begin{pmatrix} w_{JM}^{-\frac{1}{2}}(t)&0\\0&w_{JM}^{\frac{1}{2}}(t)\end{pmatrix} Y_{JM}(x,t) ~~\text{ and }~~ \left(p,q\right)=\left(z_{JM},y_{JM}\right).\eeqq
Here we used $t \frac{d}{dt} \ln w_{JM} = - \frac{(\theta_0+\theta_\infty)t}{z_{JM}}-2t y_{JM}+\theta_\infty$.

\item The Lax pair for Painlev\'e IV proposed by Jimbo and Miwa is given by
\begin{eqnarray*}
 A_{JM}(x,t)&=& x \sigma_3 
 + \begin{pmatrix} t & u_{JM} \\ \frac{2(z_{JM}-\theta_0-\theta_\infty) }{u_{JM}}& -t \end{pmatrix} 
 + \frac{1}{x} \begin{pmatrix} -z_{JM}+\theta_0 & - \frac{u_{JM} y_{JM}}{2} \\ 
 \frac{2z_{JM}(z_{JM}-2\theta_0)}{u_{JM} y_{JM}} & z_{JM} - \theta_0 \end{pmatrix}  \\
 B_{JM}(x,t)&=& x \sigma_3 + \begin{pmatrix} 0 & u_{JM} \\ \frac{2(z_{JM}-\theta_0-\theta_\infty)}{u_{JM}} & 0\end{pmatrix}
 \end{eqnarray*}

Our Lax pair \eqref{Lax-hbar-is-1} can be obtained from the former Lax pair by the gauge transformation
\beqq \tilde{\Psi}(x,t)=\begin{pmatrix}u_{JM}^{-\frac{1}{2}}&0\\ 0&u_{JM}^\frac{1}{2}\end{pmatrix}Y_{JM}(x,t)~~\text{and}~~
(p,q)=\left(-\frac{2z_{JM}}{y_{JM}},\frac{y_{JM}}{2}\right).
\eeqq
Here we used $\frac{d}{dt} \ln u_{JM} = - y_{JM} -2t$.

\item 
The Lax pair for Painlev\'e V proposed by Jimbo and Miwa is
\begin{eqnarray*}
A_{JM}(x,t)&=& \frac{t}{2} \sigma_3 
+ \frac{1}{x} \begin{pmatrix} z_{JM} + \frac{\theta_0}{2} & -u_{JM} (z_{JM} + \theta_0) \\ 
\frac{z_{JM}}{u_{JM}} & - z_{JM} - \frac{\theta_0}{2} \end{pmatrix} \\ 
& & 
 + \frac{1}{x-1}  \begin{pmatrix} -z_{JM} - \frac{\theta_0+\theta_\infty}{2} & 
 u_{JM} y_{JM} \left( z_{JM} + \frac{\theta_0-\theta_1+\theta_\infty}{2} \right)  \\[+.3em]  
 -\frac{1}{u_{JM} y_{JM}}  \left( z_{JM} + \frac{\theta_0+\theta_1+\theta_\infty}{2} \right)  & 
 z_{JM} + \frac{\theta_0+\theta_\infty}{2} \end{pmatrix}  \\
B_{JM}(x,t)&=& \frac{x}{2} \sigma_3 \\ 
& & \hspace{-4.em} + \frac{1}{t} 
\begin{pmatrix} 0 & u_{JM} \left( z_{JM} + \theta_0 - y_{JM} \left( z_{JM} + \frac{\theta_0-\theta_1+\theta_\infty}{2} \right) \right) \\
\frac{1}{u_{JM}} \left( z_{JM}  - \frac{1}{y_{JM}} \left( z_{JM} + \frac{\theta_0+\theta_1+\theta_\infty}{2} \right) \right)  & 0   
\end{pmatrix}
\end{eqnarray*}

We note here a typo in the Lax pair proposed in \cite{JMII} where $B_{JM}(x,t)$ lacks the $x$ factor. 
We can obtain the Lax pair \eqref{Lax-hbar-is-1} for Painlev\'e V from this one by the gauge transformation
\beqq \tilde{\Psi}(x,t)=\begin{pmatrix}u_{JM}^{-\frac{1}{2}}&0\\ 0&u_{JM}^\frac{1}{2}\end{pmatrix}Y_{JM}(x,t)
~~\text{and}~~(p,q)=\left( z_{JM}y_{JM},\frac{1}{y_{JM}} \right).
\eeqq
Here we used
$t\frac{d}{dt} \ln u_{JM} = - 2z_{JM}-\theta_0+y_{JM} \left(z_{JM} + \frac{\theta_0-\theta_1+\theta_\infty}{2}  \right) 
+ \frac{1}{y_{JM}} \left(z_{JM} + \frac{\theta_0+\theta_1+\theta_\infty}{2}  \right)$.

\item 
The Lax pair for Painlev\'e VI proposed by Jimbo and Miwa is
\beqq A_{JM}(x,t)=\frac{\left(A_{0}\right)_{JM}}{x} +\frac{\left(A_{1}\right)_{JM}}{x-1}+\frac{\left(A_{t}\right)_{JM}}{x-t} \text{ and } B(x,t)=-\frac{ \left(A_{t}\right)_{JM}}{x-t},\eeqq
where the matrices $\left(A_{0}\right)_{JM}$, $\left(A_{1}\right)_{JM}$ and $\left(A_{t}\right)_{JM}$ are defined by
\beaa \left(A_0\right)_{JM}&=&\begin{pmatrix} (z_0)_{JM}+\theta_0&-u_{JM}(z_0)_{JM}\\ u_{JM}^{-1}((z_0)_{JM}+\theta_0)&-(z_0)_{JM}\end{pmatrix}\cr
\left(A_1\right)_{JM}&=&\begin{pmatrix} (z_1)_{JM}+\theta_1&-v_{JM}(z_1)_{JM}\\ v_{JM}^{-1}((z_1)_{JM}+\theta_1)&-(z_1)_{JM}\end{pmatrix}\cr
\left(A_t\right)_{JM}&=&\begin{pmatrix} (z_t)_{JM}+\theta_t&-w_{JM}(z_t)_{JM}\\ w_{JM}^{-1}((z_t)_{JM}+\theta_t)&-(z_t)_{JM}\end{pmatrix}
\eeaa
with
\beqq u_{JM}=\frac{k_{JM}y_{JM}}{t (z_0)_{JM}} \,\,,\,\,v_{JM}=-\frac{k_{JM}(y_{JM}-1)}{(t-1) (z_1)_{JM}} \,\,,\,\, w_{JM}=\frac{k_{JM}(y_{JM}-t)}{t(t-1) (z_t)_{JM}}. \eeqq
We also set 
\[
(A_\infty)_{JM}=- \left( (A_0)_{JM}+(A_1)_{JM}+(A_t)_{JM} \right) =\begin{pmatrix} \frac{\theta_\infty-\theta_0-\theta_1-\theta_t}{2} &0\\0 & -\frac{\theta_\infty+\theta_0+\theta_1+\theta_t}{2}\end{pmatrix}.
\]
Our Lax pair \eqref{Lax-hbar-is-1} is connected to the previous Lax pair via a gauge transformation
\beqq \tilde{\Psi}(x,t)=x^{-\frac{\theta_0}{2}}(x-1)^{-\frac{\theta_1}{2}}(x-t)^{-\frac{\theta_t}{2}}\begin{pmatrix} k_{JM}(t)^{-\frac{1}{2}}&0\\0&k_{JM}(t)^{\frac{1}{2}}\end{pmatrix}Y_{JM}(x,t)\eeqq
and the identifications
\beqq z_0=(z_0)_{JM} \,\,,\,\, z_1=(z_1)_{JM}, z_t=(z_t)_{JM} \text{  and  } (p,q)=(z_{JM},y_{JM}).\eeqq
Here we used $t\frac{d}{dt} \ln k_{JM} = (\theta_\infty - 1) \frac{y_{JM}-1}{t(t-1)}$.
\end{itemize}

%%%%%%%%%%%%%%%%%%%%%%%%%%%%%%%%%%%%%%%%%%%%
\subsection{Introduction of $\hbar$ by rescaling}
Here we modify the previous Lax pair \eqref{Lax-hbar-is-1} by introducing $\hbar$ in the Lax system. Introducing $\hbar$ in this way may appear arbitrary, but for all six cases we can find the $\hbar$-dependent version given in \eqref{P1}--\eqref{P6} by a proper rescaling. 
Let us put tilde's on the variables $x$, $t$ etc.\, in the system \eqref{Lax-hbar-is-1}, 
and perform a rescaling of the form
\beq \label{Scaling}(\td{t},\td{x},\td{q},\td{p},\td{\theta}_i)=(\hbar^{\delta_t}t,\hbar^{\delta_x}x,\hbar^{\delta_q}q,\hbar^{\delta_p}p,\hbar^{\delta_i}\theta_i) \text{ and }\td{\Psi}=\begin{pmatrix} \hbar^{\delta_\Psi} &0\\0&\hbar^{-\delta_\Psi}\end{pmatrix}\Psi \eeq
with suitable exponents. Then the system satisfied by $\Psi$ in the new variables $x$ and $t$ is nothing but the $\hbar$-dependent versions of the systems with Lax matrices \eqref{P1}-\eqref{P6}. Note that as soon as the general form of the rescaling \eqref{Scaling} is chosen, the choice of exponents is almost unique if we impose that the resulting system can be treated by the WKB method. The explicit choice of the exponents are given as follows: 

\begin{itemize}
\item (Painlev\'e I) 
\beq \label{ScalingP1}(\td{t},\td{x},\td{q},\td{p})=\left(\hbar^{-\frac{4}{5}}t,\hbar^{-\frac{2}{5}}x,\hbar^{-\frac{2}{5}}q,\hbar^{-\frac{3}{5}}p\right) \text{ and } \td{\Psi}= \begin{pmatrix} \hbar^{-\frac{1}{10}}&0\\ 0 &\hbar^{\frac{1}{10}}\end{pmatrix}\Psi. \eeq
\item (Painlev\'e II) 
\beq \label{ScalingP2}(\td{t},\td{x},\td{q},\td{p},\td{\theta})= \left(\hbar^{-\frac{2}{3}}t, \hbar^{-\frac{1}{3}}x,\hbar^{-\frac{1}{3}}q, \hbar^{-\frac{2}{3}}p, \hbar^{-1}\theta\right) \text{ and } \td{\Psi}= \begin{pmatrix} \hbar^{-\frac{1}{6}}&0\\ 0 &\hbar^{\frac{1}{6}}\end{pmatrix}\Psi. \eeq
\item (Painlev\'e III) 
\beq \label{ScalingP3} (\td{t},\td{x},\td{q},\td{p},\td{\theta}_0,\td{\theta}_\infty)=\left(\hbar^{-1} t, x,q, \hbar^{-1} p, \hbar^{-1} \theta_0,\hbar^{-1}\theta_\infty\right) \text{ and } \td{\Psi}=\Psi. \eeq
\item (Painlev\'e IV) 
\beq \label{ScalingP4} (\td{t},\td{x},\td{q},\td{p},\td{\theta}_0,\td{\theta}_\infty)=\left(\hbar^{-\frac{1}{2}}t, \hbar^{-\frac{1}{2}}x,\hbar^{-\frac{1}{2}}q, \hbar^{-\frac{1}{2}} p, \hbar^{-1}\theta_0,\hbar^{-1}\theta_\infty\right) \text{ and } \td{\Psi}= \begin{pmatrix} \hbar^{\frac{1}{4}}&0\\ 0 &\hbar^{-\frac{1}{4}}\end{pmatrix}\Psi. \eeq
\item (Painlev\'e V) 
\beq\label{ScalingP5} (\td{t},\td{x},\td{q},\td{p},\td{\theta}_0,\td{\theta}_1,\td{\theta}_\infty)=\left(\hbar^{-1} t, x,q, \hbar^{-1} p, \hbar^{-1}\theta_0,\hbar^{-1} \theta_1,\hbar^{-1}\theta_\infty\right) \text{ and } \td{\Psi}=\Psi. \eeq 
\item (Painlev\'e VI)
\begin{multline} \label{ScalingP6} 
(\td{t},\td{x},\td{q},\td{z}_0,\td{z}_1,\td{z}_t,\td{\theta}_0,\td{\theta}_1,\td{\theta}_t,\td{\theta}_\infty)=\left(t,x,q,\hbar^{-1} z_0,\hbar^{-1} z_1, \hbar^{-1} z_t, \hbar^{-1}\theta_0,\hbar^{-1}\theta_1,\hbar^{-1} \theta_t,\hbar^{-1}\theta_\infty\right) \\ 
\text{ and } \td{\Psi}= \begin{pmatrix} \hbar^{-\frac{1}{2}}&0\\ 0 &\hbar^{\frac{1}{2}}\end{pmatrix}\Psi.
\end{multline}
\end{itemize}

In this way, we observe that the introduction of a parameter $\hbar$ is equivalent to considering a suitable scaling limit of the problem in which the phase space parameters, $x$, $t$ and monodromy parameters $\theta_\ast$ are sent to infinity in a certain way. 

%%%%%%%%%%%%%%%%%%%%%%%%%%%%%%%%%%%%%%%%%%%%
\section{\label{AppendixDerivationPainleve} Derivation of the Painlev\'e equations from the Lax pairs}
The introduction of the $\hbar$ parameter in the Lax pair after rescaling slightly modifies the various equations obtained from the compatibility equation 
\beqq \hbar \partial_t \mathcal{D}(x,t) -\hbar\partial_x \mathcal{R}(x,t)+\left[\mathcal{D}(x,t),\mathcal{R}(x,t)\right]=0.\eeqq
In this section, we present those modifications as well as their consequences on the final Painlev\'e equations in Table \ref{table:P-eq}.
\subsection{Painlev\'e I}
The compatibility equation for \eqref{P1} gives the following system of equations:
\beq\label{CompatibilityP1} \hbar \dot{p}=6q^2+t \text{ and } \hbar \dot{q}=p.\eeq
Then, it is trivial to deduce that $q(t)$ satisfy the Painlev\'e ${\rm I}$ equation: 
\beqq \hbar^2\ddot{q}=6q^2+t.\eeqq

\subsection{Painlev\'e II}
The compatibility equation for \eqref{P2} gives the following system of equations:
\beq\label{CompatibilityP2} \hbar \dot{p}=-2qp-\theta \text{ and } \hbar \dot{q}=p+q^2+\frac{t}{2}.\eeq
Taking the derivative of the first equation and inserting it back into the second equation gives that $q(t)$ satisfies the Painlev\'e ${\rm II}$ equation:
\beqq \hbar^2\ddot{q}=2q^3+tq+\frac{\hbar}{2}-\theta.\eeqq

\subsection{Painlev\'e III}
The compatibility equation for \eqref{P3} gives the following system of equations:
\beq\label{CompatibilityP3} \hbar \dot{p}=\frac{1}{t}\left[ -4qp^2-p(-4tq+2\theta_\infty-\hbar)+ t(\theta_0+\theta_\infty) \right] \text{ and } \hbar \dot{q}=\frac{1}{t}\left[4q^2p-2tq^2+q(2\theta_\infty-\hbar)+2t\right].\eeq
Extracting $p(t)$ from the second equation and inserting it back into the first one gives that $q(t)$ satisfies the Painlev\'e ${\rm III}$ equation:
\beqq \hbar^2\ddot{q}=\frac{\hbar^2}{q}\dot{q}^2-\frac{\hbar^2}{t}\dot{q}+\frac{4}{t}\left(\theta_0 q^2-\theta_\infty+\hbar\right)+4 q^3-\frac{4}{q}.\eeqq

\subsection{Painlev\'e IV}
The compatibility equation for \eqref{P4} gives the following system of equations:
\beq\label{CompatibilityP4} \hbar \dot{p}=-p^2-4pq-2tp-2(\theta_0+\theta_\infty) \text{  and  } \hbar \dot{q}=2(pq+q^2+tq+\theta_0).\eeq
Extracting $p$ from the second equation and inserting it back into the first equation gives (after substantial computations) that $q(t)$ satisfies the Painlev\'e ${\rm IV}$ equation
\beqq \hbar^2\ddot{q}=\frac{\hbar^2}{2q}\dot{q}^2+2\left(3q^3+4tq^2+\left(t^2-2\theta_\infty+\hbar\right)q-\frac{\theta_0^2}{q}\right).\eeqq

\subsection{Painlev\'e V}
The compatibility equation for \eqref{P5} gives the following system of equations:
\beaa &&t\hbar\frac{d (pq)}{dt}=-q(q^2-1)p^2+\left(-q^2\frac{3\theta_0+\theta_1+\theta_\infty}{2}+\frac{\theta_0-\theta_1+\theta_\infty}{2}\right)p-\frac{q\theta_0(\theta_0+\theta_1+\theta_\infty)}{2}\cr
 &&2t\hbar \dot{q}(2pq+\theta_0-\theta_1+\theta_\infty)-4t\hbar q\frac{d (pq)}{dt}=4p^2q^2(3q-1)(q-1)\cr
&&+4pq\left( (4\theta_0+2\theta_\infty)q^2-q(t+4\theta_0+2\theta_1+3\theta_\infty)+\theta_0-\theta_1+\theta_\infty\right)\cr
&&+q^2(5\theta_0^2-\theta_1^2+\theta_\infty^2+6\theta_0\theta_\infty)-2q(\theta_0-\theta_1+\theta_\infty)(t+2\theta_0+\theta_\infty)+(\theta_0-\theta_1+\theta_\infty)^2.\cr
\eeaa
This is equivalent to
\bea\label{CompatibilityP5} t\hbar\dot{q}&=&2q(q-1)^2p+\frac{3\theta_0+\theta_1+\theta_\infty}{2}q^2-(t+2\theta_0+\theta_\infty)q+\frac{\theta_0-\theta_1+\theta_\infty}{2}\cr
t\hbar\dot{p}&=&-(3q^2-4q+1)p^2+\left(-(3\theta_0+\theta_1+\theta_\infty)q+t+2\theta_0+\theta_\infty\right)p -\frac{1}{2}\theta_0(\theta_0+\theta_1+\theta_\infty). \nonumber \\\eea
Extracting $p$ from the first equation and inserting it back into the second gives that $q(t)$ satisfies the Painlev\'e ${\rm V}$ equation
\beaa \hbar^2\ddot{q}&=&\left(\frac{1}{2q}+\frac{1}{q-1}\right)(\hbar\dot{q})^2-\hbar^2\frac{\dot{q}}{t}+\frac{(q-1)^2}{t^2}\left(\alpha q+\frac{\beta}{q}\right)+\frac{\gamma q}{t}+\frac{\delta q(q+1)}{q-1}\cr
&\text{with}& \,\, \alpha=\frac{(\theta_0-\theta_1-\theta_\infty)^2}{8}\,,\,\beta=-\frac{(\theta_0-\theta_1+\theta_\infty)^2}{8}\,\,,\,\,\gamma=\theta_0+\theta_1-\hbar \text{ and } \delta=-\frac{1}{2}.\cr
\eeaa

\subsection{Painlev\'e VI}
Following the various steps proposed in \cite{JMII}, one can follow the introduction of the $\hbar$ parameter in the Painlev\'e $6$ system. We find
\bea\label{CompatibilityP6} \hbar t(t-1)\dot{q}&=&2q(q-1)(q-t)p -\theta_0(q-1)(q-t)-\theta_1q(q-t)-(\theta_t-\hbar)q(q-1)\cr
\hbar t(t-1)\dot{p}&=&(-3q^2+2q(t+1)-t)p^2+\left( (2q-t-1)\theta_0+(2q-t)\theta_1+(2q-1)(\theta_t-\hbar)\right)p\cr
&&-\frac{1}{4}(\theta_0+\theta_1+\theta_t-\theta_\infty)(\theta_0+\theta_1+\theta_t+\theta_\infty-2\hbar).
\eea
Extracting $p$ from the first equation and inserting it back into the first one gives that $q$ satisfies the Painlev\'e ${\rm VI}$ equation:
\bea\hbar^2\ddot{q}&=&\frac{\hbar^2}{2}\left(\frac{1}{q}+\frac{1}{q-1}+\frac{1}{q-t}\right)\dot{q}^2-\hbar^2\left(\frac{1}{t}+\frac{1}{t-1}+\frac{1}{q-t}\right)\dot{q}\cr
&&+\frac{q(q-1)(q-t)}{t^2(t-1)^2}\left[\alpha+\beta \frac{t}{q^2}+\gamma \frac{t-1}{(q-1)^2}+\delta \frac{t(t-1)}{(q-t)^2}\right],
\eea
where the parameters are:
\beqq \alpha=\frac{1}{2}(\theta_\infty-\hbar)^2\,,\, \beta=-\frac{\theta_0^2}{2}\,,\, \gamma=\frac{\theta_1^2}{2} \text{ and } \delta=\frac{\hbar^2-\theta_t^2}{2}.\eeqq

\section{Spectral curve for Painlev\'e V and Painlev\'e VI\label{AppendixSpecP56}}
In this appendix, we present the computations required to obtain the various results in Section \ref{subsection:properties-of-spectral-curves} regarding the spectral curves of the Painlev\'e V and Painlev\'e VI cases. 
\subsection{Spectral curve for Painlev\'e V}
Projecting the compatibility equation \eqref{eq:compatibility} for Painlev\'e V onto the leading order $\hbar^0$, we find that $p_0$ and $q_0$ must obey
\bea \label{P5eq1}0&=&p_0\left(p_0+\frac{\theta_0-\theta_1+\theta_\infty}{2q_0}\right)-(p_0q_0+\theta_0)\left(p_0q_0+\frac{\theta_0+\theta_1+\theta_\infty}{2}\right)\cr
0&=&t-2p_0(q_0-1)^2+(q_0-1)\left(\frac{\theta_0-\theta_1+\theta_\infty}{2q_0}-\frac{3\theta_0+\theta_1+\theta_\infty}{2}\right).
\eea
The determinant of $\mathcal{D}^{(0)}(x,t)$ is given by
\footnotesize{\bea \label{DeterminantP5} \Delta&=&\frac{t^2}{4x^2(x-1)^2}\left(x(x-1)-\frac{\theta_\infty}{t}x-\frac{2p_0q_0+\theta_0}{t}\right)^2 \cr
&&+\frac{1}{x^2(x-1)^2}\left(-(x-1)(p_0q_0+\theta_0)+x\left(p_0+\frac{\theta_0-\theta_1+\theta_\infty}{2q_0}\right)\right)\left(p_0q_0(x-1)-q_0x\left(p_0q_0+\frac{\theta_0+\theta_1+\theta_\infty}{2}\right)\right).\cr\eea}\normalsize{}
From the first equation of \eqref{P5eq1}, it is easy to see that the second term of \eqref{DeterminantP5} admits a double zero $Q_0(t)$ given by
\beq \label{rrr}Q_0=-\frac{p_0}{p_0(q_0-1)+\frac{\theta_0+\theta_1+\theta_\infty}{2}}=\frac{p_0q_0+\theta_0}{p_0q_0+\theta_0-\left(p_0+\frac{\theta_0-\theta_1+\theta_\infty}{2q_0}\right)}.\eeq
The second equation of \eqref{P5eq1} shows that $Q_0$ is also a zero of $x(x-1)-\frac{\theta_\infty}{t}x-\frac{2p_0 q_0+\theta_0}{t}$. Consequently, we also get that
\beq Q_0=\frac{1}{2}\left(1+\frac{\theta_\infty}{t}\right)+\frac{1}{2}\sqrt{\left(1+\frac{\theta_\infty}{t}\right)^2-\frac{4(2p_0q_0+\theta_0)}{t}}.\eeq
Solving for $p_0$ in the second equation of \eqref{P5eq1} shows from \eqref{rrr} that we also have
\bea \label{Valuea} Q_0&=&\frac{-(3\theta_0+\theta_1+\theta_\infty)(q_0-1)^2+2(t-\theta_0-\theta_1)(q_0-1)+2t}{(q_0-1)\left((\theta_0-\theta_1-\theta_\infty)(q_0-1)^2-2(t+\theta_\infty)(q_0-1)-2t\right)}\cr
&=& \frac{q_0\left((\theta_0-\theta_1-\theta_\infty)(q_0-1)^2+2(t-\theta_0-\theta_1)(q_0-1)+2t\right)}{(q_0-1)\left((\theta_0-\theta_1-\theta_\infty)(q_0-1)^2+2(t-\theta_\infty)(q_0-1)+2t\right)}.\eea
The last two expressions are equivalent since $q_0$ satisfies \eqref{Q0P5}. Therefore, we may write the determinant as
\beq \Delta=-\frac{t^2(x-Q_0)^2(x^2-Sx+P)}{4x^2(x-1)^2},\eeq
where at the moment $S$ and $P$ are the remaining unknown parameters of $\Delta$. In fact, they are given by
\footnotesize{\bea S&=&\frac{1}{2tq_0}\left[-(\theta_0-\theta_1-\theta_\infty)q_0^2+2(\theta_\infty+t)q_0+\theta_0-\theta_1+\theta_\infty\right]\cr
P&=&\frac{-1}{16t^2q_0^2(q_0-1)^2}\left[(\theta_0-\theta_1-\theta_\infty)q_0^3-(3\theta_0-3\theta_1-\theta_\infty+2t)q_0^2+(3\theta_0-3\theta_1+\theta_\infty-2t)q_0-\theta_0+\theta_1-\theta_\infty\right].\cr
\eea}\normalsize{}

Alternatively, by considering the asymptotics of $x$ at $0, 1,$ and $\infty$, we obtain the coefficients ($Q_0,S,P$). Using the form of the matrix $\mathcal{D}(x,t)$, we have that $\det\mathcal{D}\underset{x\to 0}{\sim}-\frac{\theta_0^2}{4x^2}$, $\det\mathcal{D}(x,t)\underset{x\to 1}{\sim}-\frac{\theta_1^2}{4(x-1)^2}$ and $\det\mathcal{D}(x,t)\underset{x\to \infty}{=}-\frac{t^2}{4}+\frac{t\theta_\infty}{2x}+O\left(\frac{1}{x^2}\right)$. Therefore, the parameters $(Q_0,S,P)$ of the spectral curve are characterized by the following system of equations:
\bea \label{syss}t^2Q_0^2P&=&\theta_0^2\cr
t^2(1-Q_0)^2(1-S+P)&=&\theta_1^2\cr
t(2Q_0-2+S)&=&2\theta_\infty.
\eea
This system can be solved, and we get that the double zero $Q_0$ of $\Delta(x,t)$ must satisfy the algebraic equation
\bea\label{last} 0&=&2t^2Q_0^5-t(5t+2\theta_\infty)Q_0^4+4t(t+\theta_\infty)Q_0^3-( (t+\theta_\infty)^2-(\theta_0^2-\theta_1^2+\theta_\infty^2))Q_0^2-2\theta_0^2Q_0+\theta_0^2\cr
0&=&Q_0^2(Q_0-1)^2(2Q_0-1)t^2-2\theta_\infty Q_0^2(Q_0-1)^2t+(Q_0(\theta_0+\theta_1)-\theta_0)(Q_0(\theta_0-\theta_1)-\theta_0). \nonumber \\
\eea
This provides a direct evolution of the double zero $Q_0(t)$ in terms of $t$. Note that this evolution is completely independent of $q_0$ and $p_0$ and only depends on the monodromy parameters. We now need to rule out possible non-generic cases:
\begin{itemize}\item The double zero may never equal $0$ or $1$. Indeed, in that case, \eqref{last} implies that $\theta_0^2=0$ and $-\theta_1^2=0$ respectively.
\item The simple zeros may never equal $0$ or $1$. Indeed, in that case, \eqref{syss} implies that $\theta_0=0$ or $\theta_1=0$ respectively.
\item The simple zeros may never coincide. Indeed, in that case, we would have $P=d^2$ and $S=2d$ in \eqref{syss}. Solving the first and third equations leads to $Q_0+d=1+\frac{\theta_\infty}{t}$ and $dQ_0=\frac{\epsilon_0 \theta_0}{t}$. Inserting, these relations into the second equation $t^2(Q_0-1)^2(d-1)^2=\theta_1^2$ is equivalent to $\frac{1}{t}\prod_{(\epsilon_0,\epsilon_1)\in \{\pm 1\}^2}\left(\theta_\infty+\epsilon_0\theta_0+\epsilon_1\theta_1\right)=0$. Since $t\neq0$, we get that the simple zeros never coincide as soon as
\beq \theta_\infty+\epsilon_0\theta_0+\epsilon_1\theta_1\neq 0 \text{ for all choices of } (\epsilon_0,\epsilon_1)\in \{\pm 1\}^2.\eeq
These conditions are exactly the same as those requiring that Painlev\'e $5$ is non-degenerate.
\item Also, we need to rule out the possibility that one of the simple zero becomes equal to the double zero $Q_0$. 
%Let us first observe that the singular times correspond to:
%\beqq A_{V}(q_0,t)=0 \text{ and } \frac{\partial A_{V}}{\partial q_0}(q_0,t)=0\eeqq
%These two equations provides second order algebraic equations in $t$. By a suitable linear combination we can obtain a first order algebraic equation in $t$ that can be solved explicitly. We find:
%\beqq t_{\text{sing}}=-\frac{(q-1)^2\left((\theta_0-\theta_1-\theta_\infty)^2q^4+2(\theta_0-\theta_1-\theta_\infty)^2q^3-2(\theta_0-\theta_1+\theta_\infty)^2q-(\theta_0-\theta_1+\theta_\infty)^2\right)}{8q^3(\theta_0+\theta_1)}\eeqq
%Inserting this expression back into $A_{V}(q_0,t)=0$ or $\frac{\partial A_{V}}{\partial q_0}(q_0,t)=0$ is equivalent to say that the singular values for $q_0$ correspond to the solution of the algebraic equation:
%\beaa \label{QSingP5} 0&=&(\theta_0-\theta_1-\theta_\infty)^4q_{\text{sing}}^9+3(\theta_0-\theta_1-\theta_\infty)^4q_{\text{sing}}^8+8(\theta_0-\theta_1-\theta_\infty)^2(\theta_0^2+6\theta_0\theta_1+\theta_1^2-\theta_\infty^2)q_{\text{sing}}^6\cr
%&&-6(\theta_0-\theta_1-\theta_\infty)^2(\theta_0-\theta_1+\theta_\infty)^2q_{\text{sing}}^5+6(\theta_0-\theta_1-\theta_\infty)^2(\theta_0-\theta_1+\theta_\infty)^2q_{\text{sing}}^4\cr
%&&-8(\theta_0-\theta_1+\theta_\infty)^2(\theta_0^2+6\theta_0\theta_1+\theta_1^2-\theta_\infty^2)q_{\text{sing}}^3 -3(\theta_0-\theta_1+\theta_\infty)^4q_{\text{sing}}-(\theta_0-\theta_1+\theta_\infty)^4\cr
%&\overset{\text{def}}{=}&R_9(q_{\text{sing}})
%\eeaa
Let us observe that we have
\beqq Q_0=-\frac{ (3\theta_0+\theta_1+\theta_\infty)q_0^2-2(2\theta_0+\theta_\infty+t)q_0+\theta_0-\theta_1+\theta_\infty}{(\theta_0-\theta_1-\theta_\infty)q_0^2-2(\theta_0-\theta_1+t)q_0+\theta_0-\theta_1+\theta_\infty}.\eeqq
Moreover, assuming that $Q_0$ is a triple zero of $\Delta(x)$ is equivalent to saying that $\Delta''(Q_0)=0$. This provides a polynomial relation $P(q_0,t)$ between $q_0$ and $t$. This polynomial is originally of degree $4$ in $t$ and $8$ in $q_0$. However, since $q_0$ and $t$ are related by a polynomial relation $E_5(q_0,t)=0$ (cf. \eqref{Q0P5}) which is of degree $2$ in $t$, requiring $P(q_0,t)=0$ is equivalent to require that $\gcd(P,E_5)(q_0,t)=0$ therefore leading to a new polynomial relation $R(q_0,t)=0$ of degree $1$ in $t$. Let us denote $t=g(q_0)$ the corresponding solution where $z\mapsto g(z)$ is a rational function of $z$ whose expression is explicit (we omit it here due to its length). Hence we have proved that the spectral curve admits a triple zero if and only if $t=g(q_0)$ with $g$ an explicit rational function. We can then insert $t=g(q_0)$ into \eqref{Q0P5} and we get that
\beqq 4(q_0^2-1)^3q_0^2\left((\theta_0+\theta_1+\theta_\infty)q_0+ \theta_0-\theta_1+\theta_\infty\right)^2R_9(q_0)=0,\eeqq
where $R_9(z)$ is given by \eqref{SingP5}.
It is obvious to see that $q_0=-\frac{\theta_0-\theta_1+\theta_\infty}{\theta_0+\theta_1+\theta_\infty}$ does not lead to an admissible solution since it corresponds to $t=0$. Consequently, we recognize here that $Q_0$ is a triple zero implies that $t$ is a singular time. The converse can also easily be verified, and we have that $Q_0$ is a triple zero if and only if $t$ is a singular time.
\end{itemize}

\subsection{Spectral curve for Painlev\'e VI}
The computation of the spectral curve for Painlev\'e VI leads to

\beq \label{SpecCurveP6} Y^2=\frac{\theta_\infty^2(x-q_0)^2 P_2(x)}{4(x-t)^2x^2(x-1)^2} \,\text{ with } P_2(x)=x^2+\left(-1-\frac{\td{\theta}_0^2t^2}{q_0^2}+\frac{\td{\theta}_1^2(t-1)^2}{(q_0-1)^2}\right)x+\frac{\td{\theta}_0^2t^2}{q_0^2}.\eeq
Note that we can factorize $\theta_\infty^2$ and rewrite the spectral curve only in terms of reduced monodromy parameters defined by $\td{\theta}_i=\frac{\theta_i}{\theta_\infty}$ for $i\in\{0,1,t\}$. Moreover, $P_2(x)$ also admits a more symmetric formulation. Using \eqref{Q0P6} we can reformulate it into
\beq \label{SpecCurve2} Y^2=\frac{\theta_\infty^2(x-q_0)^2 P_2(x)}{4(x-t)^2x^2(x-1)^2} \,\text{ with } P_2(x)=x^2+\left(-\frac{\td{\theta}_0^2t(t+1)}{q_0^2}+\frac{\td{\theta}_1^2t(t-1)}{(q_0-1)^2}-\frac{\td{\theta}_t^2t(t-1)}{(q_0-t)^2}\right)x+\frac{\td{\theta}_0^2t^2}{q_0^2}.\eeq
Equivalently, it is defined by the following $3$ conditions
\beq \label{Poly2} P_2(0)=\frac{\td{\theta}_0^2t^2}{q_0^2} \,\,,\,\, P_2(1)=\frac{(t-1)^2\td{\theta}_1^2}{(q_0-1)^2}\text{ and, } P_2(t)=\frac{t^2(t-1)^2\td{\theta}_t^2}{(q_0-t)^2}.\eeq
So that
\beq \label{P2bis}P_2(x)=\frac{\td{\theta}_0^2t}{q_0^2}(x-1)(x-t)-\frac{(t-1)\td{\theta}_1^2}{(q_0-1)^2}x(x-t)+\frac{t(t-1)\td{\theta}_t^2}{(q_0-t)^2}x(x-1).\eeq
We now discuss the general form of the spectral curve and its possible degeneracies.
\begin{itemize} \item The spectral curve is generically of genus $0$ since the numerator admits a double zero at $x=q_0$.
\item From \eqref{Poly2}, we observe that $P_2(x)$ cannot have zeros at $x\in \{0,1,t\}$ when the monodromy parameters $\theta_0, \theta_1$ and $\theta_t$ are non-vanishing. Additionally, $P_2(x)$ cannot have a double zero since in that case, it would lead to the fact that $1\pm \frac{\td{\theta}_0 t}{q_0}\pm \frac{\td{\theta}_1 (t-1)}{q_0-1}=0$ in contradiction with \eqref{Q0P6} when the monodromy parameters are non-vanishing.
\item The simple zeros of $P_2(x)$ can never coincide. Indeed, if that were the case, \eqref{SpecCurveP6} implies that
\beqq 1+\epsilon_0\frac{\theta_0t}{\theta_\infty q_0}+\epsilon_1\frac{\theta_1(t-1)}{\theta_\infty(q_0-1)}=0 \text{ with } \epsilon_0,\epsilon_1\in\{-1,+1\}. \eeqq
Extracting $t$ from this equation and inserting it back into the algebraic equation satisfied by $q_0$ \eqref{Q0P6} leads to
\small{\beqq \theta_\infty^2\left(\epsilon_0\theta_0+\epsilon_1\theta_1\right)\left(\left(\epsilon_0\theta_0+\epsilon_1\theta_1+\theta_\infty\right)^2-\theta_t^2\right)\left(q_0+\frac{\epsilon_0\theta_0}{\theta_\infty}\right)\left(q_0-1-\frac{\epsilon_1\theta_1}{\theta_\infty}\right)\left(q_0-\frac{\epsilon_0\theta_0}{\epsilon_0\theta_0+\epsilon_1\theta_1}\right)=0.\eeqq}\normalsize{} 
Note also that $q_0=\frac{\epsilon_0\theta_0}{\epsilon_0\theta_0+\epsilon_1\theta_1}$ is equivalent to $t=\infty$ or $\theta_\infty+\epsilon_0\theta_0+\epsilon_1\theta_1=0$. Similarly, $q_0=-\frac{\epsilon_0\theta_0}{\theta_\infty}$ is equivalent to $t=1$ and $q_0=1+\frac{\epsilon_1\theta_1}{\theta_\infty}$ is equivalent to $t=1$, both of which have been ruled out before. Consequently, as soon as
\beq \theta_0^2\neq \theta_1^2 \,,\,\theta_\infty+\epsilon_0\theta_0+\epsilon_1\theta_1\neq 0 \text{ and }\, \theta_\infty+\epsilon_0\theta_0+\epsilon_1\theta_1+\epsilon_t\theta_t\neq0 \eeq
for any choice of the signs $(\epsilon_0,\epsilon_1,\epsilon_t)\in\{-1,+1\}^3$, we can rule out that the simple zeros coincide at any time. This corresponds to the assumptions made for non-singular monodromy parameters.
\item The zeros of $P_2(x)$ can never coincide with $q_0$. Indeed, if that were the case, \eqref{P2bis} implies 
\beqq \frac{\td{\theta}_0^2t}{q_0^3}-\frac{(t-1)\td{\theta}_1^2}{(q_0-1)^3}+\frac{t(t-1)\td{\theta}_t^2}{(q_0-t)^3}=0,\eeqq
which precisely corresponds to a singular time \eqref{SingP6}. 
\end{itemize}

\section{Proof of the parity property \label{AppendixParityProof}}
We prove in this section that the $\hbar$ series expansion of the correlation functions $W_n(x_1,\dots,x_n)$ only involves powers of $\hbar$ of the same parity as $n$ (i.e. we explain why we have an exponent $2g$ instead of $g$ in \eqref{Exponents}). In order to do this, we use Proposition $3.3$ of \cite{BBEnew}, which provides a sufficient criteria to obtain the $\hbar\leftrightarrow -\hbar$ symmetry. We recall their proposition here:

\begin{proposition}[Proposition $3.3$ of \cite{BBEnew}] Let us denote by $\dagger$ the operator switching $\hbar$ into $-\hbar$. If there exists an invertible matrix $\Gamma(t)$ independent of $x$ such that
\beq\label{eq:L-and-Gamma} 
\Gamma^{-1}(t) \mathcal{D}^t(x,t)\Gamma(t)=\mathcal{D}^\dagger(x,t),\eeq
then the correlation functions $W_{n}$ satisfy
\beq \forall \, n\geq 1\,:\, W_{n}^{\dagger} = (-1)^{n} W_{n}.\eeq 
\end{proposition}

In particular, if this proposition is satisfied, then it automatically follows that the $\hbar$ expansion of a given function $W_n(x_1,\dots,x_n)$ may only involve powers of $\hbar$ with the same parity (given by the parity of $n$). Therefore, it suffices to prove the existence of a suitable matrix $\Gamma(t)$ in our six Painlev\'e cases. Since we know from \eqref{ParityP1}, \eqref{ParityP2}, \eqref{ParityP3}, \eqref{ParityP4}, \eqref{ParityP5}, and \eqref{ParityP6} the expression of $(p^\dagger,q^\dagger)$ in terms of $(p,q)$ it is straightforward to compute the various $\Gamma(t)$ matrices. 

\begin{theorem}[Parity Property] In all six Painlev\'e cases, there exists a matrix $\Gamma_J(t)$ ($J={\rm I}, \dots, {\rm VI}$) such that
\beqq  \label{Gamma}\Gamma_J^{-1}(t) \mathcal{D}_J^t(x,t)\Gamma_J(t)=\mathcal{D}_J^\dagger(x,t).\eeqq
\end{theorem}

The corresponding matrices are the following:
\begin{itemize}
\item (Painlev\'e I) $\Gamma_{\rm I}(t)=\begin{pmatrix}0&1\\1&0 \end{pmatrix}$
\item (Painlev\'e II)  $\Gamma_{\rm II}(t)=\begin{pmatrix}-2p&0\\0&1 \end{pmatrix}$.
\item (Painlev\'e III)  $\Gamma_{\rm III}(t)=\begin{pmatrix}-\frac{p-t}{t}&0\\0&1 \end{pmatrix}$.
\item (Painlev\'e IV)  $\Gamma_{\rm IV}(t)=\begin{pmatrix}-2(pq+\theta_0+\theta_\infty)&0\\0&1 \end{pmatrix}$.
\item (Painlev\'e V)  $\Gamma_{\rm V}(t)=\begin{pmatrix}-\frac{pq}{pq+\theta_0}&0\\0&1 \end{pmatrix}$.
\item (Painlev\'e VI)  $\Gamma_{\rm VI}(t)=\begin{pmatrix}-\frac{t^2 z_0(z_0+\theta_0)}{q}+\frac{(t-1)^2z_1(z_1+\theta_1)}{q-1}&0\\0&1 \end{pmatrix}$.
\end{itemize}  
% We remark that the determination of $\Gamma(t)$ is not unique. Indeed, it is determined up to a global constant since multiplying $\Gamma(t)$ by a constant does not change the product $\Gamma^{-1}(t) \mathcal{D}^t(x,t)\Gamma(t)$. 
% To fix this degree of freedom, we chose to fix one entry to $1$ (usually the entry $(2,2)$ except for Painlev\'e I). We also observe that in all six cases $\Gamma^t(t)=\Gamma(t)$ and $\Gamma^\dagger(t)=\Gamma(t)$. 

\section{Proof of the pole structure\label{AppendixProofPole}}

We prove in this section that the correlation functions defined through the determinantal formulas only have poles (as functions of $x$) at the branch points of the spectral curve (i.e., simple zeros of the function $E_J(x)$). In particular, we show that there is no singularities at the even zeros of $E_J(x)$ or at the poles of the matrices $\DD_J(x)$ and $\RR_J(x)$ (i.e. $x=0$, $x=1$ and/or $x=t$, depending on the label of the Painlev\'e equation). In this appendix, we will omit the $t$-dependence in notations since we only discuss the pole structure in $x$-variables. Also, we sometimes drop the subscript $J$ representing the type of the Painlev\'e equation for simplicity.

First, we note that this condition is not trivial because $E(x)$ admits a double zero in all six cases. The idea of the proof follows the same spirit as the one proposed in Appendix B of \cite{P2}. It consists in two steps:
\begin{enumerate} 
\item Compute the matrix $M^{(0)}(x)$ and observe that it is regular at the poles of $\DD(x)$ and $\RR(x)$ or at the even zeros of $E(x)$.  
\item Derive a recursive relation between $M^{(k)}(x)$ and lower orders $M^{(j)}(x)$ with $0\leq j\leq k-1$ and their time derivatives, and check that $M^{(k)}(x)$ has singularities only at the branch points of the spectral curve.
\end{enumerate}

Differential equations on $\Psi(x)$ defining the Lax pairs turns into the following system for $M(x)$:
\beq \label{EquationM} \hbar \partial_x M(x)=\left[\mathcal{D}(x),M(x)\right] \text{ and } \hbar \partial_t M(x)=\left[\mathcal{R}(x),M(x)\right].\eeq
These equations will give a way to compute all orders $M^{(k)}(x)$.

\subsection{Computation of $M^{(0)}(x)$}
In full generality, the matrix $M^{(0)}(x)$ is characterized by the following set of equations:
\beq \left[\mathcal{D}^{(0)}(x),M^{(0)}(x)\right]=0 \text{ or } \left[\mathcal{R}^{(0)}(x),M^{(0)}(x)\right]=0,\eeq
$\Tr M^{(0)}(x)=1$ and $\det M^{(0)}(x)=0$. Note that using the matrix $\mathcal{D}^{(0)}(x)$ or $\mathcal{R}^{(0)}(x)$ in the last equation is completely equivalent since the differential systems of the Lax pair are compatible (meaning that the system of equations is overdetermined but remain compatible). Thus, if we denote $M^{(0)}(x)=\begin{pmatrix} m_{1,1}& m_{1,2}\\ m_{2,1}& 1-m_{1,1}\end{pmatrix}$ a possible minimal set of equations is given by:
\bea  0&=&\mathcal{R}^{(0)}_{2,1}m_{1,2}-\mathcal{R}^{(0)}_{1,2}m_{2,1}\cr
0&=&(2m_{1,1}-1)\mathcal{R}_{1,2}^{(0)}-2\mathcal{R}^{(0)}_{1,1}m_{1,2}\cr
0&=&m_{1,1}(1-m_{1,1})-m_{1,2}m_{2,1}.
\eea
This system can be solved explicitly by
\beq \label{M0}M^{(0)}(x)=\begin{pmatrix}\frac{1}{2}+\frac{\mathcal{R}^{(0)}_{1,1}(x)}{2\sqrt{-\det \mathcal{R}^{(0)}(x)}}& \frac{\mathcal{R}^{(0)}_{1,2}(x)}{2\sqrt{-\det \mathcal{R}^{(0)}(x)}}\\
\frac{\mathcal{R}^{(0)}_{2,1}(x)}{2\sqrt{-\det \mathcal{R}^{(0)}(x)}}&\frac{1}{2}-\frac{\mathcal{R}^{(0)}_{1,1}(x)}{2\sqrt{-\det \mathcal{R}^{(0)}(x)}}\end{pmatrix}.
\eeq
Note that one can replace the matrix $\mathcal{R}^{(0)}$ by $\mathcal{D}^{(0)}$ without changing the solution. From the definition of the Lax pairs, the entries $\mathcal{R}^{(0)}_{i,j}(x)$ may be singular at some finite $x$, but that is only the case for Painlev\'e III and VI. In those cases, the poles of $\mathcal{R}^{(0)}_{i,j}(x)$ cancel out with the determinants of $\mathcal{R}^{(0)}(x)$ that we give below. Consequently, the matrix $M^{(0)}(x)$ is singular only at the points where the determinant $\mathcal{R}^{(0)}(x)$ vanishes. It is long but straightforward computations to get these determinants in all six Painlev\'e cases:

\begin{itemize}
\item 
(Painlev\'e I) \quad $$\det \mathcal{R}_{\rm I}^{(0)}(x)=-(x+2q_0).$$ 
\item
(Painlev\'e II)  \quad $$\det \mathcal{R}_{\rm II}^{(0)}(x)=-\frac{1}{4}\left(x^2+2q_0x+q_0^2+\frac{\theta}{q_0}\right).$$ 
\item 
(Painlev\'e III) \quad 
$$\det \mathcal{R}_{\rm III}^{(0)}(x)=-\frac{(q_0x-1)^2\left((\theta_\infty-\theta_0q_0^2)x^2-2xq_0(\theta_\infty q_0^2-\theta_0)+q_0^2(\theta_\infty-\theta_0 q_0^2)\right)}{4x^2q_0^2(\theta_\infty-\theta_0q_0^2)}.$$ 
\item 
(Painlev\'e IV) \quad 
$$\det \mathcal{R}_{\rm IV}^{(0)}(x)=q_0^2\left(x^2+2(q_0+t)x+\frac{\theta_0^2}{q_0^2}\right).$$ 
\item
(Painlev\'e V) \quad 
$$\det \mathcal{R}_{\rm V}^{(0)}(x)=-\frac{1}{4}(x-Q_1)(x-Q_2),$$ 
where $Q_1$ and $Q_2$ are the simple zeros of the spectral curve. 
\item  
(Painlev\'e VI) \quad 
$$\det \mathcal{R}_{\rm VI}^{(0)}(x)=-\frac{(q_0-t)^2\theta_\infty^2P_2(x)}{4t^2(t-1)^2(x-t)^2}, $$ 
where $P_2(x)$ is the monic polynomial of degree $2$ appearing in the spectral curve.
\end{itemize}

One can observe that, except for the case $J = {\rm III}$, these determinants only vanish at the simple zeros of $E(x)$ (i.e., the branch points of the spectral curve), but never vanish at the double zero of $E(x)$. In the case of Painlev\'e III, there is an additional zero at $x = \frac{1}{q_0}$, but it does not cause any problem since $x=\frac{1}{q_0}$ is not a singularity of $\DD^{(0)}$ nor a zero of $\det \DD^{(0)}$. (See also Remark \ref{remark:determinant-PIII}.) Consequently, we get that in all six Painlev\'e cases, the matrix $M^{(0)}(x)$ is only singular at the branch points of the spectral curve. From the expression \eqref{M0}, direct computations from the definition \eqref{alternative} of $W_2^{(0)}(x_1,x_2)$ show that in all six cases we have $W_2^{(0)}(x(z_1),x(z_2))dx(z_1)dx(z_2)=\frac{dz_1dz_2}{(z_1-z_2)^2}$.

\begin{remark} \label{remark:determinant-PIII}
Since $\det \mathcal{R}_{\rm III}^{(0)}$ has a double zero at $x = \frac{1}{q_0}$ in addition to the simple zeros of $E_{\rm III}(x)$, it seems to cause a singularity of $M^{(0)}(x)$ in view of \eqref{M0}. However, this point $x = \frac{1}{q_0}$ never becomes a singular point of $M^{(k)}$ for any $k \ge 0$ because of the original definition \eqref{defM} of $M(x)$ and the fact that the coefficients of the WKB solution \eqref{eq:WKB-solution} are singular only at zeros and poles of $E_{\rm III}(x)$. (Although $E_{\rm III}(x)$ has a double zero at $x = - \frac{1}{q_0}$ as in Table \ref{table:spectral-curve-equation}, it never vanishes at $x = \frac{1}{q_0}$ in question. There is an important sign difference here.)  This is consistent with the fact that all entries of $R_{\rm III}^{(0)}(x)$ vanish at $x = \frac{1}{q_0}$ (which can be checked by a straightforward calculation), and we see that the expression \eqref{M0} is in fact holomorphic at $x = \frac{1}{q_0}$. 
\end{remark}

\begin{remark} \label{remark:advantage-of-Lax-pair}
One could replace $\mathcal{R}^{(0)}(x)$ by $\mathcal{D}^{(0)}(x)$ everywhere in \eqref{M0} and still obtain the same expression for $M^{(0)}(x)$. However, by doing so, we see that the discussion about a possible singularity at a double zero of $E(x)$ is not obvious because the denominator is vanishing there. One would have to prove that for any of the four entries, the numerator also vanishes at the double zero, which is far from obvious. This is one of advantages of the existence of the time evolution in the Lax pair. 
%The pole structure of $M^{(0)}(x)$ is invariant under admissible gauge transformations. Indeed gauge transformations \eqref{Gauge2bis} of the form $\td{\Psi}=U(t)\Psi$ do not change $\det \mathcal{R}^{(0)}(x)$ nor $\Tr \mathcal{R}^{(0)}(x)$ (so that \eqref{M0} still holds) and mixes the entries of $\mathcal{R}(x)$ with $x$-independent factors. Consequently from \eqref{M0} we get that the $x$ pole structure of $M^{(0)}(x)$ is unchanged. In the case of gauge transformations of type \eqref{Gauge1bis} then $M(x)$ is invariant so the pole structure of $M^{(0)}(x)$ is invariant too. Note that \eqref{M0} does not hold anymore since we used the fact that $\Tr \mathcal{R}(x)=0$ to derive it and it is no longer true when we perform gauge transformations of type \eqref{Gauge1bis}.
\end{remark}

\subsection{Recursive system for higher orders}
Let us first start with the observation that the entries $\mathcal{R}^{(k)}_{i,j}(x)$ for $k\geq 1$ are polynomial of $x$ in all Painlev\'e cases except for Painlev\'e III and VI. Indeed, these entries are only singular at $x=0$ for Painlev\'e III and at $x=t$ for Painlev\'e VI. Let us now look at order $\hbar^k$ with $k\geq 1$ of \eqref{EquationM} and in $\det M(x)=0$. We have that
\bea \label{EQQ}&&\left[ \mathcal{R}^{(0)}(x), M^{(k)}(x)\right]=\partial_t M^{(k-1)}(x)-\underset{i=0}{\overset{k-1}{\sum}} \left[\mathcal{R}^{(k-i)}(x),M^{(i)}(x)\right]\cr
&&M^{(k-1)}(x)_{1,1}\left(1-2M^{(k-1)}(x)_{1,1}\right)-M^{(0)}(x)_{2,1}M^{(k)}(x)_{1,2}-M^{(0)}(x)_{1,2}M^{(k)}(x)_{2,1}\cr
&&=\underset{i=1}{\overset{k-1}{\sum}}\left( M^{(i)}(x)_{1,1}M^{(k-i)}(x)_{1,1}+M^{(i)}(x)_{1,2}M^{(k-i)}(x)_{2,1}\right).
\eea 
The first matrix equation provides two independent scalar equations and thus we have a $3\times 3$ linear system that can be written in the following matrix form:
\scriptsize{\beq\begin{pmatrix} 0&-\mathcal{R}^{(0)}_{2,1}&\mathcal{R}^{(0)}_{1,2}\\
-2\mathcal{R}^{(0)}_{1,2}&2\mathcal{R}^{(0)}_{1,1}&0\\
2M^{(0)}_{1,1}-1&M^{(0)}_{2,1}&M^{(0)}_{1,2}\end{pmatrix}
\begin{pmatrix} M^{(k)}(x)_{1,1}\\ M^{(k)}(x)_{1,2}\\ M^{(k)}(x)_{2,1}\end{pmatrix}=\begin{pmatrix} \partial_t M^{(k-1)}(x)_{1,1}-\underset{i=0}{\overset{k-1}{\sum}}\left[\mathcal{R}^{(k-i)}(x),M^{(i)}(x)\right]_{1,1}\\ 
\partial_t M^{(k-1)}(x)_{1,2}-\mathcal{R}_{1,2}^{(k)}-\underset{i=0}{\overset{k-1}{\sum}} \left[\mathcal{R}^{(k-i)}(x),M^{(i)}(x)\right]_{1,2}\\
\underset{i=1}{\overset{k-1}{\sum}}\left( M^{(i)}(x)_{1,1}M^{(k-i)}(x)_{1,1}+M^{(i)}(x)_{1,2}M^{(k-i)}(x)_{2,1}\right)\end{pmatrix}.
\eeq}\normalsize{}
Using the exact expression for $M^{(0)}(x)$ we have
\scriptsize{\beq \label{rec} \begin{pmatrix} 0&-\mathcal{R}^{(0)}_{2,1}&\mathcal{R}^{(0)}_{1,2}\\
-2\mathcal{R}^{(0)}_{1,2}&2\mathcal{R}^{(0)}_{1,1}&0\\
\mathcal{R}^{(0)}_{1,1}&\frac{1}{2}\mathcal{R}^{(0)}_{2,1}&\frac{1}{2}\mathcal{R}^{(0)}_{1,2}\end{pmatrix}
\begin{pmatrix} M^{(k)}(x)_{1,1}\\ M^{(k)}(x)_{1,2}\\ M^{(k)}(x)_{2,1}\end{pmatrix}=\begin{pmatrix} \partial_t M^{(k-1)}(x)_{1,1}-\underset{i=0}{\overset{k-1}{\sum}}\left[\mathcal{R}^{(k-i)}(x),M^{(i)}(x)\right]_{1,1}\\ 
\partial_t M^{(k-1)}(x)_{1,2}-\underset{i=0}{\overset{k-1}{\sum}} \left[\mathcal{R}^{(k-i)}(x),M^{(i)}(x)\right]_{1,2}\\
\sqrt{-\det \mathcal{R}^{(0)}}\underset{i=1}{\overset{k-1}{\sum}}\left( M^{(i)}(x)_{1,1}M^{(k-i)}(x)_{1,1}+M^{(i)}(x)_{1,2}M^{(k-i)}(x)_{2,1}\right)\end{pmatrix}.
\eeq}\normalsize{}

Note in particular that the $3\times 3$ matrix on the l.h.s. does not depend on the order $k$; we only consider the $\hbar^0$ terms. In general, inverting a matrix may create poles at the zeros of the determinant of the matrix (this is obvious if one uses the definition of the inverse using the matrix of cofactors). However, in our case we have
\beq \label{Ident}\det \begin{pmatrix} 0&-\mathcal{R}^{(0)}_{2,1}&\mathcal{R}^{(0)}_{1,2}\\
-2\mathcal{R}^{(0)}_{1,2}&2\mathcal{R}^{(0)}_{1,1}&0\\
\mathcal{R}^{(0)}_{1,1}&\frac{1}{2}\mathcal{R}^{(0)}_{2,1}&\frac{1}{2}\mathcal{R}^{(0)}_{1,2}\end{pmatrix}=2\mathcal{R}_{1,2}^{(0)}(x) \det \mathcal{R}^{(0)}(x).\eeq
Therefore, the inverse of the matrix will only have singularities at the zeros of the former determinant and at the singularities of the entries of $\mathcal{R}^{(0)}$. We have also seen earlier that the entries of $\mathcal{R}^{(0)}$ are regular, except for Painlev\'e III and VI. In exceptional cases, one has that the zeros of $\left[2\mathcal{R}_{1,2}^{(0)}(x) \det \mathcal{R}^{(0)}(x)\right]^{-1}$ cancel out the poles of the entries of $\mathcal{R}^{(0)}$. Also, as we noted before, $\det \RR^{(0)}(x)$ only vanishes at the branch points of the spectral curve, as we wish. (To be precise, $\det \RR^{(0)}_{\rm III}(x)$ has an additional zero, but it does not cause a problem as we explained in Remark \ref {remark:determinant-PIII}.) Thus, the only singularities that may arise now are at the zeros of $\RR^{(0)}_{1,2}(x)$. Again, from the definition, we see that this term is actually independent of $x$ and it doesn't vanish for any Painlev\'e cases, except for Painlev\'e I and III. For Painlev\'e I, $\RR^{(0)}_{1,2}(x)$ vanishes at a branch point, as we wish.  For Painlev\'e III, $\RR^{(0)}_{1,2}(x)$ has a unique zero exactly at $x = \frac{1}{q_0}$ which never becomes a singularity of $M^{(k)}(x)$ as we explained in Remark  \ref {remark:determinant-PIII}.
Thus, one can proceed to a simple recursion to prove that $M^{(k)}(x)$ only has poles at branch points of the spectral curve.

\begin{theorem} The function $M^{(k)}(x)$ only has poles (as a function of $x$) at the branch points of the spectral curve for any $k\geq 0$.
\end{theorem}

From the last theorem and the alternative definition of the correlation functions \eqref{alternative}, it is then obvious that the correlation functions $W_n^{(g)}$ only have poles at the branch points of the spectral curve for $(g,n) \neq (0,1), (0,2)$, and $W^{(0)}_2$ only has a double pole along the diagonal and no other poles.

%%%%%%%%%%%%%%%%%%%%%%%%%%%%%%%%%
%%%%%%%%%%%%%%%%%%%%%%%%%%%%%%%%%
\section{Proof of the $O(\hbar^{n-2})$-property \label{AppendixLeadingOrder}}

We will keep the same notational rule; that is, we will omit $t$-dependence in the argument of functions. 

Let us start the proof by introducing some notions similar to \cite{BBEnew} and \cite{Deter}. We remind the reader that determinantal formulas \eqref{DeterFormm} have been introduced so that they satisfy a set of equations known as loop equations. These loop equations (also known as Schwinger-Dyson equations) originate in random matrix theory where they are crucial. We recall here the main result of \cite{Deter}:

\begin{proposition}[Theorem $2.9$ of \cite{Deter}]\label{Loop} Let us define the following functions:
\bea \label{defP}
P_1(x)&=&\frac{1}{\hbar^2}\det {\mathcal{D}}(x)\nonumber \\[+.3em]
P_2(x;x_2)&=&
\frac{1}{\hbar} 
{\Tr}\left( \frac{\mathcal{D}(x)-{\mathcal{D}}(x_2)-(x-x_2){\mathcal{D}}'(x_2)}{(x-x_2)^2}M(x_2)\right)\nonumber \\[+.3em]
Q_{n+1}(x;L_n)  
& = & %\hspace{-6.em}  = 
\frac{1}{\hbar}\sum_{\sigma\in S_n} 
\frac{{\Tr}\left( \mathcal{D}(x)M(x_{\sigma(1)})\dots 
M(x_{\sigma(n)})\right)}{(x-x_{\sigma(1)})
(x_{\sigma(1)}-x_{\sigma(2)})\dots 
(x_{\sigma(n-1)}-x_{\sigma(n)})(x_{\sigma(n)}-x)} \cr
P_{n+1}(x;L_n)&=&
(-1)^n\left[Q_{n+1}(x;L_n)-\sum_{j=1}^n 
\frac{1}{x-x_j}\Res_{x'\to x_j} Q_{n+1}(x',L_n)\right],
\eea
where we denote by $L_n$ the set of variables $\{x_1,\dots,x_n\}$. Then the correlation functions satisfy 
\beq \label{eq:first-loop-equation} P_1(x)=W_2(x,x)+W_1(x)^2, \eeq
and for $n\geq 1$:
\bea\label{ind1}
0&=&P_{n+1}(x;L_n)+W_{n+2}(x,x,L_n)+2W_1(x)W_{n+1}(x,L_n)\nonumber \\[+.3em]
&& + \sum_{J \subset L_n, J\notin\{ \emptyset,L_n\}} W_{1+|J|}(x,J)W_{1+n-|J|}(x,L_n\setminus J) \nonumber \\[+.3em]
&&+ \sum_{j=1}^n \frac{d}{dx_j} \frac{ W_n(x,L_n\setminus x_j)-W_n(L_n)}{x-x_j}.
\eea
Moreover $P_{n+1}(x;L_n)$ is a rational function of $x$ whose poles are at the poles of $\mathcal{D}(x)$.
\end{proposition} 

The equations \eqref{eq:first-loop-equation} and \eqref{ind1} are called loop equations. As we will see, this proposition and a subtle induction are sufficient to prove that $W_n$ is at least of order $\hbar^{n-2}$ as developed in \cite{P2}. Let us now analyze the different possible poles of $P_{n+1}(x,L_n)$ using Proposition \ref{Loop}. We have a key theorem:

\begin{theorem}[Pole Structure of $P_{n+1}(x,L_n)$]\label{PoleTheo} For any of the six Painlev\'e cases we have the following:
\begin{itemize}
\item (Painlev\'e I and II) $P_{n+1}(x,L_n)$ does not depend on $x$.
\item (Painlev\'e III) $P_{n+1}(x,L_n)=\dfrac{\td{P}_{n+1}(L_n)}{x^2}$.
\item (Painlev\'e IV) $P_{n+1}(x,L_n)=\dfrac{\td{P}_{n+1}(L_n)}{x}$.
\item (Painlev\'e V) $P_{n+1}(x,L_n)=\dfrac{\td{P}_{n+1}(L_n)}{x(x-1)}$.
\item (Painlev\'e VI) $P_{n+1}(x,L_n)=\dfrac{\td{P}_{n+1}(L_n)}{x(x-1)(x-t)}$.
\end{itemize}
In all relevant cases, $\td{P}_{n+1}(L_n)$ is a rational function independent of $x$.
\end{theorem}

\proof{The proof of the previous theorem is based on the evaluation of the different orders of singularity of $P_{n+1}(x,L_n)$ at the finite possible singularities and at $x=\infty$ from definition \eqref{defP}.
\begin{itemize}
\item (Painlev\'e I and II) $\mathcal{D}(x)$ does not have singularities at finite $x$, and therefore $P_{n+1}(x,L_n)$ is a polynomial of $x$. However, from its definition \eqref{defP}, we see that $P_{n+1}(x,L_n)\underset{x\to \infty}{=}O(1)$ so that it cannot depend on $x$.
\item (Painlev\'e III) $\mathcal{D}(x)$ has a double pole at $x=0$ so that $P_{n+1}(x,L_n)$ is a rational function of $x$ with only a possible double pole at $x=0$ and a pole at $x=\infty$. Moreover, from its definition \eqref{defP}, we see that $P_{n+1}(x,L_n)\underset{x\to \infty}{=}O\left(\frac{1}{x^2}\right)$. Therefore, the only possible case is that $P_{n+1}(x,L_n)=\frac{\td{P}_{n+1}(L_n)}{x^2}$.
\item (Painlev\'e IV) $\mathcal{D}(x)$ has a simple pole singularity at $x=0$ and from \eqref{defP}, the behavior of $P_{n+1}(x,L_n)$ is of the form $P_{n+1}(x,L_n)\underset{x\to \infty}{=}O\left(\frac{1}{x}\right)$. Therefore, the only possible case is that $P_{n+1}(x,L_n)=\frac{\td{P}_{n+1}(L_n)}{x}$.
\item (Painlev\'e V) $\mathcal{D}(x)$ has a simple pole singularity at $x=0$ and $x=1$. From \eqref{defP}, the behavior at infinity of $P_{n+1}(x,L_n)$ is of the form $P_{n+1}(x,L_n)\underset{x\to \infty}{=}O\left(\frac{1}{x^2}\right)$. Consequently, the only possible case is that $P_{n+1}(x,L_n)=\frac{\td{P}_{n+1}(L_n)}{x(x-1)}$
\item (Painlev\'e VI) $\mathcal{D}(x)$ has a simple pole singularity at $x=0$, $x=1$ and $x=t$. From \eqref{defP}, the behavior at infinity of $P_{n+1}(x,L_n)$ is of the form $P_{n+1}(x,L_n)\underset{x\to \infty}{=}O\left(\frac{1}{x^3}\right)$. Consequently, the only possible case is that $P_{n+1}(x,L_n)=\frac{\td{P}_{n+1}(L_n)}{x(x-1)(x-t)}$.
\end{itemize}
}

As in \cite{P2}, the last theorem is sufficient to prove by induction that the leading order of $W_n(x_1,\dots,x_n)$ is at least of order $\hbar^{n-2}$. As we will see, the induction is very similar in the six cases and the previous theorem is used only at very specific places. Let us define the following statement:
\beq \mathcal{P}_k \,:\,  
W_j(x_1,\dots,x_j) \text{ is at least of order } \hbar^{k-2} \quad \text{for $j \ge k$}. \eeq
Recall that the correlation functions has a formal series expansion of the form 
\[
W_1(x) = \sum_{k=-1}^{\infty} w_1^{(k)}(x)  \hbar^k,\quad
W_n(x_1,\dots,x_n) = \sum_{k = 0}^\infty w_n^{(k)}(x_1,\dots,x_n) \hbar^k ~~~~(n \ge 2).
\]
Here we slightly changed the label of the coefficients of $W_{1}$ from \eqref{eq:correlation-series-pre} so that 
\begin{equation} \label{eq:w1-1-appendix}
w_1^{(-1)}(x) = \sqrt{E(x)}.
\end{equation}
The statement $\mathcal{P}_k$ is obviously true for $k=1$ and $k=2$ from the definitions. 

Let us assume that the statement $\mathcal{P}_i$ is true for all $i\leq n$. Now we look at the loop equations \eqref{ind1}. By the induction assumption, we have that the last two terms are at least of order $\hbar^{n-2}$. Indeed, in the sum we have terms of order $\hbar^{1+|J|-2+1+n-|J|-2}=\hbar^{n-2}$. Moreover, we also have from the same assumption that $W_{n+2}(x,x,L_n)$ is also of order at least $\hbar^{n-2}$ (since $n+2\geq n$). Therefore, $W_{n+1}(x,L_n)$ is at least of order $\hbar^{n-2}$, and by considering the coefficients of the $\hbar^{n-3}$ in \eqref{ind1} we have
\beq \label{eqsys} 0=P_{n+1}^{(n-3)}(x;L_n)+2w_1^{(-1)}(x)w_{n+1}^{(n-2)}(x,L_n). \eeq
If we assume that $w_{n+1}^{(n-2)}(x,L_{n})\neq 0$, then we have
\beq w_{n+1}^{(n-2)}(x,L_{n})=\frac{P_{n+1}^{(n-3)}(x;L_{n})}{2w_1^{(-1)}(x)}. \eeq
Using the explicit expression of $w_1^{(-1)}(x)$ (see \eqref{eq:w1-1-appendix} and Table \ref{table:spectral-curve-equation}),
we have the following in our six cases:
\begin{itemize}
\item (Painlev\'e I)
\beqq w_{n+1}^{(n-2)}(x,L_{n})=\frac{P_{n+1}^{(n-3)}(L_{n})}{4(x-q_0)\sqrt{x+2q_0} }. \eeqq
\item (Painlev\'e II)
\beqq w_{n+1}^{(n-2)}(x,L_{n})=\frac{P_{n+1}^{(n-3)}(L_{n})}{2(x-q_0)\sqrt{x^2+2q_0x+q_0^2+\frac{\theta}{q_0}} }.\eeqq
\item (Painlev\'e III)
\beqq w_{n+1}^{(n-2)}(x,L_{n})=\frac{\sqrt{q_0(q_0^4-1)}P_{n+1}^{(n-3)}(L_{n})}{\sqrt{t}(q_0x+1)\sqrt{(\theta_\infty-\theta_0q_0^2)x^2-2xq_0(\theta_\infty q_0^2-\theta_0)+q_0^2(\theta_\infty-\theta_0 q_0^2)}}.\eeqq
\item (Painlev\'e IV)
\beqq w_{n+1}^{(n-2)}(x,L_{n})=\frac{P_{n+1}^{(n-3)}(L_{n})}{2(x-q_0)\sqrt{x^2+2(q_0+t)x+\frac{\theta_0^2}{q_0^2}} }. \eeqq
\item (Painlev\'e V)
\beqq w_{n+1}^{(n-2)}(x,L_{n})=\frac{P_{n+1}^{(n-3)}(L_{n})}{t(x-Q_0)\sqrt{(x-Q_1)(x-Q_2)} }. \eeqq
\item (Painlev\'e VI)
\beqq w_{n+1}^{(n-2)}(x,L_{n})=\frac{P_{n+1}^{(n-3)}(L_{n})}{\theta_\infty (x-q_0) \sqrt{x^2+\left(-1-\frac{\theta_0^2t^2}{\theta_\infty^2 q_0^2}+\frac{\theta_1^2(t-1)^2}{\theta_\infty^2(q_0-1)^2}\right)x+\frac{\theta_0^2t^2}{\theta_\infty^2q_0^2}} }.\eeqq
\end{itemize}

Now, we observe that in all cases, we get that $w_{n+1}^{(n-2)}(x,L_{n})$ must have a simple pole (as a function of $x$) at the double zero of $E(x)$ (i.e. $q_0$ for Painlev\'e ${\rm I,II,IV,VI}$ and $-\frac{1}{q_0}$ for Painlev\'e III and $Q_0$ for Painlev\'e V). This is in contradiction with the pole structure of the correlation functions proved in Appendix \ref{AppendixProofPole}. Consequently, we must have $w_{n+1}^{(n-2)}(x,L_{n})=0$. This proves that $W_{n+1}(x,L_n)$ is at least of order $\hbar^{n-1}$. 

We now need to prove the same statement for higher correlation functions. Let us prove it by a second induction by defining
\beq \td{\mathcal{P}}_i \,:\, W_i(x_1,\dots,x_i) \text{ is of order at least } \hbar^{n-1}. \eeq
We prove $\td{\mathcal{P}}_i$ for all $i\geq n+1$ by induction. We just proved it for $i=n+1$ so initialization is done. Let us assume that $\td{\mathcal{P}}_j$ is true for all $j$ satisfying $n+1\leq j\leq i_0$. We look at the loop equation:
\bea \label{ind2} 0&=&P_{i_0+1}(x;L_{i_0})+W_{i_0+2}(x,x,L_{i_0})+2W_1(x)W_{i_0+1}(x,L_{i_0})\cr
&&+ \sum_{J \subset L_{i_0}, J\notin\{ \emptyset,L_{i_0}\}} W_{1+|J|}(x,J)W_{1+i_0-|J|}(x,L_{i_0}\setminus J)\cr
&&+\sum_{j=1}^{i_0} \frac{d}{dx_j} \frac{ W_{i_0}(x,L_{i_0}\setminus x_j)-W_{i_{0}}(L_{i_0})}{x-x_j}. 
\eea
By assumption on $\td{\mathcal{P}}_{i_0}$, the last sum with the derivatives contains terms of order at least $\hbar^{n-1}$. In the sum involving the subsets of $L_{i_0}$, it is straightforward to see that the terms are all of order at least $\hbar^{n-1}$. Indeed, as soon as one of the indices is greater than $n+1$, the assumption $\td{\mathcal{P}}_i$ for $n+1\leq i\leq i_0$ tells us that this term is already at order at least $\hbar^{n-1}$. Since the second factor of the product is at least of order $\hbar^0$, it does not decrease the order. Now, if both factors have indices strictly lower than $n+1$, then the assumption of $\mathcal{P}_j$  for all $j\leq n$ tell us that the order of the product is at least of $\hbar^{|J|+1-2+1+i_0-|J|-2}=\hbar^{i_0-2}$ which is greater than $n-1$ since $i_0\geq n+1$. Additionally, by induction on $\mathcal{P}_n$ we know that $W_{i_0+1}(x,L_{i_0})$ is at least of order $\hbar^{n-2}$ as well as $W_{i_0+2}(x,x,L_{i_0})$. Consequently, looking at order $\hbar^{n-3}$ in \eqref{ind2} gives
\beq \label{eqsys2} 0=P_{i_0+1}^{(n-3)}(x;L_{i_0})+2w_1^{(-1)}(x)w_{i_0+1}^{(n-2)}(x,L_{i_0}). \eeq
We can apply a similar reasoning as the one developed for \eqref{eqsys}. If we assume $w_{i_0+1}^{(n-2)}(x,L_{i_0})\neq 0$, then we have
\beq w_{i_0+1}^{(n-2)}(x,L_{i_0})=\frac{P_{i_0+1}^{(n-3)}(x;L_{i_0})}{2w_1^{(-1)}(x)}. \eeq
In our six cases, we get the following:
\begin{itemize}\item (Painlev\'e I)
\beqq w_{i_0+1}^{(n-2)}(x,L_{i_0})=\frac{P_{i_0+1}^{(n-3)}(L_{i_0})}{4(x-q_0)\sqrt{x+2q_0} }.\eeqq
\item (Painlev\'e II)
\beqq w_{i_0+1}^{(n-2)}(x,L_{i_0})=\frac{P_{i_0+1}^{(n-3)}(L_{i_0})}{2(x-q_0)\sqrt{x^2+2q_0x+q_0^2+\frac{\theta}{q_0}} }.\eeqq
\item (Painlev\'e III)
\beqq w_{i_0+1}^{(n-2)}(x,L_{i_0})=\frac{\sqrt{q_0(q_0^4-1)}P_{i_0+1}^{(n-3)}(L_{i_0})}{\sqrt{t}(q_0x+1)\sqrt{(\theta_\infty-\theta_0q_0^2)x^2-2xq_0(\theta_\infty q_0^2-\theta_0)+q_0^2(\theta_\infty-\theta_0 q_0^2)}}.\eeqq
\item (Painlev\'e IV)
\beqq w_{i_0+1}^{(n-2)}(x,L_{i_0})=\frac{P_{i_0+1}^{(n-3)}(L_{i_0})}{2(x-q_0)\sqrt{x^2+2(q_0+t)x+\frac{\theta_0^2}{q_0^2}} }.\eeqq
\item (Painlev\'e V)
\beqq w_{i_0+1}^{(n-2)}(x,L_{i_0})=\frac{P_{i_0+1}^{(n-3)}(L_{i_0})}{t(x-Q_0)\sqrt{(x-Q_1)(x-Q_2)} }.\eeqq
\item (Painlev\'e VI)
\beqq w_{i_0+1}^{(n-2)}(x,L_{i_0})=\frac{P_{i_0+1}^{(n-3)}(L_{i_0})}{\theta_\infty (x-q_0) \sqrt{x^2+\left(-1-\frac{\theta_0^2t^2}{\theta_\infty^2 q_0^2}+\frac{\theta_1^2(t-1)^2}{\theta_\infty^2(q_0-1)^2}\right)x+\frac{\theta_0^2t^2}{\theta_\infty^2q_0^2}} }.\eeqq
\end{itemize}

In all cases, we obtain that $w_{i_0+1}^{(n-2)}(x,L_{i_0})$ must have a simple pole (as a function of $x$) at the double zero of $E(x)$ in contradiction with the pole structure of the correlation functions proved in Appendix \ref{AppendixProofPole}. Consequently, we must have $w_{i_0+1}^{(n-2)}(x,L_{i_0})=0$. In particular, it means that $W_{i_0+1}(x,L_{i_0})$ (which by assumption of $\mathcal{P}_n$ was already known to be at least of order $\hbar^{n-2}$) is at least of order $\hbar^{n-1}$, making the induction on $\td{\mathcal{P}}_{i_0}$. Hence, by induction, we have proved that, $\forall\, i \geq n+1$, $\td{\mathcal{P}}_{i}$ holds which proves that $\mathcal{P}_{n+1}$ is true. Then, by induction, we have just proved that $\mathcal{P}_{n}$ holds for $n \ge 1$. In other words, we have proved the leading order condition of the topological type property in our six cases.

\section{Computation of the free energies $F_J^{(g)}$\label{AppendixSymplecticInvariants}}

\subsection{Computation of $F_J^{(0)}$}
The computation of $F_J^{(0)}$ requires specific computations detailed in \cite{EO}. We find the following results:
\begin{itemize}
\item (Painlev\'e I) $d\omega(z)=y(z)dx(z)$ has one singularity at $z=\infty$ (pole of $x(z)$ of order $4$ in the language of \cite{EO}). The temperature $t_\infty$ is vanishing and we find (see \cite{EO, IS})
\beq F^{(0)}_{\rm I}=\frac{48\,q_0^5}{5}.  \eeq
Note that we have $\dot{q}_0=-\frac{1}{12\,q_0}$ so that $-\frac{d}{dt} F_{\rm I}^{(0)}=4q_0^3$ in agreement with $\frac{d}{dt}\tau^{(0)}_{\rm I}=4q_0^3$. 
\item 
(Painlev\'e II) $d\omega(z)=y(z)dx(z)$ has two singularities at $z=0$ and $z=\infty$. We get (see \cite{P2})
\beq F_{\rm II}^{(0)}=\frac{4\theta q_0^3}{3}-\frac{\theta^2}{4}+\frac{\theta^3}{24q_0^3}- \frac{\theta^2}{2}\ln\left( -\frac{\theta}{4q_0}\right).\eeq
We verify that $-\frac{d}{dt}F_{\rm II}^{(0)}=\frac{\theta(8q_0^3-\theta)}{8q_0^2}=\frac{d}{dt}\tau_{\rm II}^{(0)}$.
\item 
(Painlev\'e III) $d\omega(z) = y(z)dx(z)$ has two additional singularities at simple zeros of $x(z)$. These singularities are poles of order $2$. We find
\bea F^{(0)}_{\rm III}&=&\frac{1}{4}\theta_0\theta_\infty\ln\left(\frac{q_0^2+1}{q_0^2-1}\right)+\frac{\theta_0^2}{8}\ln \left(\frac{q_0^2(q_0^2\theta_0-\theta_\infty)^2}{q_0^4-1}\right)+\frac{\theta_\infty^2}{8}\ln \left(\frac{(q_0^2\theta_0-\theta_\infty)^2}{q_0^2(q_0^4-1)}\right)\cr
&&+ \frac{3\theta_0^2-3\theta_\infty^2-2\theta_0\theta_\infty q_0^2+(5\theta_\infty^2-\theta_0^2)q_0^4-2\theta_0\theta_\infty q_0^6}{(q_0^4-1)^2}.
\eea
We verify that $-\frac{d}{dt}F_{\rm III}^{(0)}=\frac{(\theta_0^2-\theta_\infty^2)-4\theta_0\theta_\infty q_0^2+4(\theta_\infty^2+\theta_0^2)q_0^4-4\theta_0\theta_\infty q_0^6+(\theta_\infty^2-\theta_0^2)q_0^8}{4q_0(q_0^4-1)(q_0^2\theta_0-\theta_\infty)}=\frac{d}{dt}\tau_{\rm III}^{(0)}$.
\item (Painlev\'e IV) $d\omega(z)=y(z)dx(z)$ has four singularities at $z=0$ and $z=\infty$ (poles of $x(z)$ of order $2$) as well as zeros of $x(z)$ (poles of $Y(z)$). We get
\bea F^{(0)}_{\rm IV}&=&\frac{(3q_0^4-(8\theta_0+\theta_\infty)q_0^2+2\theta_0^2)\sqrt{q_0^4+2\theta_\infty q_0^2+\theta_0^2}}{2q_0^2}-\frac{3q_0^4}{2}+5\theta_0q_0^2+\frac{\theta_0^3}{q_0^2}\cr
&&+\frac{(\theta_0^2+\theta_\infty^2)}{2}\ln \left(q_0^2+\theta_\infty^2+\sqrt{q_0^4+2\theta_\infty q_0^2+\theta_0^2}\right)-2\theta_0(\theta_0-\theta_\infty)\ln q_0\cr
&&-\theta_0\theta_\infty\ln \left( 2\theta_0^2+2\theta_\infty q_0^2+2\theta_0\sqrt{q_0^4+2\theta_\infty q_0^2+\theta_0^2}\right).
\eea
We can verify that $-\frac{d}{dt}F_{\rm IV}^{(0)}=\frac{2(\theta_0-q_0^2)(q_0^2-\theta_0-\sqrt{q_0^4+2\theta_\infty q_0^2+\theta_0^2)}}{q_0}=\frac{d}{dt}\tau_{\rm IV}^{(0)}$.
\item (Painlev\'e V) In this case, it is easier to express all quantities in terms of the double zero $Q_0$ of the spectral curve, and we have that $d\omega(z)=y(z)dx(z)$ has six singularities at $z=0$ and $z=\infty$ and at the two conjugate zeros of $x(z)$ and $x(z)-1$. Tedious computations give
\footnotesize{\bea F^{(0)}_{\rm V}&=&(\theta_0^2+\theta_1^2+\theta_\infty^2)\left(\frac{1}{2}\ln 2-\frac{1}{8}\ln \left(\prod_{\pm}\left(1\pm \frac{\theta_0}{tQ_0}\pm \frac{\theta_1}{t(Q_0-1)}\right)\right)\right)+\frac{\theta_0^2}{2}\ln \theta_0+\frac{\theta_1^2}{2}\ln \theta_1-\frac{\theta_0^2}{2}\ln Q_0\cr
&&-\frac{\theta_1^2}{2}\ln(Q_0-1)-\frac{\theta_0^2+\theta_1^2}{2}\ln t+\frac{\theta_0\theta_\infty}{4}\ln\left(\frac{\left(1- \frac{\theta_0}{tQ_0}+ \frac{\theta_1}{t(Q_0-1)}\right)\left(1- \frac{\theta_0}{tQ_0}- \frac{\theta_1}{t(Q_0-1)}\right)}{ \left(1+ \frac{\theta_0}{tQ_0}+ \frac{\theta_1}{t(Q_0-1)}\right)\left(1+ \frac{\theta_0}{tQ_0}- \frac{\theta_1}{t(Q_0-1)}\right)}\right)\cr
&&+ \frac{\theta_1\theta_\infty}{4}\ln\left(\frac{\left(1+ \frac{\theta_0}{tQ_0}- \frac{\theta_1}{t(Q_0-1)}\right)\left(1- \frac{\theta_0}{tQ_0}- \frac{\theta_1}{t(Q_0-1)}\right)}{ \left(1+ \frac{\theta_0}{tQ_0}+ \frac{\theta_1}{t(Q_0-1)}\right)\left(1- \frac{\theta_0}{tQ_0}+ \frac{\theta_1}{t(Q_0-1)}\right)}\right)\cr
&&+ \frac{\theta_0\theta_1}{4}\ln\left(\frac{\left(1- \frac{\theta_0}{tQ_0}+ \frac{\theta_1}{t(Q_0-1)}\right)\left(1+ \frac{\theta_0}{tQ_0}- \frac{\theta_1}{t(Q_0-1)}\right)}{ \left(1+ \frac{\theta_0}{tQ_0}+ \frac{\theta_1}{t(Q_0-1)}\right)\left(1- \frac{\theta_0}{tQ_0}- \frac{\theta_1}{t(Q_0-1)}\right)}\right)\cr
&&-\frac{\theta_0^2}{2Q_0}+\frac{\theta_1^2}{2(Q_0-1)}+\frac{1}{4}t\theta_\infty \left(1+ \frac{\theta_0^2}{t^2Q_0^2}- \frac{\theta_1^2}{t^2(Q_0-1)^2}\right)-\frac{t^2}{32}\prod_{\pm}\left(1\pm \frac{\theta_0}{tQ_0}\pm \frac{\theta_1}{t(Q_0-1)}\right), \nonumber \\
\eea}\normalsize{}

\noindent
where the product $\prod_{\pm}$ is to be taken on the four possible choices of signs. Moreover, we get from the Lax pair
\footnotesize{\beq \label{tau0P5}\frac{d}{dt}\tau_{\rm V}^{(0)}=\frac{(\theta_0+\theta_1-\theta_\infty)q_0^2+\theta_0+\theta_1+\theta_\infty}{4(q_0^2-1)}-\frac{\left((\theta_0-\theta_1-\theta_\infty)q_0^2-\theta_0+\theta_1-\theta_\infty\right)\left((\theta_0-\theta_1-\theta_\infty)q_0-\theta_0+\theta_1-\theta_\infty\right)}{8t(q_0+1)q_0}.\eeq}\normalsize{}
In order to compare it with the expression of $F^{(0)}_{\rm V}$, we observe that we have
\beq \label{q0Q0}q_0=\frac{Q_0^2(\theta_0-\theta_1)-\theta_0(2Q_0-1)+Q_0(Q_0-1)((2Q_0-1)t-\theta_\infty)}{Q_0(Q_0-1)(\theta_0-\theta_1-\theta_\infty)},\eeq
and following \eqref{Q0P5}, $(t,Q_0)$ satisfies the following algebraic equation:
\beq \label{Q0t} Q_0^2(Q_0-1)^2(2Q_0-1)t^2-2\theta_\infty t Q_0^2(Q_0-1)^2+(Q_0(\theta_0+\theta_1)-\theta_0)(Q_0(\theta_0-\theta_1)-\theta_0)=0.\eeq
Consequently, we can express $\frac{d}{dt}\tau_{\rm V}^{(0)}$ by \eqref{tau0P5} and \eqref{q0Q0} in terms of $t$ and $Q_0$ and observe using \eqref{Q0t} that $-\frac{d}{dt}F_{\rm V}^{(0)}=\frac{d}{dt}\tau_{\rm V}^{(0)}$.
\item (Painlev\'e VI) The long computation is detailed in Section \ref{app:Fg-PVI} as illustrations of the method. We find
\bea \label{F0final}F_{\rm VI}^{(0)}&=&\frac{\theta_0^2+\theta_1^2+\theta_t^2+\theta_\infty^2}{2}\ln 2-\frac{\theta_0^2+\theta_1^2+\theta_t^2}{2}\ln\theta_\infty+\frac{\theta_0^2}{2}\ln \theta_0+\frac{\theta_1^2}{2}\ln\theta_1+\frac{\theta_t^2}{2}\ln \theta_t\cr
&&-\frac{\theta_0^2}{2}\ln q_0-\frac{\theta_1^2}{2}\ln(q_0-1)-\frac{\theta_t^2}{2}\ln(q_0-t)+\frac{i\pi}{4}(\theta_0\theta_1+\theta_0\theta_t+\theta_1\theta_t)\cr
&&+\left(\frac{\theta_0^2+\theta_t^2}{2}-\frac{\theta_0^2+\theta_1^2+\theta_\infty^2+\theta_t^2}{12}\right)\ln t+\left(\frac{\theta_1^2+\theta_t^2}{2}-\frac{\theta_0^2+\theta_1^2+\theta_\infty^2+\theta_t^2}{12}\right)\ln(t-1)\cr
&&-\left(\frac{\theta_0^2+\theta_1^2+\theta_\infty^2+\theta_t^2}{24}+\frac{\theta_0\theta_\infty}{8}+\frac{\theta_1\theta_\infty}{8}+\frac{\theta_0\theta_1}{4}\right) \ln\left(1+\frac{\theta_0 t}{\theta_\infty q_0}+\frac{\theta_1(t-1)}{\theta_\infty (q_0-1)}\right)\cr
&&-\left(\frac{\theta_0^2+\theta_1^2+\theta_\infty^2+\theta_t^2}{24}-\frac{\theta_0\theta_\infty}{8}-\frac{\theta_1\theta_\infty}{8}+\frac{\theta_0\theta_1}{4}\right) \ln\left(1-\frac{\theta_0 t}{\theta_\infty q_0}-\frac{\theta_1(t-1)}{\theta_\infty (q_0-1)}\right)\cr
&&-\left(\frac{\theta_0^2+\theta_1^2+\theta_\infty^2+\theta_t^2}{24}+\frac{\theta_0\theta_\infty}{8}-\frac{\theta_1\theta_\infty}{8}-\frac{\theta_0\theta_1}{4}\right) \ln\left(1+\frac{\theta_0 t}{\theta_\infty q_0}-\frac{\theta_1(t-1)}{\theta_\infty (q_0-1)}\right)\cr
&&-\left(\frac{\theta_0^2+\theta_1^2+\theta_\infty^2+\theta_t^2}{24}-\frac{\theta_0\theta_\infty}{8}+\frac{\theta_1\theta_\infty}{8}-\frac{\theta_0\theta_1}{4}\right) \ln\left(1-\frac{\theta_0 t}{\theta_\infty q_0}+\frac{\theta_1(t-1)}{\theta_\infty (q_0-1)}\right)\cr
&&-\left(\frac{\theta_0^2+\theta_1^2+\theta_\infty^2+\theta_t^2}{24}+\frac{\theta_0\theta_\infty}{8}-\frac{\theta_t\theta_\infty}{8}-\frac{\theta_0\theta_t}{4}\right)\ln\left(1+\frac{\theta_0}{\theta_\infty q_0}+\frac{\theta_t(t-1)}{\theta_\infty (q_0-t)}\right)\cr
&&-\left(\frac{\theta_0^2+\theta_1^2+\theta_\infty^2+\theta_t^2}{24}-\frac{\theta_0\theta_\infty}{8}+\frac{\theta_t\theta_\infty}{8}-\frac{\theta_0\theta_t}{4}\right)\ln\left(1-\frac{\theta_0}{\theta_\infty q_0}-\frac{\theta_t(t-1)}{\theta_\infty (q_0-t)}\right)\cr
&&-\left(\frac{\theta_0^2+\theta_1^2+\theta_\infty^2+\theta_t^2}{24}+\frac{\theta_0\theta_\infty}{8}+\frac{\theta_t\theta_\infty}{8}+\frac{\theta_0\theta_t}{4}\right)\ln\left(1+\frac{\theta_0}{\theta_\infty q_0}-\frac{\theta_t(t-1)}{\theta_\infty (q_0-t)}\right)\cr
&&-\left(\frac{\theta_0^2+\theta_1^2+\theta_\infty^2+\theta_t^2}{24}-\frac{\theta_0\theta_\infty}{8}-\frac{\theta_t\theta_\infty}{8}+\frac{\theta_0\theta_t}{4}\right)\ln\left(1-\frac{\theta_0}{\theta_\infty q_0}+\frac{\theta_t(t-1)}{\theta_\infty (q_0-t)}\right)\cr
&&-\left(\frac{\theta_0^2+\theta_1^2+\theta_\infty^2+\theta_t^2}{24}-\frac{\theta_1\theta_\infty}{8}-\frac{\theta_t\theta_\infty}{8}+\frac{\theta_1\theta_t}{4}\right)
\ln\left(1+\frac{\theta_1}{\theta_\infty (q_0-1)}+\frac{\theta_t t}{\theta_\infty (q_0-t)}\right)\cr
&&-\left(\frac{\theta_0^2+\theta_1^2+\theta_\infty^2+\theta_t^2}{24}+\frac{\theta_1\theta_\infty}{8}+\frac{\theta_t\theta_\infty}{8}+\frac{\theta_1\theta_t}{4}\right)
\ln\left(1-\frac{\theta_1}{\theta_\infty (q_0-1)}-\frac{\theta_t t}{\theta_\infty (q_0-t)}\right)\cr
&&-\left(\frac{\theta_0^2+\theta_1^2+\theta_\infty^2+\theta_t^2}{24}-\frac{\theta_1\theta_\infty}{8}+\frac{\theta_t\theta_\infty}{8}-\frac{\theta_1\theta_t}{4}\right)
\ln\left(1+\frac{\theta_1}{\theta_\infty (q_0-1)}-\frac{\theta_t t}{\theta_\infty (q_0-t)}\right)\cr
&&-\left(\frac{\theta_0^2+\theta_1^2+\theta_\infty^2+\theta_t^2}{24}+\frac{\theta_1\theta_\infty}{8}-\frac{\theta_t\theta_\infty}{8}-\frac{\theta_1\theta_t}{4}\right)
\ln\left(1-\frac{\theta_1}{\theta_\infty (q_0-1)}+\frac{\theta_t t}{\theta_\infty (q_0-t)}\right).\nonumber \\
\eea
We can verify that $-\frac{d}{dt}F_{\rm VI}^{(0)}=\frac{d}{dt}\tau_{\rm VI}^{(0)}$ with
\footnotesize{\beq \label{dtau0P6}\frac{d}{dt}\tau_{\rm VI}^{(0)}=\frac{\left(\theta_0^2-(\theta_0^2-\theta_1^2+\theta_\infty^2-\theta_t^2)q_0+(\theta_\infty^2-\theta_t^2)q_0^2\right)t(t-2q_0)+q_0^2\left(\theta_0^2+\theta_t^2-(\theta_1^2-\theta_0^2-\theta_\infty^2-\theta_t^2)q_0+\theta_\infty^2q_0^2\right)}{4t(t-1)q_0(q_0-1)(q_0-t)}.\eeq}\normalsize{}
\end{itemize}

\subsection{Computation of $F_J^{(1)}$}

\underline{For Painlev\'e I (one branch point case)}:

In the case of a parametrization of the form \eqref{ParametrizationP1} $x(z)=z^2+a$ and $y(z)=zg(x(z))$ (i.e. $y^2(x)=(x-a)g(x)^2$ with $g$ a rational function that does not vanish at $x=a$) with only one branch point at $x=a$ (i.e. $z=0$), the formula proposed by Eynard and Orantin for $F^{(1)}$ in \cite{EO} reduces to
\beq F_{\rm I}^{(1)}=\frac{1}{24}\ln( y'(0))=\frac{1}{24}\ln g(a).\eeq
For Painlev\'e ${\rm I}$, we find ($g(x)=2(x-q_0)$ and $a=-2q_0$)
\beq F_{\rm I}^{(1)}=\frac{1}{24}\ln(-6q_0) \text{ and } \frac{d}{dt}\tau_{\rm I}^{(1)}=-\frac{d}{dt} F_{\rm I}^{(1)}=-\frac{1}{288q_0^2}=\frac{1}{48t}.\eeq
We have used here the fact that $6q_0^2+t=0$ to simplify quantities. Thus we have verified that $\frac{d}{dt}\tau_{\rm I}^{(1)}=-\frac{d}{dt}F_{\rm I}^{(1)}$.

\bigskip \noindent
\underline{For Painlev\'e II -- VI (two branch points cases)}:

In the case of a parametrization of the form \eqref{eq:par-rep2} $x(z)=\frac{a+b}{2}+\frac{b-a}{4}\left(z+\frac{1}{z}\right)$ and $y(x)=\frac{(b-a)(z-1)(z+1)}{4z}g(x(z))$ (i.e. $y^2(x)=(x-a)(x-b)g(x)^2$) with $g(x)$ a rational function in $x$ not vanishing at $x=a$ and $x=b$, the formula proposed by Eynard and Orantin for $F^{(1)}$ in \cite{EO} reduces to
\beq F^{(1)}=\frac{1}{24}\ln \left(-(b-a)^4g(a)g(b)\right)
\quad (\text{note that  } \tau_{\text{Berg}}=(b-a)^{\frac{1}{4}}).
\eeq
It is also in agreement with the formula presented in \cite{F1}. Note that the previous notion $\tau_{\text{Berg}}$ of Bergman $\tau$-function was introduced by D. Korotkin in the resolution of Riemann-Hilbert problems and Schlesinger problems in \cite{KorotkinNew} and in the context of integrable systems associated to Hurwitz spaces in \cite{KorotkinNew2}. Then, it is straightforward to compute the values of $F^{(1)}$ in all six cases by inserting the values of $a$, $b$ and $g(x)$ in the previous formula. We find the following

\begin{itemize}
\item (Painlev\'e II)
\beq F_{\rm II}^{(1)}=\frac{1}{24}\ln \left(-16\theta^2\left(4+\frac{\theta}{q_0^3}\right)\right) \text{ and }\frac{d}{dt}\tau_{\rm II}^{(2)}=-\frac{d}{dt}F_{\rm II}^{(1)}=-\frac{\theta q_0}{8(4q_0^3+\theta)^2},\eeq
where we used $t=-2q_0^2+\frac{\theta}{q_0}$ to remove $t$ from all previous quantities.
\item (Painlev\'e III)
\bea F_{\rm III}^{(1)}&=&\frac{1}{24}\ln\left( \frac{4(\theta_\infty^2-\theta_0^2)^2(\theta_0q_0^6-3\theta_0q_0^4+3\theta_0q_0^2-\theta_\infty)}{(\theta_\infty-\theta_0 q_0^2)^3}\right)\cr
\frac{d}{dt}\tau_{\rm III}^{(2)}&=&-\frac{d}{dt}F^{(1)}_{\rm III}=-\frac{(\theta_\infty^2-\theta_0^2)q_0^3(q_0^4-1)^2}{2(\theta_\infty-\theta_0 q_0^2)(\theta_0q_0^6-3\theta_0q_0^4+3\theta_0q_0^2-\theta_\infty)^2},
\eea
where we used $t=\frac{q_0(\theta_\infty-\theta_0 q_0^2)}{q_0^4-1}$ to remove $t$ from all previous quantities.
\item (Painlev\'e IV)
\bea F_{\rm IV}^{(1)}&=&\frac{1}{24}\ln \left( -\frac{16(\theta_0-q_0^2-tq_0)^2(\theta_0+q_0^2+tq_0)^2(3q_0^4+2q_0^3t+\theta_0^2)}{\theta_0^2q_0^4}\right)\cr
\frac{d}{dt}\tau_{\rm IV}^{(1)}&=&-\frac{d}{dt}F^{(1)}_{\rm IV}=\frac{(\theta_0-q_0^2-tq_0)(\theta_0+q_0^2+tq_0)q_0^3}{4(3q_0^4+2q_0^3t +\theta_0^2)^2},
\eea
where we used $\theta_\infty=\frac{3q_0^4+4q_0^3t+tq_0^2-\theta_0^2}{2q_0^2}$ to remove $\theta_\infty$ from all previous quantities.
\item (Painlev\'e V)
\beq F_{\rm V}^{(1)}=\frac{1}{24} \ln \left[-\frac{4\left((\theta_0-\theta_1-\theta_\infty)q_0^2+2(t+\theta_1-\theta_0)q_0+\theta_0-\theta_1+\theta_\infty\right)^2P_4(t)}{(q_0-1)^2\left((q_0-1)^4\left(\theta_0+\theta_\infty-\theta_1-(\theta_0-\theta_1-\theta_\infty)q_0\right)^2-4q_0^2(q_0+1)^2t^2\right)^2} \right],\eeq
where $P_4(t)=a_4t^4+a_3t^3+a_2t^2+a_1t+a_0$ is given by
\footnotesize{\bea a_4&=&16q_0^4(6q_0+q_0^2+1)\cr
a_3&=&32(q_0-1)q_0^3\left( (\theta_0+\theta_\infty-\theta_1)+(3\theta_\infty+\theta_0-5\theta_1)q_0+(3\theta_\infty+\theta_1-5\theta_0)q_0^2+(-\theta_0+\theta_1+\theta_\infty)q_0^3\right)\cr
a_2&=&8q_0^2(q_0-1)^2\Big( 3(-\theta_0-\theta_\infty+\theta_1)^2+4(-\theta_0-\theta_\infty+\theta_1)(2\theta_0-\theta_\infty+2\theta_1)q_0+(2\theta_0^2+2\theta_1^2+28\theta_0\theta_1+10\theta_\infty^2)q_0^2\cr
&&+(-4(2\theta_0-\theta_\infty+2\theta_1)(-\theta_0+\theta_1+\theta_\infty))q_0^3+3(-\theta_0+\theta_1+\theta_\infty)^2q_0^4  \Big)\cr
a_1&=&8q_0(q_0-1)^3\left((\theta_0+\theta_\infty-\theta_1)+(-\theta_0+\theta_1+\theta_\infty)q_0\right)^2\cr
&& \left((\theta_0+\theta_\infty-\theta_1)+(-3\theta_0-\theta_\infty-\theta_1)q_0+(-\theta_\infty-\theta_0-3\theta_1)q_0^2+(-\theta_0+\theta_1+\theta_\infty)q_0^3\right)\cr
a_0&=&(q_0-1)^6\left( (\theta_0+\theta_\infty-\theta_1)+(-\theta_0+\theta_1+\theta_\infty)q_0 \right).
\eea}\normalsize{}
One can verify with tedious computations and \eqref{Q0P5} that $\frac{d}{dt}\tau_{\rm V}^{(1)}=-\frac{d}{dt}F^{(1)}_{\rm V}$ holds.
\item (Painlev\'e VI)
\bea \hspace{-1.em}  F_{\rm VI}^{(1)}&=&-\frac{1}{12}\left(\ln 2+\ln\theta_0+\ln\theta_1+\ln \theta_\infty+\ln \theta_t\right)\nonumber \\[+.3em]
&&-\frac{1}{9}\ln t-\frac{1}{9}\ln(t-1)+\frac{1}{12}\ln q_0+\frac{1}{12}\ln (q_0-1)+\frac{1}{12}\ln(q_0-t)\cr
&&+\frac{1}{24}\ln\left(\theta_\infty^2q_0^2-q_0\left(\frac{\theta_0^2t(t+1)}{q_0^2}-\frac{\theta_1^2t(t-1)}{(q_0-1)^2}+\frac{\theta_t^2t(t-1)}{(q_0-t)^2}\right)+\frac{\theta_0^2t^2}{q_0^2}\right)\cr
&&+\frac{1}{36}\ln\left(\prod_{\pm} \left(\theta_\infty\pm \frac{\theta_0t}{q_0}\pm \frac{\theta_1(t-1)}{q_0-1}\right)\left(\theta_\infty\pm \frac{\theta_0}{q_0}\pm \frac{\theta_t(t-1)}{q_0-t}\right)\left(\theta_\infty\pm \frac{\theta_1}{q_0-1}\pm \frac{\theta_t t}{q_0-t}\right) \right), \nonumber \\
\eea
where the product indexed by $\pm$ indicates that we must take all possible choice of signs. One can verify with tedious computations and \eqref{Q0P6} that $\frac{d}{dt}\tau_{\rm VI}^{(1)}=-\frac{d}{dt}F^{(1)}_{\rm VI}$ holds.
\end{itemize}

\subsection{Higher genus symplectic invariants}
For the simplest Painlev\'e equations, we can compute the first orders $F_J^{(2)}$, $F_J^{(3)}$, etc. of the topological recursion depending on the complexity of the spectral curve. We find
\begin{itemize}
\item (Painlev\'e I)
\beq F^{(2)}_{\rm I}=-\frac{7}{207360\,q_0^5} \,\,,\,\, F^{(6)}_{\rm I}=-\frac{245}{429981696\,q_0^{10}}.\eeq
(See also \cite{EO, IS})
One can verify 
\[
\frac{d}{dt}\tau^{(2)}_{\rm I}=\frac{7}{497664\,q_0^7}
=-\frac{d}{dt} F_{\rm I}^{(2)}
\] and 
\[
\frac{d}{dt}\tau^{(3)}_{\rm I}=\frac{1225}{2579890176\,q_0^{12}}
=-\frac{d}{dt} F_{\rm I}^{(3)}.
\] 
% and $\frac{d}{dt}\tau^{(g)}_{\rm I}=-\frac{d}{dt} F_{\rm I}^{(g)}$. 
Here we used $\dot{q}_0=-\frac{1}{12\,q_0}$.

\item (Painlev\'e II)
\bea
F_{\rm II}^{(2)}&=&{\frac {1}{480}}\,{\frac { \left( 
2048\,{q_0}^{12}+2560\,\theta\,{q_0}^{9}+1280\,{\theta}^{2}{q_0}^{6}
+1020\,{\theta}^{3}{q_0}^{3}-45\,{\theta}^{4} 
\right) {q_0}^{3}}{{\theta}^{2}\left( 4\,{q_0}^{3}+\theta \right) ^{5}}}  \cr
F_{\rm II}^{(3)}&=&-\frac {q_0^6}{4032\theta^4 \left( \theta+4\,q_0^3 \right) ^{10}}\, \Big(4194304\,q_0^{24}+10485760\,\theta\,q_0^{21}+11796480\,\theta^2q_0^{18}\cr 
&& \hspace{-5.em}
+7864320\,\theta^3q_0^{15}+3440640\,\theta^4q_0^{12}
-5694528\,\theta^5q_0^9+5232752\,\theta^6q_0^6
-510412\,\theta^7q_0^3+3969\,\theta^8 \Big). \nonumber \\
\eea
(See also \cite{P2}.) Using $\dot{q}_0=-\frac{q_0^2}{4q_0^3+\theta}$, 
one can verify that $\frac{d}{dt}\tau^{(g)}_{\rm II}=-\frac{d}{dt} F_{\rm II}^{(g)}$ for $g=2$ and $g=3$.
\item 
(Painlev\'e III) \beq F_{\rm III}^{(2)}=\frac{Q_{30}(q_0)}{240(\theta_0^2-\theta_\infty^2)(-\theta_0q_0^6+3\theta_\infty q_0^4-3\theta_0 q_0^2+\theta_\infty)^5}\eeq
where $Q_{30}(q_0)$ is a even polynomial in $q_0$ of degree $30$:
\scriptsize{\bea Q_{30}(q_0)&=&\theta_\infty^3(10\theta_\infty^2\theta_0^2+7\theta_\infty^4-\theta_0^4)+(-15\theta_\infty^2\theta_0(10\theta_\infty^2\theta_0^2+7\theta_\infty^4-\theta_0^4))q_0^2+15\theta_\infty(-8\theta_0^6+46\theta_\infty^4\theta_0^2+9\theta_\infty^6+65\theta_\infty^2\theta_0^4)q_0^4\cr
&&+(-5\theta_\infty^2\theta_0(205\theta_\infty^4+341\theta_0^4+910\theta_\infty^2\theta_0^2))q_0^6+(-15\theta_\infty(-195\theta_0^6-652\theta_\infty^2\theta_0^4-631\theta_\infty^4\theta_0^2+22\theta_\infty^6))q_0^8\cr
&&+(-3\theta_0(5212\theta_\infty^2\theta_0^4+8837\theta_\infty^4\theta_0^2+1494\theta_\infty^6+473\theta_0^6))q_0^{10}+5\theta_\infty(534\theta_\infty^6+3893\theta_\infty^4\theta_0^2+9884\theta_\infty^2\theta_0^4+1705\theta_0^6)q_0^{12}\cr
&&+(-15\theta_0(13\theta_0^6+3465\theta_\infty^4\theta_0^2+2604\theta_\infty^2\theta_0^4+782\theta_\infty^6))q_0^{14}+15\theta_\infty(3465\theta_\infty^2\theta_0^4+782\theta_0^6+13\theta_\infty^6+2604\theta_\infty^4\theta_0^2)q_0^{16}\cr
&&+(-5\theta_0(1705\theta_\infty^6+534\theta_0^6+3893\theta_\infty^2\theta_0^4+9884\theta_\infty^4\theta_0^2))q_0^{18}+3\theta_\infty(1494\theta_0^6+473\theta_\infty^6+5212\theta_\infty^4\theta_0^2+8837\theta_\infty^2\theta_0^4)q_0^{20}\cr
&&+(-15\theta_0(631\theta_\infty^2\theta_0^4+652\theta_\infty^4\theta_0^2-22\theta_0^6+195\theta_\infty^6))q_0^{22}+5\theta_\infty\theta_0^2(205\theta_0^4+910\theta_\infty^2\theta_0^2+341\theta_\infty^4)q_0^{24}\cr
&&+15\theta_0(8\theta_\infty^6-65\theta_\infty^4\theta_0^2-9\theta_0^6-46\theta_\infty^2\theta_0^4)q_0^{26}+(-15\theta_\infty\theta_0^2(-10\theta_\infty^2\theta_0^2+\theta_\infty^4-7\theta_0^4))q_0^{28}\cr
&&+\theta_0^3(-10\theta_\infty^2\theta_0^2+\theta_\infty^4-7\theta_0^4)q_0^{30}.
\eea}\normalsize{}
Using $\dot{q}_0=\frac{(q_0^4-1)^2}{\theta_0q_0^6-3\theta_\infty q_0^4+3\theta_0q_0^2-\theta_\infty}$,
one can verify that $-\frac{d}{dt} F_{\rm III}^{(2)}=-\frac{d}{dt}\tau^{(2)}_{\rm III}$.

\item (Painlev\'e IV)
\beq F^{(2)}_{\rm IV}=-\frac{q_0^4Q_9(t,q_0)}{960\theta_0^2\left(3q_0^4+2tq_0^3+\theta_0^2\right)^5\left((tq_0+q_0^2)^2-\theta_0^2\right)^2}\eeq
with
\bea Q_9(t,q_0)&=&243q_0^{24}-603q_0^{20}\theta_0^2+353q_0^4\theta_0^{10}-16\theta_0^{12}-3474q_0^{16}\theta_0^4+1962q_0^{12}\theta_0^6-2561q_0^8\theta_0^8\cr
&&+q_0^3(1782q_0^{20}-1593q_0^{16}\theta_0^2+8406q_0^8\theta_0^6-4762q_0^4\theta_0^8+91\theta_0^{10}-16212q_0^{12}\theta_0^4)t\cr
&&+2q_0^6(6690q_0^4\theta_0^6+582q_0^{12}\theta_0^2-1525\theta_0^8+2889q_0^{16}-16188q_0^8\theta_0^4)t^2\cr
&&-q_0^5(589\theta_0^8-9569q_0^{12}\theta_0^2-10872q_0^{16}-10299q_0^4\theta_0^6+35655q_0^8\theta_0^4)t^3\cr
&&+3q_0^8(1289\theta_0^6+5303q_0^8\theta_0^2+4361q_0^{12}-7785q_0^4\theta_0^4)t^4\cr
&&+q_0^7(10442q_0^{12}-9120q_0^4\theta_0^4+545\theta_0^6+13365q_0^8\theta_0^2)t^5\cr
&&+4q_0^{10}(1382q_0^8+1573q_0^4\theta_0^2-491\theta_0^4)t^6-q_0^9(175\theta_0^4-1591q_0^4\theta_0^2-1872q_0^8)t^7\cr
&&+2q_0^{12}(184q_0^4+85\theta_0^2)t^8+  32q_0^{15} t^9.
\eea
Using $\dot{q}_0=-\frac{q_0^3(t+2q_0)}{3q_0^4+2tq_0^3+\theta_0^2}$, 
one can verify that $-\frac{d}{dt} F_{\rm IV}^{(2)}=\frac{d}{dt}\tau^{(2)}_{\rm IV}$.

\end{itemize}

\subsection{Details for the computation $F^{(0)}_{\rm VI}$}\label{app:Fg-PVI}

\subsubsection{Spectral curve}
The spectral curve for $\PVI$ is given by 
\beq  y^2=\frac{\theta_\infty^2(x-q_0)^2 P(x)}{4x^2(x-1)^2(x-t)^2} \,\text{ with } P(x)=x^2+\left(-1-\frac{\td{\theta}_0^2t^2}{q_0^2}+\frac{\td{\theta}_1^2(t-1)^2}{(q_0-1)^2}\right)x+\frac{\td{\theta}_0^2t^2}{q_0^2}, 
\eeq
where we remind the reader that the reduced monodromy parameters are $\td{\theta}_i=\frac{\theta_i}{\theta_\infty}$. Note that $P(x)$ also admits a more symmetric formulation. Using the algebraic equation satisfied by $q_0$, we can reformulate
\beq  P(x)=x^2+\left(-\frac{\td{\theta}_0^2t(t+1)}{q_0^2}+\frac{\td{\theta}_1^2t(t-1)}{(q_0-1)^2}-\frac{\td{\theta}_t^2t(t-1)}{(q_0-t)^2}\right)x+\frac{\td{\theta}_0^2t^2}{q_0^2}. 
\eeq
Equivalently, it is also defined by the following $3$ conditions:
\beq 
P(0)=\frac{\td{\theta}_0^2t^2}{q_0^2} \,\,,\,\, 
P(1)=\frac{(t-1)^2\td{\theta}_1^2}{(q_0-1)^2}\text{ and } 
P(t)=\frac{t^2(t-1)^2\td{\theta}_t^2}{(q_0-t)^2}. 
\eeq

Since the spectral curve is of genus $0$, we can parametrize it globally. Let us define the following parametrizations:
\bea \label{definitions} x(z)&=&\frac{a+b}{2}+\frac{b-a}{4}\left(z+\frac{1}{z}\right) \text{ where by definition } P(x)=(x-a)(x-b),\cr
x(z)&=&0+\frac{b-a}{4z}(z-z_{0,+})(z-z_{0,-}) \text{ where by definition } x(z_{0,+})=x(z_{0,-})=0,\cr
x(z)&=&1+\frac{b-a}{4z}(z-z_{1,+})(z-z_{1,-}) \text{ where by definition } x(z_{1,+})=x(z_{1,-})=1,\cr
x(z)&=&t+\frac{b-a}{4z}(z-z_{t,+})(z-z_{t,-}) \text{ where by definition } x(z_{t,+})=x(z_{t,-})=t,\cr
x(z)&=&q_0+\frac{b-a}{4z}(z-z_{q_0,+})(z-z_{q_0,-}) \text{ where by definition } x(z_{q_0,+})=x(z_{q_0,-})=q_0.\cr
\eea
Thus, the branch points are located at $z=\pm 1$ and we can rewrite the spectral curve and the 1-form $w=ydx$:
\bea \label{w} 
y(z) & = &\frac{2\theta_\infty (z-z_{q_0,+})(z-z_{q_0,-})(z^2-1)z}{(b-a)(z-z_{0,+})(z-z_{0,-})(z-z_{1,+})(z-z_{1,-})(z-z_{t,+})(z-z_{t,-})}, \\[+.3em]
w(z) & = & y(z)dx(z) \\ 
& = & \frac{\theta_\infty (z-z_{q_0,+})(z-z_{q_0,-})(z^2-1)^2}{2z(z-z_{0,+})(z-z_{0,-})(z-z_{1,+})(z-z_{1,-})(z-z_{t,+})(z-z_{t,-})}dz. \nonumber
\eea
The important point is to notice that $w(z)$ is a meromorphic 1-form with simple poles only at $z\in\{0,z_{i,\pm}, \infty\}$ with $i\in\{0,1,t\}$. Analyzing the different residues at these points and using \eqref{SpecCurve2}, we get that
\begin{multline} \label{w2}  
w(z)=\biggl(\frac{\theta_\infty}{2z}-\frac{\theta_0}{2(z-z_{0,+})}+\frac{\theta_0}{2(z-z_{0,-})}+\frac{\theta_1}{2(z-z_{1,+})}-\frac{\theta_1}{2(z-z_{1,-})}-\frac{\theta_t}{2(z-z_{t,+})}+\frac{\theta_t}{2(z-z_{t,-})}\biggr)dz. 
\end{multline}
Note that in particular, the ambiguity between $z_{i,+}$ and $z_{i,-}$ is settled by the choice of the sign in the previous residues. We adopted for convenience a sign difference for $\theta_1$. In fact, one can verify that equations \eqref{w} and \eqref{w2} are equivalent to the algebraic equation for $q_0(t)$. Let us compute the first symplectic invariant $F^{(0)}$ for our spectral curve.

\subsubsection{Computation of $F_{\rm VI}^{(0)}$}
Following the Eynard-Orantin framework \cite{EO}, we observe that $w(z)$ has $8$ singular points: $z\in\{0, \infty, z_{0,+} ,z_{0,-}, z_{1,+}, z_{1,-} ,z_{t,+}, z_{t,-}\}$. In the language of \cite{EO}, $z=0$ and $z=\infty$ are type $1$ singular points since they are simple poles of $x(z)$ (but not of $y(z)$). On the other hand, the remaining six singular points fall into the type $2$ category (i.e. poles of $y(z)$). Hence, the local coordinates are
\bea 
z_\alpha(p)= \left\{
\begin{array}{ll}
x(p) & \text{ for } \alpha\in\{0,\infty\}, \\
\displaystyle
\frac{1}{x(p)-x(\alpha)} 
& \text{ for } \alpha\in\{z_{0,\pm}, z_{1,\pm}, z_{t,\pm}\}.
\end{array}
 \right. \nonumber
\eea
The temperatures are given from \eqref{w2}:
\begin{center}
  \begin{tabular}{|c||c|c|c|c|c|c|c|c|} \hline
    $\alpha$ & $0$ & $\infty$ & $z_{0,+}$ & $z_{0,-}$ & 
    $z_{1,+}$ & $z_{1,-}$ & $z_{t,+}$ & $z_{t,-}$   \\ \hline 
    $t_{\alpha}$ & $\frac{\theta_\infty}{2}$ & $-\frac{\theta_\infty}{2}$ 
    & $-\frac{\theta_0}{2}$ & $\frac{\theta_0}{2}$ 
    & $\frac{\theta_1}{2}$  & $-\frac{\theta_1}{2}$ 
    & $-\frac{\theta_t}{2}$ & $\frac{\theta_t}{2}$   \\ \hline
  \end{tabular}
\end{center}
We can also verify that the potential $V_\alpha(p)$ is trivial for any pole $\alpha$. 
In order to compute the $\mu_\alpha$'s we observe that
\bea&& \text{For~~} i\in\{0,1,t\}\,:\, \frac{d z_{\alpha_i}(p)}{z_{\alpha_i}(p)}=\frac{dp}{p}-\frac{dp}{p-z_{i,+}}-\frac{dp}{p-z_{i,-}}.\cr
&& \text{For }z=0: \frac{dz_\alpha(p)}{z_\alpha(p)}=-\frac{1}{p}+d_p\left(\ln (1+p^2+\frac{2(a+b)p}{b-a}) \right).\cr
&&\text{For }z=\infty: \frac{dz_\alpha(p)}{z_\alpha(p)}=\frac{1}{p}+d_p\left(\ln (1+\frac{1}{p^2}+\frac{2(a+b)}{(b-a)p}) \right).
\eea
Then, a tedious computation from the Eynard-Orantin formula (\cite[\S 4.2.2]{EO}) shows that we obtain
\bea F^{(0)}_{\rm VI}&=&\frac{\theta_0^2+\theta_1^2-\theta_\infty^2+\theta_t^2}{8}\ln\frac{(b-a)^2}{16} \cr
&&+\frac{1}{8}\Big[\theta_0^2\ln\frac{(z_{0,+}-z_{0,-})^4}{z_{0,+}z_{0,-}} + 
\theta_1^2\ln\frac{(z_{1,+}-z_{1,-})^4}{z_{1,+}z_{1,-}}+\theta_t^2\ln\frac{(z_{t,+}-z_{t,-})^4}{z_{t,+}z_{t,-}}\Big]\cr
&&+\frac{1}{4}\Big[\theta_0\theta_\infty \ln\frac{z_{0,+}}{z_{0,-}} -\theta_1\theta_\infty \ln\frac{z_{1,+}}{z_{1,-}}+\theta_t\theta_\infty\ln\frac{z_{t,+}}{z_{t,-}}\Big]\cr
&&+\frac{1}{4}\Big[\theta_0\theta_1\ln \frac{(z_{1,+}-z_{0,+})(z_{1,-}-z_{0,-})}{(z_{1,+}-z_{0,-})(z_{0,+}-z_{1,-})}+\theta_1\theta_t\ln \frac{(z_{1,+}-z_{t,+})(z_{1,-}-z_{t,-})}{(z_{1,+}-z_{t,-})(z_{t,+}-z_{1,-})}\cr
&& -\theta_0\theta_t\ln \frac{(z_{t,+}-z_{0,+})(z_{t,-}-z_{0,-})}{(z_{t,+}-z_{0,-})(z_{0,+}-z_{t,-})}\Big].
\eea
Since $z_{i,+}z_{i,-}=1$ holds for any $i\in \, \{0,1,t\}$, the expression for $F_{\rm VI}^{(0)}$ simplifies into 
\bea \label{F0P6}F^{(0)}_{\rm VI}&=&\frac{\theta_0^2+\theta_1^2-\theta_\infty^2+\theta_t^2}{8}\ln\frac{(b-a)^2}{16} \cr
&&+\frac{1}{8}\Big[\theta_0^2\ln \frac{(z_{0,+}^2-1)^4}{z_{0,+}^4} + 
\theta_1^2\ln \frac{(z_{1,+}^2-1)^4}{z_{1,+}^4}+\theta_t^2\ln\frac{(z_{t,+}^2-1)^4}{z_{t,+}^4}\Big]\cr
&&+\frac{1}{4}\Big[\theta_0\theta_\infty \ln z_{0,+}^2 -\theta_1\theta_\infty \ln z_{1,+}^2+\theta_t\theta_\infty\ln z_{t,+}^2\Big]\cr
&&+\frac{1}{4}\Big[\theta_0\theta_1\ln \left(-\frac{(z_{1,+}-z_{0,+})^2}{(1-z_{1,+}z_{0,+})^2}\right)+\theta_1\theta_t\ln \left(-\frac{(z_{1,+}-z_{t,+})^2}{(1-z_{1,+}z_{t,+})^2}\right)\cr
&& -\theta_0\theta_t\ln \left(-\frac{(z_{t,+}-z_{0,+})^2}{(1-z_{t,+}z_{0,+})^2}\right)\Big].
\eea
Ingenious computations shows that we have
\bea \label{ab} ab&=&
\frac{\theta_0^2t^2}{\theta_\infty^2q_0^2}, \cr
a+b&=&1+\frac{\theta_0^2t^2}{\theta_\infty^2q_0^2}-\frac{\theta_1^2(t-1)^2}{\theta_\infty^2 (q_0-1)^2}=t\left[1+\frac{\theta_0^2}{\theta_\infty^2q_0^2}-\frac{\theta_t^2(t-1)^2}{\theta_\infty^2(q_0-t)^2} \right]\cr
&=&t+1+\frac{(t-1)\theta_1^2}{(q_0-1)^2\theta_\infty^2}-\frac{t^2(t-1)\theta_t^2}{(q_0-t)^2\theta_\infty^2}.
\eea
Consequently, using $(b-a)^2=(a+b)^2-4ab$ 
we find three equivalent expressions for $(b-a)^2$:
\bea \label{BminusA} (b-a)^2&=&\underset{\pm}{\prod} \left(1\pm\frac{\theta_0t}{\theta_\infty q_0}\pm\frac{\theta_1(t-1)}{\theta_\infty(q_0-1)}\right)\cr
&=&t^2\underset{\pm}{\prod}\left(1\pm\frac{\theta_0}{\theta_\infty q_0}\pm\frac{\theta_t(t-1)}{\theta_\infty (q_0-t)}\right)\cr
&=&(t-1)^2\underset{\pm}{\prod}\left(1\pm \frac{\theta_1}{\theta_\infty(q_0-1)}\pm \frac{\theta_t t}{\theta_\infty(q_0-t)} \right).
\eea
Here, each product is taken over the four possibilities for the signs. We now define non-ambiguously $z_{0,+}, z_{1,+}$ and $z_{t,+}$, from the definition \eqref{definitions}. We have
\beq \label{res1} z_{i,+}+\frac{1}{z_{i,+}}=\frac{2(2i-a-b)}{b-a} \text{ for } i\in \{0,1,t\}.\eeq
Moreover, the points $z_{0,+}, z_{1,+}$ and $z_{t,+}$ are defined such that
\beq \Res_{z\to z_{0,+}} w(z)dz=-\frac{\theta_0}{2}\,\,,\,\,\Res_{z\to z_{1,+}} w(z)dz=\frac{\theta_1}{2}\,\,,\,\,\Res_{z\to z_{t,+}} w(z)dz=-\frac{\theta_t}{2},\eeq
where
\beq w(z)=\frac{\theta_\infty (x(z)-q_0)(b-a)^2(z^2-1)^2}{32z^3x(z)(x(z)-1)(x(z)-t)}.\eeq
Computing the various residues gives that:
\bea \label{res2} z_{0,+}-\frac{1}{z_{0,+}}&=&
\frac{4\theta_0 t}{\theta_\infty q_0(b-a)}\cr
 z_{1,+}-\frac{1}{z_{1,+}}&=&\frac{4\theta_1(t-1)}{\theta_\infty(q_0-1)(b-a)}\cr
z_{t,+}-\frac{1}{z_{t,+}}&=&\frac{4\theta_t t(t-1)}{\theta_\infty (q_0-t)(b-a)}.
\eea
Combining \eqref{res1} with \eqref{res2} gives
\bea z_{0,+}&=& \frac{1}{b-a}\left(\frac{2\theta_0 t}{\theta_\infty q_0}-a-b\right)\cr
z_{1,+}&=&\frac{1}{b-a}\left(\frac{2\theta_1 (t-1)}{\theta_\infty (q_0-1)}+2-a-b\right)\cr
z_{t,+}&=&\frac{1}{b-a}\left(\frac{2\theta_t t(t-1)}{\theta_\infty (q_0-t)}+2t-a-b\right).
\eea
We can now replace $a+b$ using expression given in \eqref{ab}. We find the following values for $z_{i,+}$:
\bea \label{zzzz} z_{0,+}&=&-\frac{1}{b-a}\left(1-\frac{\theta_0 t}{\theta_\infty q_0}-\frac{\theta_1 (t-1)}{\theta_\infty (q_0-1)}\right)\left(1-\frac{\theta_0 t}{\theta_\infty q_0}+\frac{\theta_1 (t-1)}{\theta_\infty (q_0-1)}\right)\cr
z_{0,+}&=&-\frac{t}{b-a}\left(1-\frac{\theta_0}{\theta_\infty q_0}-\frac{\theta_t (t-1)}{\theta_\infty (q_0-t)}\right)\left(1-\frac{\theta_0}{\theta_\infty q_0}+\frac{\theta_t (t-1)}{\theta_\infty (q_0-t)}\right)\cr
z_{1,+}&=&\frac{1}{b-a}\left(1+\frac{\theta_0 t}{\theta_\infty q_0}+\frac{\theta_1 (t-1)}{\theta_\infty (q_0-1)}\right)\left(1-\frac{\theta_0 t}{\theta_\infty q_0}+\frac{\theta_1 (t-1)}{\theta_\infty (q_0-1)}\right)\cr
z_{1,+}&=&-\frac{t-1}{b-a}\left(1-\frac{\theta_1}{\theta_\infty(q_0-1)}+\frac{t\theta_t}{\theta_\infty(q_0-t)}\right)\left(1-\frac{\theta_1}{\theta_\infty(q_0-1)}-\frac{t\theta_t}{\theta_\infty(q_0-t)}\right)\cr
z_{t,+}&=&\frac{t}{b-a}\left(1+\frac{\theta_0}{\theta_\infty q_0}+\frac{\theta_t (t-1)}{\theta_\infty(q_0-t)}\right)\left(1-\frac{\theta_0}{\theta_\infty q_0}+\frac{\theta_t (t-1)}{\theta_\infty(q_0-t)}\right)\cr
z_{t,+}&=&\frac{t-1}{b-a}\left(1+\frac{\theta_1}{\theta_\infty(q_0-1)}+\frac{\theta_t t}{\theta_\infty (q_0-t)}\right)\left(1-\frac{\theta_1}{\theta_\infty(q_0-1)}+\frac{\theta_t t}{\theta_\infty (q_0-t)}\right).
\eea 

We now regroup these results for the computation of $F^{(0)}$ given by \eqref{F0P6}. We first observe that the terms involving $\theta_i^2$'s are given by
\begin{center}
\begin{tabular}{ |c | c | }
\hline
Term for $\theta_0^2$& $\frac{1}{8}\ln\left( \frac{16 \,\theta_0^4 t^4}{\theta_\infty^4q_0^4\underset{\pm}{\prod}\left(1\pm \frac{\theta_0 t}{\theta_\infty q_0}\pm \frac{\theta_1(t-1)}{\theta_\infty(q_0-1)}\right)}\right)$\\
\hline
Term for $\theta_1^2$& $\frac{1}{8}\ln\left( \frac{16 \,\theta_1^4 (t-1)^4}{\theta_\infty^4(q_0-1)^4\underset{\pm}{\prod}\left(1\pm \frac{\theta_0 t}{\theta_\infty q_0}\pm \frac{\theta_1(t-1)}{\theta_\infty(q_0-1)}\right)}\right)$\\
\hline
Term for $\theta_t^2$& $\frac{1}{8}\ln\left( \frac{16\, \theta_t^4 t^4(t-1)^4}{\theta_\infty^4(q_0-t)^4\underset{\pm}{\prod}\left(1\pm \frac{\theta_0 t}{\theta_\infty q_0}\pm \frac{\theta_1(t-1)}{\theta_\infty(q_0-1)}\right)}\right)$\\
\hline
Term for $\theta_\infty^2$&$-\frac{1}{8}\ln\left(\frac{1}{16}\underset{\pm}{\prod}\left(1\pm \frac{\theta_0 t}{\theta_\infty q_0}\pm \frac{\theta_1(t-1)}{\theta_\infty(q_0-1)}\right)\right)$\\
\hline
\end{tabular}
\end{center}
Additionally, we can obtain the cross-terms (we looked for the expression minimizing the appearance of $\theta_t$ in order to match it more easily with \eqref{dtau0P6}, and tried to obtain a symmetric expression in $(\theta_0,\theta_1,\theta_t)$, one can easily take each contribution symmetrical relatively to these variables using \eqref{zzzz}):
\begin{center}
\begin{tabular}{ |c | c | }
\hline
Term for $\theta_0\theta_\infty$& $\frac{1}{4}\ln \frac{\left(1-\frac{\theta_0 t}{\theta_\infty q_0}-\frac{\theta_1 (t-1)}{\theta_\infty (q_0-1)}\right)\left(1-\frac{\theta_0 t}{\theta_\infty q_0}+\frac{\theta_1 (t-1)}{\theta_\infty (q_0-1)}\right)}{\left(1+\frac{\theta_0 t}{\theta_\infty q_0}+\frac{\theta_1 (t-1)}{\theta_\infty (q_0-1)}\right)\left(1+\frac{\theta_0 t}{\theta_\infty q_0}-\frac{\theta_1 (t-1)}{\theta_\infty (q_0-1)}\right)} $\\
\hline
Term for $\theta_1\theta_\infty$& $\frac{1}{4}\ln \frac{\left(1-\frac{\theta_0 t}{\theta_\infty q_0}-\frac{\theta_1 (t-1)}{\theta_\infty (q_0-1)}\right)\left(1+\frac{\theta_0 t}{\theta_\infty q_0}-\frac{\theta_1 (t-1)}{\theta_\infty (q_0-1)}\right)}{\left(1+\frac{\theta_0 t}{\theta_\infty q_0}+\frac{\theta_1 (t-1)}{\theta_\infty (q_0-1)}\right)\left(1-\frac{\theta_0 t}{\theta_\infty q_0}+\frac{\theta_1 (t-1)}{\theta_\infty (q_0-1)}\right)} $\\
\hline
Term for $\theta_t\theta_\infty$& $\frac{1}{4}\ln \frac{ \left(1+\frac{\theta_0}{\theta_\infty q_0}+\frac{\theta_t (t-1)}{\theta_\infty(q_0-t)}\right)\left(1-\frac{\theta_0}{\theta_\infty q_0}+\frac{\theta_t (t-1)}{\theta_\infty(q_0-t)}\right)}{\left(1-\frac{\theta_0}{\theta_\infty q_0}-\frac{\theta_t (t-1)}{\theta_\infty(q_0-t)}\right)\left(1+\frac{\theta_0}{\theta_\infty q_0}-\frac{\theta_t (t-1)}{\theta_\infty(q_0-t)}\right)}  $\\
\hline
Term for $\theta_0\theta_1$&$\frac{1}{4}\ln\left(- \frac{\left(1+\frac{\theta_0 t}{\theta_\infty q_0}-\frac{\theta_1 (t-1)}{\theta_\infty (q_0-1)}\right)\left(1-\frac{\theta_0 t}{\theta_\infty q_0}+\frac{\theta_1 (t-1)}{\theta_\infty (q_0-1)}\right)}{\left(1+\frac{\theta_0 t}{\theta_\infty q_0}+\frac{\theta_1 (t-1)}{\theta_\infty (q_0-1)}\right)\left(1-\frac{\theta_0 t}{\theta_\infty q_0}-\frac{\theta_1 (t-1)}{\theta_\infty (q_0-1)}\right)} \right) $\\
\hline
Term for $\theta_0\theta_t$& $\frac{1}{4}\ln \left(- \frac{\left(1+\frac{\theta_0 t}{\theta_\infty q_0}+\frac{\theta_1 (t-1)}{\theta_\infty (q_0-1)}\right)\left(1-\frac{\theta_0 t}{\theta_\infty q_0}-\frac{\theta_1 (t-1)}{\theta_\infty (q_0-1)}\right)}{\left(1+\frac{\theta_0 t}{\theta_\infty q_0}-\frac{\theta_1 (t-1)}{\theta_\infty (q_0-1)}\right)\left(1-\frac{\theta_0 t}{\theta_\infty q_0}+\frac{\theta_1 (t-1)}{\theta_\infty (q_0-1)}\right)} \right) $\\
\hline
Term for $\theta_1\theta_t$& $\frac{1}{4}\ln\left(- \frac{\left(1+\frac{\theta_1}{\theta_\infty (q_0-1)}-\frac{\theta_t t}{\theta_\infty(q_0-t)}\right)\left(1-\frac{\theta_1}{\theta_\infty (q_0-1)}+\frac{\theta_t t}{\theta_\infty(q_0-t)}\right) }{\left(1+\frac{\theta_1}{\theta_\infty (q_0-1)}+\frac{\theta_t t}{\theta_\infty(q_0-t)}\right)\left(1-\frac{\theta_1}{\theta_\infty (q_0-1)}-\frac{\theta_t t}{\theta_\infty(q_0-t)}\right)}\right)  $\\
\hline
\end{tabular}
\end{center}

We now have all the ingredients to compute explicitly $F_{\rm VI}^{(0)}$ from \eqref{F0P6} and we find \eqref{F0final}.

It is then easy to verify that
\beq \frac{d}{dt}F_{\rm VI}^{(0)}(t,q_0)=\dot{q}_0\frac{\partial}{\partial q_0}F_{\rm VI}^{(0)}(t,q_0)+\frac{\partial }{\partial t} F_{\rm VI}^{(0)}(t,q_0)=-\frac{d}{dt}\dot{\tau}_0. \eeq

\begin{remark}
Computation of $F_{\rm VI}^{(1)}$ follows the general topological recursion as presented in \cite{EO} (see Definition \ref{def:symp-inv}). 
Computations rapidly become impossible to handle with a standard laptop to simplify expressions. However, we could verify explicitly that the following identify holds:
\beq \frac{d}{dt}F_{\rm VI}^{(1)}(t,q_0)=
\frac{\partial}{\partial t}F_{\rm VI}^{(1)}(t,q_0)+\dot{q}_0\frac{\partial}{\partial q_0}F_{\rm VI}^{(1)}(t,q_0)=-\frac{d}{dt}\tau_{\rm VI}^{(1)}. \eeq
Unfortunately our final expression for $F_{\rm VI}^{(1)}(t,q_0)$ is several pages long and presents no particular interest but for the fact that its derivative recovers $-\frac{d}{dt}\tau_{\rm VI}^{(1)}$. Hence we do not reproduce it here.
\end{remark}

%%%%%%%%%%%%%%%%%%%%%%%%%%%%%%%%%
%%%%%%%%%%%%%%%%%%%%%%%%%%%%%%%%%

\end{document}